\documentclass[11pt,a4paper]{article}
\usepackage{amssymb}\usepackage{graphicx}
\usepackage{amsmath}
\usepackage{tikz}
\usepackage{pgflibraryarrows}
\usepackage{pgflibrarysnakes}

\newcommand\qed{\hfill$\sqcap\kern-7.5pt\hbox{$\sqcup$}$}

\newcommand{\NN}{\mathbb{N}}
\newcommand{\RR}{\mathbb{R}}

\newtheorem{theo}{Theorem}
\newtheorem{prop}[theo]{Proposition}
\newtheorem{lem}[theo]{Lemma}

\newtheorem{rem}[theo]{Remark}

\newcommand{\re}{R}
\renewcommand{\Re}{\re}
\newcommand{\im}{I}
\renewcommand{\Im}{\im}
\newcommand{\beqn}{\begin{equation}}
\newcommand{\eeqn}{\end{equation}}
\newcommand{\bear}{\begin{eqnarray}}
\newcommand{\eear}{\end{eqnarray}}
\newcommand{\bean}{\begin{eqnarray*}}
\newcommand{\eean}{\end{eqnarray*}}

\begin{document}
\title{Local well posedness for a linear coagulation equation.}
\maketitle

\vspace{0.3cm}

\begin{center}
M. Escobedo\footnotemark[1] and J. J. L. Vel\'azquez \footnotemark[2]
\end{center}
\footnotetext[1]{Departamento de Matem\'aticas, Universidad del
Pa{\'\i}s Vasco, Apartado 644, E--48080 Bilbao, Spain. E-mail: {\tt
mtpesmam@lg.ehu.es}} 
\footnotetext[2]{ICMAT (CSIC-UAM-UC3M-UCM) Facultad de Matem\'aticas, Universidad Complutense. E--28040 Madrid, Spain. E-mail~: {\tt
JJ\_Velazquez@mat.ucm.es}}

\vspace{0.7cm}
\section{Introduction}

\setcounter{equation}{0}
\setcounter{theo}{0}

\bigskip

This paper is part of a program to study the well posedness of classical
solutions after the gelation time for a general class of coagulation
equations that behave asymptotically as homogeneous kernels.

The classical coagulation equation reads:
\begin{equation}
\partial_{t}f\left( x,t\right) =\frac{1}{2}\int_{0}^{x}K\left( x-y,y\right)
f\left( x-y,t\right) f\left( y,t\right) dy-\int_{0}^{\infty }K\left(
x,y\right) f\left( x,t\right) f\left( y,t\right) dy   \label{E1}
\end{equation}

This equation can be though as describing the distribution of sizes for a
set of particles with size $x$ that aggregate with particles of size $y,$
independently distributed, with a rate $K\left( x,y\right) .$

It is well known that for kernels $K\left( x,y\right) $ that behave
asymptotically for large $x,y$ as $\left( xy\right) ^{\frac{\lambda}{2}}$
with $1<\lambda<2,$ the solutions of (\ref{E1}) exhibit the phenomenon known
as ``gelation'', that means that the first moment of $f,$ that is formally
preserved for the solutions of (\ref{E1}), is not any longer preserved after
some finite time $t^{\ast},$ due to the fact that a macroscopic fraction of
particles ``escapes '' to infinitely large sizes (cf. \cite{Le}).

\bigskip

A detailed description of the asymptotics of the function $f\left(
x,t\right) $ as $x\rightarrow \infty $ for solutions exhibiting gelation
behaviour is still lacking, except for the case $K\left( x,y\right) =x\cdot y
$ where (\ref{E1}) can be explicitly solved using integral transform (cf. \cite{EZH}, \cite{MP1}, \cite{MP2}) . In
order to obtain more information on the asymptotics of the solutions
of\thinspace\ (\ref{E1}) for more general kernels, we have studied in a
detailed way in \cite{EV} the fundamental solution of the linearized problem
obtained considering the evolution of functions $f$ that are close to $\bar{f%
}\left( x\right) =Ax^{-\frac{3+\lambda }{2}}$ with kernel $K\left(
x,y\right) =\left( x\, y\right) ^{\frac{\lambda }{2}},\;1<\lambda <2.$
The reason for considering the evolution near such power law, is that the
interpretation of this distribution in the particle setting above
corresponds to a continuous transport of particles emanating from the origin
and being transported towards $x=\infty $ at a constant rate (cf. \cite{EV}). This power law was obtained in \cite{vDE} where explicit particular solutions of the discrete coagulation equations yielding loss of mass have been constructed. 

In order to
understand solutions of the coagulation equation (\ref{E1}) yielding a
``flux'' of particles towards infinity, it is natural to linearize around $%
\bar{f}\left( x\right) ,$ in order to clarify how such transport of
particles could take place. We will denote the linearized problem considered
in \cite{EV} as:
\begin{equation}
g_{t}=L\left[ g\right] \;\;,\;\;\;g\left( 0\right) =g_{0}  \label{E2}
\end{equation}
where

\bear
\label {S1Eopele}
L(g)&=&\int_0^{x/2}\left((x-y)^{-3 /2}-x^{-3
/2} \right)y^{\lambda /2}g(y)dy \nonumber \\ \nonumber \\ 
& + &\int_0^{x/2}\left((x-y)^{\lambda /2}g(x-y)-x^{\lambda /2}g(x)
\right)y^{-3/2}dy \nonumber \\ \nonumber \\
& - &x^{-3/2}\int_{x/2}^{\infty }y^{\lambda /2}g(y)dy-2\sqrt 2 x^{(\lambda -1)/2}g(x).
\eear

A technical difficulty that arises in the study of the linearization of (\ref
{E1}) around $\bar{f}\left( x\right) $ is due to the fact that this function
is singular near $x=0.$ This has several relevant consequences. First, the
resulting linearized operator becomes singular near $x=0,$ and as a
consequence, it has regularizing effects that cannot be expected to take
place for the original problem (\ref{E1}). As a matter of fact, the
linearized operator considered in \cite{EV} behaves, locally near a given
value of $x$, as:
\bean
g_{t}=-\left( -D_{xx}\right) ^{\frac{1}{4}}g+\hbox{higher order terms}
\eean

In particular, this problem can be considered a nonlocal parabolic equation
whose generator has Fourier symbol $-\sqrt{2ik}.$ Clearly, the regularity
properties associated to this problem can be expected to be very different
from the ones associated to (\ref{E1}). 

On the other hand, from the physical
point of view, the solution $\bar{f}\left( x\right) $ is associated to the
presence of a constant flux of particles leaving from the origin, as it can
be seen in \cite{EV}. Bounded solutions of (\ref{E1}) do not have such a
constant flux of particles with size $x=0.$

However, in spite of these differences between the linearized problem (\ref
{E2}) and (\ref{E1}) there are good reasons for studying (\ref{E2}) in order
to understand particle fluxes towards infinity for the nonlinear equation (%
\ref{E1}). The main one is that solutions of (\ref{E1}) yielding particle
fluxes towards $x=\infty$ can be expected to behave as one of the solutions $%
\bar{f}$ as indicated above for each given time. Moreover, the problem (\ref
{E2}) can be solved explicitly using the methods in \cite{BZ} due to its good properties under rescalings. Moreover  it is possible to derive detailed estimates of the
corresponding fundamental solution in all the regions of the space-time $%
\left( x,t\right)$ (cf. \cite{EV}).

Nevertheless, in order to avoid the shortcomings of (\ref{E2}) as an
approximation of (\ref{E1}), it would be more convenient to study the
linearization of (\ref{E1}) near a smooth bounded function $f_{0}\left(
x\right) $ that behaves asymptotically as $Ax^{-\frac{3+\lambda }{2}}$ as $%
x\rightarrow \infty .$ The resulting problem would be:
\begin{equation}
g_{t}=\mathcal{L}\left( g\right) \;\;,\;\;g\left( 0\right) =g_{0}  \label{E3}
\end{equation}
with:
\bean
&&\mathcal{L}\left( g\right) =\int_{0}^{x}\left( x-y\right) ^{\frac{\lambda }{2%
}}f_{0}\left( x-y\right) x^{\frac{\lambda }{2}}g\left( y\right) dy-x^{\frac{%
\lambda }{2}}f_{0}\left( x\right) \int_{0}^{\infty }y^{\frac{\lambda }{2}%
}g\left( y\right) dy-\\
&& \hskip 10cm -x^{\frac{\lambda }{2}}g\left( x\right) \int_{0}^{\infty
}y^{\frac{\lambda }{2}}f_{0}\left( y\right) dy
\eean

Problem (\ref{E3}) is in some sense closer to (\ref{E1}) than (\ref{E2}).
Indeed, (\ref{E3}) does not have regularizing properties at any $x>0$. On the other hand,
bounded solutions of (\ref{E3}) yield a zero flux of particles from the
origin.

Unfortunately, the solution of (\ref{E3}) cannot be obtained explicitly
as it has been made in \cite{EV}. Moreover, to prove even local solvability in
time of (\ref{E3}) is not an easy task due to the presence of the integral
term $\int_{0}^{\infty }y^{\frac{\lambda }{2}}g\left( y\right) dy$. In the
absence of this term the local solvability of (\ref{E3}) could be easily
obtained using a fixed point argument. However, the presence of this
integral term makes this problem much harder to solve.

\bigskip

The key idea that will be used in this paper is to solve (\ref{E3})
approximating it by means of (\ref{E2}) for $x\rightarrow\infty$.  The operator on the right hand side can be thought as an operator
having half derivative at $x=\infty.$ As a consequence, the equation (\ref
{E3}) has some kind of ``smoothing effects'' for $x=\infty.$ The presence of
these regularizing effects is more clear in the equation (\ref{E2}).
Nevertheless, this last equation has regularizing effects for all the values of $x$. Therefore, to approximate the regularizing effects of (\ref{E3}%
) by means of those of (\ref{E2}) is something that must be given a precise
meaning and it will be given in this paper. Regularising effects in kinetic equations with singular kernels have been previously obtained for Boltzmann equation cf. \cite{D} and \cite{V}.

\bigskip

There is another feature of the approximation of (\ref{E3}) using (\ref{E2})
that is worth mentioning. As indicated above, the function $\bar{f}\left(
x\right) $ can be thought as the source of a flux of particles coming from $%
x=0$ that are transported towards $x=\infty.$ On the contrary, the funtion $%
f_{0}\left( x\right) $ does not provide any flux of particles from the
origin, although it is associated to a flux of particles transported towards 
$x=\infty.$ If we rewrite these two functions using the change of variables $%
F=R^{\frac{3+\lambda}{2}}f\left( R\xi\right) ,$ with $\xi$ of order one and $%
R\rightarrow\infty,$ it follows that $\bar{f}\left( x\right) =Ax^{-\frac
{3+\lambda}{2}}$ becomes $\bar{F}\left( \xi\right) =A\xi^{-\frac{3+\lambda }{%
2}}$\ and $f_{0}\left( x\right) $ becomes $F_{0}\left( \xi\right) =R^{\frac{%
3+\lambda}{2}}f_{0}\left( R\xi\right) .$ Notice that $F_{0}\left( \xi\right)
\rightarrow\bar{F}\left( \xi\right) $ as $R\rightarrow\infty,$ for all $%
\xi>0.$ Such a convergence fails for $\xi\rightarrow0$ or, more precisely,
for $x$ of order one. Actually that is the region where (\ref{E3}) cannot be
approximated by (\ref{E2}). This region can be considered as containing a
``boundary layer'' where the boundedness of $f_{0}$ plays a role, and where
the absence of particle fluxes and regularizing effects for (\ref{E3}) are
seen. The analysis of this paper can be thought as the development of the
mathematical techniques to handle such a boundary layer effects, as well as
the proof of the fact that the dynamics of (\ref{E3}) can be approximated by
means of the singular problem (\ref{E2}) at least for times of order one.

Let us remark that to solve the problems (\ref{E2}),\ (\ref{E3}) is
equivalent to the solution of suitable problems with sources and vanishing
initial data. Indeed, suppose that $\tilde{g}=g\left( x,t\right) ,$ is a
smooth function satisfying $\tilde{g}\left( x,0\right) =g_{0}\left( x\right)
.$ Let us define $g=h-\tilde{h}.$ Then:
\begin{equation}
h_{t}=L\left[ h\right] +\tilde{\mu}\;\;,\;\;\;h\left( 0\right) =0  \label{E5}
\end{equation}
\begin{equation}
h_{t}=\mathcal{L}\left( h\right) +\mu \;\;,\;\;h\left( 0\right) =0
\label{E6}
\end{equation}
with $\tilde{\mu}=L\left[ \tilde{g}\right] -\tilde{g}_{t},\;\mu =L\left[ g%
\right] -g_{t}.\;$

\bigskip 

\bigskip 

The method that we will use in this paper to solve (\ref{E3}) makes use of a
classical continuation method. More precisely, we will embed  (\ref{E6}) into
the family of problems:
\begin{equation}
h=\left( 1-\theta \right) L\left[ h\right] +\theta \mathcal{L}\left(
h\right) +\mu \;\;,\;\;\;h\left( 0\right) =0\;\;,\;\;\theta \in \left[ 0,1%
\right]   \label{S8E10-2}
\end{equation}

The problem (\ref{S8E10-2}) can be explicitly solved for $\theta =0$ using
the fundamental solution in \cite{EV}. Suppose that (\ref{S8E10-2}) can be solved
for $\theta =\theta ^{\ast }\in \left[ 0,1\right) .$ We will show that (\ref
{S8E10-2}) can be solved for $\theta >\theta ^{\ast }$ with $\left( \theta
-\theta ^{\ast }\right) $ small enough. This will allow to extend by
continuity the solution of (\ref{S8E10-2}) from $\theta =0$ to $\theta =1,$
and then to obtain a solution of (\ref{E6}).

Similar continuity methods have been extensively used in the analysis of PDE's (cf. \cite{GT, LU, vW}). 

\bigskip 

The plan of the paper is the following. In Section 2 we define our functional framework and state the main results of this paper. In Section 3 we obtain some technical auxiliary results that are needed in the proofs of the interior regularity estimates for the operator $\mathcal L$. These are later obtained in Section 4 and will provide the essential smoothness required in this paper.
Section 5 contains some estimates that provide a precise meaning to the approximation of the operator $\mathcal{L}$ by means of $%
L$. Finally, Section 6 provides the proof of the main
result of the paper, namely the local well-posedness of (\ref{E6}).
\bigskip
\section{Functional Framework and statement of the main results.}
\setcounter{equation}{0}
\setcounter{theo}{0}
We introduce now the set of initial data, $f_0\in \textbf{C}^{1, \gamma}(\RR^+)$, $\gamma \in (0, 1)$, that we shall consider in this paper.
We will assume that the function $f_0$ is close to the function
$Ax^{-(3+\lambda)/2)}$ for some constant $A\in \RR$. To this end define
\bear
\label{defhachecero}
h_0(x)=f_0(x)-Ax^{-(3+\lambda)/2)} \xi (x)
\eear
where $\xi$ is a smooth cutoff function such that $\xi(x)=1$ for $x\ge 1$ and $\xi(x)=0$ if $0\le x \le 1/2$.
We then require in all this paper that for some positive constants $B$ and $\delta$, the following condition holds :
\bear
\label{S2Ecotasregdato}
y^{\frac{3+\lambda}{2}+\delta}|h_0(y)|+y^{\frac{3+\lambda}{2}+1+\delta}|h'_0(y)|
+\sup_{R\ge 1}R^{\frac{3+\lambda}{2}+\gamma+\delta}[h_0']_{C^{\gamma}[R/2, 2R]}
+[h_0']_{C^{\gamma}(0, 1)}\le B.
\eear
All the estimates in the rest of paper will depend on the constants $A, B, \gamma$ and $\delta$. For the sake of shortedness this dependence will not be indicated.\\ \\

The operator ${\mathcal L}$ is given by:
\bear
\label{S7E1}
{\mathcal L}(g)=\int_0^x(x-y)^{\lambda/2}f_0(x-y)\, y^{\lambda/2}\, g(y)\, dy-x^{\lambda/2}\, f_0(x)\, \int_0^\infty y^{\lambda/2}\, g(y)\, dy\\-x^{\lambda/2}\, g(x)\, \int_0^\infty y^{\lambda/2}\, f_0(y).
\eear
Our first goal is to study the solutions of the Cauchy problem:
\bear
&&\frac{\partial h}{\partial \tau}={\mathcal L}(h)+\mu(\tau, x) \label{S8E9}
\\ &&h(0, x)=0 \label{S8E10}
\eear
for some initial data $h_0$ and non homogeneous term  $\mu$. \\ 
We shall also use repeatedly  the following ``localised version'' of this equation. To this end, for all $R>1$ fixed, let $\chi(x)\in \textbf{C}_0^\infty (0, +\infty)$ be such that:
\bear
\label{S8Echi}
\chi(x)=\left\{
\begin{array}{l}
1\quad \hbox{if}\,\,x\in \left(R-\frac{R}{8}, R+\frac{R}{8}\right)\,,\\ \\
0\quad \hbox{if}\,\,\,x\not\in \left(R-\frac{R}{4}, R+\frac{R}{4}\right)\, .
\end{array}
\right.
\eear
If we multiply the equation (\ref{S8E9}) by $\chi(x)$ and call $\tilde g = \chi(x)\, g(x)$ we obtain:
\bear
\frac{\partial \tilde g}{\partial t} & = & \mathcal {L}(\tilde g)+ \mathcal{R}(g) \label{S8E9loc}\\
 \mathcal{R}(g) & = & \chi(x)\int_0^{x/2}\left((x-y)^{\lambda/2}f_0(x-y)-x^{\lambda/2}f_0(x) \right)y^{\lambda/2}g(y)dy
\nonumber\\
&-&x^{\lambda/2}\tilde g(x)\int_{x/2}^{\infty}y^{\lambda/2}f_0(y)dy
- x^{\lambda/2}f_0(x)\chi(x)\int_{x/2}^{\infty}y^{\lambda/2}g(y)dy\nonumber \\
& + & \int_0^{x/2}\left(\chi(x)-\chi(x-y) \right)\,(x-y)^{\lambda/2}g(x-y)y^{\lambda/2}f_0(y)\, dy.
\eear

For any $p\ge 1$, $L^p$ will denote the usual Lebesgue space. For any $\sigma>0$ and any interval $I\subset (0, +\infty)$ we denote $H^\sigma(I)$ the usual Sobolev space $W^{\sigma, 2}(I)$. The corresponding norms will be denoted $||\cdot||_{L^p}$ and  $||\cdot||_{H^\sigma}$.
When dealing with functions depending on variables $x$ and $t$ we will write  $H_x^\sigma$ or $L^p_t$ in order to indicate the argument with respect to which the norm is taken. 

In order to define the functional spaces that will be needed we first introduce
\bear
N_{\infty}(h;\, t_0, R)=\left(R^{\frac{\lambda-1}{2}}\int_{t_0}^{\min(t_0+R^{-(\lambda-1)/2}, T)}||h(t)||^2_{L^\infty(R/2, 2R)}dt\right)^{1/2} \label{S3TnormaNinfty}
\eear

\bear
N_{2;\, \sigma}(h;\, t_0, R)=\left(R^{\frac{\lambda-1}{2}+2\sigma-1}\int_{t_0}^{\min(t_0+R^{-(\lambda-1)/2}, T)}||D_x^{\sigma}h(t)||^2_{L^2(R/2, 2R)}dt\right)^{1/2} \label{S3TnormaNsigma}
\eear

\bear
M_\infty (h;\, R)=\left(\int_0^{T}||h(t)||^2_{L^\infty(R/2, 2R)} dt\right)^{1/2} \label{S3TnormaMinfty}
\eear
\bear
M_{2; \sigma} (h;\, R)=\left(R^{2\sigma-1}\int_0^{T}||D^\sigma_x h(t)||^2_{L^2(R/2, 2R)} dt\right)^{1/2} \label{S3TnormaMsigma}
\eear
Then, we define the spaces:
\bear
&&||f||_{X_{q, p}(T)}=\sup_{0<R<1}R^{q}\, M_\infty (f; R) +\sup_{0\le t_0\le T}\sup_{R\ge 1}
 R^p\, N_\infty (f; t_0, R) \\
&&X_{p, q}(T)=\left\{f;  ||f||_{X_{q, p}(T)}<\infty \right\}\\
&& ||f||_{Y^\sigma_{q,p} (T)}=\sup_{0<R\le 1}R^q\, M_{2;\, \sigma}(f; R)+
\sup_{0\le t_0\le T}\sup_{R\ge 1}
 R^p\, N_{2;\, \sigma} (f; t_0, R)\\
&&Y^{\sigma}_{q, p}(T)=\left\{f;  ||f||_{Y^\sigma_{q, p}(T)}<\infty \right\}.
\eear

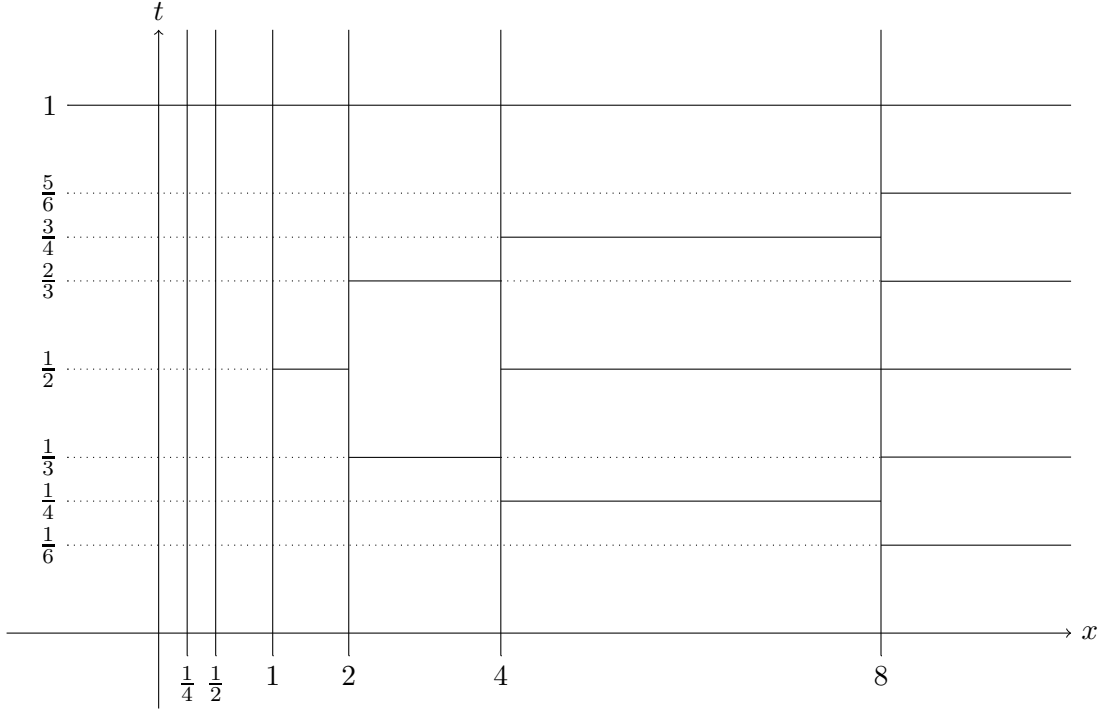
\begin{figure}
\begin{tikzpicture}
\draw[->] (-7,-5) -- (7,-5) node[right]{$x$};
\draw[->] (-5,-6) -- (-5,3) node[above]{$t$};
\draw[dotted](-6.2,-2.67) -- (-2.5, -2.67);
\draw[dotted](-0.5, -2.67)--(4.5, -2.67);
\draw(-6.2,-2.67) -- (-6.2, -2.67) node[left]{$\frac{1}{3}$};
\draw(-2.5,-2.67) -- (-0.5,-2.67);
\draw[dotted](-6.2, -0.33)--(-2.5, -0.33);
\draw[dotted](-0.5, -0.33)--(4.5, -0.33);
\draw(-6.2, -0.33)--(-6.2, -0.33) node[left]{$\frac{2}{3}$};
\draw(-2.5,-0.33) -- (-0.5,-0.33);
\draw(-6.2,2) -- (7,2);
\draw(-6.2,2) --(-6.2,2)node[left]{$1$};
\draw (-4.625,-5.3) -- (-4.625,3);
\draw (-4.625,-5.3) -- (-4.625,-5.3) node[below]{$\frac{1}{4}$};
\draw (-4.25,-5.3) -- (-4.25,3);
\draw (-4.25,-5.3) -- (-4.25,-5.3) node[below]{$\frac{1}{2}$};
\draw (-3.5,-5.3) -- (-3.5,-5.3) node[below]{$1$};
\draw(-3.5,-5.3) -- (-3.5,3);
\draw (-2.5,-5.3) -- (-2.5,3);
\draw (-2.5,-5.3)--(-2.5,-5.3)node[below]{$2$};
\draw (-0.5,-5.3) -- (-0.5,3);
\draw (-0.5,-5.3)--(-0.5,-5.3)node[below]{$4$};
\draw (4.5,-5.3) -- (4.5,3);
\draw (4.5,-5.3)--(4.5,-5.3)node[below]{$8$};
\draw[dotted](-6.2,-1.5)--(-3.5,-1.5);
\draw(-6.2,-1.5) -- (-6.2,-1.5)node[left]{$\frac{1}{2}$};
\draw(-3.5,-1.5) -- (-2.5,-1.5);
\draw(-0.5,-1.5) -- (7,-1.5);
\draw(-0.5,-3.25) -- (4.5,-3.25);
\draw[dotted](-6.2,-3.25)--(-0.5,-3.25);
\draw(-6.2,-3.25) -- (-6.2,-3.25)node[left]{$\frac{1}{4}$};
\draw(-0.5,0.25) -- (4.5,0.25);
\draw[dotted](-6.2,0.25) -- (-0.5,0.25);
\draw(-6.2,0.25) -- (-6.2,0.25)node[left]{$\frac{3}{4}$};
\draw[dotted](-6.2, 0.834)--(4.5, 0.834);
\draw(-6.2, 0.834)--(-6.2, 0.834) node[left]{$\frac{5}{6}$};
\draw(4.5,0.834) -- (7,0.834);
\draw(4.5,-0.334) -- (7,-0.334);
\draw(4.5,-3.834) -- (7,-3.834);
\draw[dotted](-6.2, -3.834)--(4.5, -3.834);
\draw(-6.2, -3.834) -- (-6.2, -3.834)node[left]{$\frac{1}{6}$};
\draw(4.5,-2.666) -- (7,-2.666);
\end{tikzpicture}
\caption{Domain decomposition for  $\lambda=1.5$, $t_0=n\, R^{-(\lambda-1)/2}$ and $R=2^n$.}
\end{figure}

We also use the following norms defined for functions $\varphi=\varphi(x)$:
\bean
\label{S8NormX}
|||\varphi|||_{q, p}=\sup_{0\le x \le 1}\{x^{q}|\varphi(x)|\}+\sup_{x>1}\{x^{p}|\varphi(x)|\}\, 
\eean
and the next one defined for functions $\psi=\psi(\cdot, t)$ and any $T>0$:
\bean
&&|||\psi|||_\sigma=\sup_{0\le t \le T}||| \psi(\cdot, t)|||_{3/2, (3+\lambda)/2}+|| \psi ||_{Y^\sigma_{3/2, (3+\lambda)/2}}.
\eean
We define the space $\mathcal E_{T; \sigma} $ as
\bean
 \mathcal E_{T; \sigma} =\{ f;  |||f|||_\sigma<\infty \}
\eean
endowed with the norm $|||\cdot |||_\sigma$. We assume in all the paper that $\sigma$ is a fixed number satisfying 
\bear
\sigma \in (1/2, 1).
\eear
In order to discharge the notation we will not write explicitly the dependence of the space $\mathcal E_{T; \sigma}$ and the norm $|||\cdot|||_{\sigma}$ on $\sigma$ unless it is needed. We will then write:
\bear
\mathcal E_{T;\, \sigma} \equiv \mathcal E_{T}; \,\,\,\,\,|||\cdot|||_{\sigma} \equiv |||\cdot|||
\eear
\\
We introduce a functional seminorm that measures in a natural way the regularising effect of the operator $\mathcal L$ as $x\to \infty$. Consider a cutoff function $\eta(x)$ defined as
\bear
\label{S8Eeta}
\eta(x)=\left\{
\begin{array}{l}
1\quad \hbox{if}\,\,x\in \left(\frac{3}{4}, \frac{5}{4}\right)\,,\\ \\
0\quad \hbox{if}\,\,\,x\not\in \left(\frac{1}{8}, 4\right)\, .
\end{array}
\right.
\eear
Given then a function $f\in \mathcal E_{T, \sigma}$, for all $R>0$ and $t_0\in [0, T]$ we define
\bear
&&F_{R, t_0}(X, \tau)=\eta (X)\, f\left(R\, X, t_0+\tau\, R^{-(\lambda-1)/2} \right) \label{S2lafuncionErre}\\
&&[f]=\sup_{R\ge 1}\sup_{0\le t_0\le T}R^{(3+\lambda)/2}\times \\
&&\hskip 0.4cm \times
\left(\int_{t_0}^{\min(t_0+R^{-(\lambda-1)/2}, T)}
 \int_\RR |\widehat F_{R, t_0}(k, \tau)|^2\,\left(1+|k|^{2\,\sigma}
\min \{|k|,\, R \}\right)\, dk\, dt
\right)^{1/2}\hskip -0.5cm . \nonumber \\
\label{S2Elcorchete}
\eear
The main results of this paper are the following
\\
\begin{theo}
\label{MainTheo}
For any $\sigma\in (1/2, 1)$, $\delta>0$ and for any $f_0$ satisfying (\ref{defhachecero}) and (\ref{S2Ecotasregdato}), there exists  $T>0$ such that for all 
$\mu \in Y^\sigma_{3/2,\, 2+\delta}$ the Cauchy problem (\ref{S8E9}) (\ref{S8E10}) has a unique solution $h$ in $\mathcal E_{T;\, \sigma}$. Moreover,
\bean
|||h|||\le C\,||\mu||_{ Y^\sigma_{3/2,\, 2+\delta}}
\eean
for some positive constant $C$ depending on $T$, $\sigma$, $\delta$ as well as $A$, $B$ and $\gamma$ in (\ref{defhachecero}) and (\ref{S2Ecotasregdato}) but not on $\mu$.
\end{theo}

\begin{theo}
\label{Main2Theo}
For any $\sigma\in (1/2, 1)$, $\delta>0$ and for any $f_0$ satisfying (\ref{defhachecero}) and (\ref{S2Ecotasregdato}), the solution of the Cauchy problem (\ref{S8E9}) (\ref{S8E10}) satisfies
\bean
[h]\le C ||\mu||_{ Y^\sigma_{3/2,\, 2+\delta}}
\eean
for some positive constant $C$ depending on $T$, $\sigma$, $\delta$ as well as $A$, $B$ and $\gamma$ in (\ref{defhachecero}) and (\ref{S2Ecotasregdato}) but not on $\mu$.
\end{theo}
Theorem \ref{Main2Theo} is a regularising effect for the solutions of (\ref{S8E9}) (\ref{S8E10}). The operator $\mathcal L$ can be thought as half a derivative as $x\to \infty$. However the solutions of (\ref{S8E9}), (\ref{S8E10}) do not gain any regularity for any finite value of $x$. The norm (\ref{S2Elcorchete}) can be thought heuristically as a measure of 
\bean
R^{\frac{3+\lambda}{2}}\left| \frac{f(x+\varepsilon)-f(x)}{\varepsilon^{1/2}}\right|
\eean
with $\varepsilon \ge 1/R$. Theorem \ref{Main2Theo} then states that for the function $h$ this quantity may be estimated by $||\mu||_{ Y^\sigma_{3/2,\, 2+\delta}}$.

We end this Section with two warning remarks. The first one is that all along the paper we are going to use freely the letters $I_1, I_2, \cdots$ and $J_1, J_2, \cdots$ to denote different integrals. These letters  will be used  in different arguments. They will be used consistently within each  argument. The second remark is that, in several arguments, we shall need to extend to a given 
interval suitable regularity estimates that have already been proved in smaller intervals. This is done following a standard and well known procedure involving decomposition of the identity and is not detailed in the paper.

\section{Interior regularity estimates for $\mathcal L$. Some technical results.}

\setcounter{equation}{0}
\setcounter{theo}{0}

In order to study the regularity properties of the solutions to the equation (\ref{S8E10-2}) we define, for all $\varepsilon $ such that $0\le \varepsilon \le 1$:
\bear
\label{S9E10000}
T_{\varepsilon, R}(f)(x) & = & \int_0^{\infty}\left(f(x)-f(x-y) \right)\Phi(y, R, \varepsilon)dy\\
\label{S9E10000Phi}
\Phi(y, R, \varepsilon)& = & \frac{\varepsilon}{y^{3/2}}+(1-\varepsilon)\, R^{(3+\lambda)/2}\, y^{\lambda/2}\, f_0(R y)
\\
\label{S9E10000bis}
\left(M_{\lambda/2}f\right)(x) & = & x^{\frac{\lambda}{2}}\, f(x).
\eear
This family of operators provides an interpolation between a half derivative operator (for $\varepsilon =1$) and the operator that we are interested in (for $\varepsilon=0$).  We will also use the operator
\bear
\label{S2Eoplambda}
\widehat{\Lambda\, \varphi}(\xi)=-\sqrt {2 \pi}\, |\xi|^{1/2}\, \widehat \varphi (\xi).
\eear 
We study  now the interior regularity properties of the linear semigroup generated by the operator $T_{\varepsilon, R} \circ M_{\lambda/2}$. 

\begin{theo}
\label{S8T3-101}
(i) Suppose that $Q  \in L^2_t(0, 1;H_x^\sigma(1/2, 2))$, $P  \in L^2_t(0, 1;H_x^{\sigma-1/2}(1/2, 2))$ with $\sigma \in (1/2, 2)$, $\kappa \in (0, 1]$ and  $f\in L^\infty((1/4, 2)\times (0, 1))\cap  L^2(0, 1; H^{1/2}(1/4, 2))\cap H^1(0, 1; L^2(1/4, 2))$ is such that
$f=0$ if $x<1/8$ or $x>4$ and 
 satisfies
\bear
\label{S8T3-101E1}
\frac{\partial f}{\partial t}=\kappa\, T_{\varepsilon, R}\left(M_{\lambda/2}\, f\right)+Q+P
\eear
for all $x\in (1/4, 2)$, $t\in (0, 1)$  and $f(x, 0)=0$. Then:
\bear
\label{S8T3-10198756}
||f||_{L^2_t(0, 1;H^\sigma_x(3/4, 5/4))}\le C\left(||Q||_{ L^2_t(0, 1;H_x^\sigma(1/2, 2))}+\frac{1}{\varepsilon\, \kappa}
||P||_{ L^2_t(0, 1;H_x^{\sigma-1/2}(1/2, 2))}+\right.\nonumber \\
\left.||f||_{L^\infty((1/4, 2)\times (0, 1))} \right)
\eear
for some positive constant $C$ independent of $\varepsilon$ and $R$.\\ \\
(ii) Suppose moreover that, for some $T_{max}>0$,  $Q  \in L^2_t(0, T_{max}; H_x^\sigma(1/2, 2))$, $P  \in L^2_t(0, T_{max};H_x^{\sigma-1/2}(1/2, 2))$,  $f\in L^\infty((1/4, 2)\times (0, T_{max}))\cap C^1_t(0, T_{max}; H^{1/2}_x(1/4, 2)))$ is such that $f=0$ if $x<1/8$ or $x>4$ and  satisfies
\bear
\label{S8T3-101E143}
\frac{\partial f}{\partial t}=T_{\varepsilon, R}\left(M_{\lambda/2}\, f\right)+Q+P-a(x, t)\, f,\qquad x\in (1/4, 2), \, t>0)\\
f(x, 0)=0\label{S8T3-101E143MAs}
\eear
for some function $a\in L^\infty({0, T_{max}; H^\sigma(1/2, 2)})$, $a\ge A>0$. Then, for all $t\in [0, T_{max}-1]$:
\bear
&&\sup_{0\le T\le T_{max}}\left(\int_{T}^{\min(T+1, T_{max})}||f(t)||^2_{H^{\sigma}(3/4, 5/4)}\, dt\right)^{1/2}\le \label{S8T3-101E143bis}\\
&&\hskip 1cm  C\sup_{0\le T\le T_{max}}\left(\int_{T}^{\min(T+1, T_{max})}||Q(t)||^2_{H^{\sigma}(
1/2,2)}\, dt\right)^{1/2} \nonumber\\
&&\hskip 1.5cm+\frac{C}{\varepsilon}
\sup_{0\le T\le T_{max}}\left(\int_{T}^{\min(T+1, T_{max})}||P(t)||^2_{H^{\sigma-1/2}(
1/2,2)}\, dt\right)^{1/2} \nonumber\\
&&\hskip 2.3cm+C\,||f||_{L^\infty((1/4, 2)\times (0, T_{max}))}\nonumber
\eear
(iii) Suppose that for some $T_{max}>0$,  $Q  \in L^2_t(0, T_{max}; H_x^\sigma(1/2, 2))$,  $f\in L^\infty((1/4, 2)\times (0, T_{max}))\cap C^1_t(0, T_{max}; H^{1/2}_x(1/4, 2)))$ is such that $f=0$ if $x<1/8$ or $x>4$ and  satisfies (\ref{S8T3-101E143})  (\ref{S8T3-101E143MAs}) with $P=0$ and $\varepsilon=0$. Then
\bear
\label{S8T3-101E143bisbis}
&&\left(\int_{T}^{\min(T+1, T_{max})}\int_{\RR}|\widehat F(k, t)|^2|k|^{2\sigma}\min\{|k|, R\}dk\right)^{1/2} \le  \nonumber \\
&&C\sup_{0\le T\le T_{max}}\left(\int_{T}^{\min(T+1, T_{max})}||Q(t)||^2_{H^{\sigma}(
1/2,2)}\, dt\right)^{1/2} 
+C\, ||f||_{L^\infty((1/4, 2)\times (0, T_{max}))} \nonumber \\
\eear
where $F(x, t)=\eta(x)\, f(x, t)$, $\eta$ is defined in (\ref{S8Eeta}) and $C$ is independent of $R$.
\end{theo}

\begin{rem} It will be used repeatedly in the paper that the condition $\sigma>1/2$ ensures that the space $H_x^{\sigma}$ is an algebra under the multiplication.
\end{rem}

\begin{rem}Roughly speaking, the part (i) of Theorem \ref{S8T3-101} provides regularity estimates for 
times $t$ of order one. While part (ii) provides regularity estimates for arbitrary long times. 
It is important to notice that in the Theorem \ref{S8T3-101}, the time $T_{max}$ can be   arbitrarily large.

\end{rem}

The proof of Theorem \ref{S8T3-101} is based on the classical freezing coefficients  method that reduces the problem to the case of a constant coefficient operator. Let us then define, for all $x_0 \in \RR^+$ the operator:
\bear
\label{S2EoperadorEse}
S_{\varepsilon, R}(t)=\exp\left[ t\, x_0^{\frac{\lambda}{2}}\, T_{\varepsilon, R}\right].
\eear
We also define the operators:
\bear
&&\widehat {T_1(\varphi)}(\xi)=\Re e W(\xi, \varepsilon, R)\widehat{\varphi}(\xi)\label{S2EoperadorTuno}\\
&&\widehat {T_2(\varphi)}(\xi)=i\, \Im m W(\xi, \varepsilon, R)\widehat{\varphi}(\xi)\label{S2EoperadorTdos}\\
&&W(\xi, \varepsilon, R)=\int_0^{\infty}\Phi(y, \varepsilon, R)\left(e^{-i\xi y}-1 \right)dy \label{S2EUvedoble}\\
&&\widetilde W(\xi)=\int_0^{\infty}\, f_0(y)\, y^{\lambda/2}\left(e^{-i\xi y}-1 \right)dy \label{S2EUvedobleTilde}
\eear
We now collect several estimates on the operators $T_1$ and $T_2$ which are used in order to obtain bounds on the operator $S_{\varepsilon, R}$.

\begin{lem}
\label{S2TUvedobles} The function $W(\xi, \varepsilon, R)$ defined in (\ref{S2EUvedoble}) may be rewritten as follows:
\bear
W(\xi, \varepsilon, R)= -\varepsilon\sqrt{2}\, \Gamma(1/2)\,(1+i sign (\xi))|\xi|^{1/2}+(1-\varepsilon) \sqrt R\, \widetilde W(\xi/R)\, ,\label{S2EUvedobleTres}
\eear
where the function $\widetilde W$ satisfies:
\bear
&&\Re e \widetilde W \le 0\,\,\hbox{ with}\,\,\,\Re e \widetilde  W=0\,\,\hbox{ if and only if}\,\, \xi=0,\label{S2EUvedobleTildeUno}
\\
&&\lim_{z\to 0}\,\,\frac{\widetilde W (z)}{(1+i sign (z))|z|^{1/2}}=-\sqrt{2}\,\Gamma(1/2)\,,\label{S2EUvedobleTildeDos}
\\
&&\lim_{z\to +\infty} \widetilde W (z) = -\int_0^\infty y^{\lambda/2}f_0(y)\, ,\label{S2EUvedobleTildeTres}
\\
&&
\label{S2cotaderivadauvedobleTilde}
|\widetilde W'(\xi)|\le \frac{C}{1+|\xi|^{1+\gamma}}\,\,\,\hbox{for all}\,\,\xi>0.
\eear
As a consequence of these properties, the function $W$ satisfies:
\bear
&&\Re e W \le 0\,\,\hbox{ with}\,\,\,\Re e W=0\,\,\hbox{ if and only if}\,\, \xi=0, \label{S2EUvedobleUno}
\eear
and is such that, for all $\varepsilon>0$ and $\xi$ fixed,
\bear
\lim_{R\to +\infty} W(\xi, \varepsilon, R)= -\sqrt{2}/2(1+i sign (\xi))|\xi|^{1/2}.\label{S2EUvedobleDos}
\eear

\end{lem}
{\bf Proof of Lemma \ref{S2TUvedobles}.} Using formulas  (\ref{S9E10000Phi}), (\ref{S2EUvedoble}) and (\ref{S2EUvedobleTilde}) properties (\ref{S2EUvedobleTildeUno})--(\ref{S2EUvedobleTildeTres}) follow. In order to prove  (\ref{S2cotaderivadauvedobleTilde})  we may write:
\bean
\widetilde W'(\xi) & = & -i\, \int_0^\infty y^{\lambda/2+1}\, f_0(y)\,e^{-i\, y\, \xi} dy=\frac{i}{\xi}\int_0^\infty \partial _y\left(y^{\lambda/2+1}\, f_0(y) \right)e^{-i\, y\, \xi}\, dy\\
& = & \frac{i}{\xi}\int_0^\infty h(y) e^{-i\, y\, \xi}\, dy\qquad \hbox{with}\,\,h(y)=\partial _y\left(y^{\lambda/2+1}\, f_0(y) \right).
\eean
Writing now
\bean
\int_0^\infty h(y) e^{-i\, y\, \xi}\, dy & = & \sum_{n=0}^\infty
\int_{\frac{2\pi\,n}{\xi}}^{\frac{2\pi\,(n+1)}{\xi}}h(y) e^{-i\, y\, \xi}\, dy\\
& = & \sum_{n=0}^\infty
\int_{\frac{2\pi\,n}{\xi}}^{\frac{2\pi\,(n+1)}{\xi}}
\left(h\left(\frac{2\pi\,n}{\xi} \right)+
{\cal O}\left(\frac{1}{|\xi|^{\gamma}}\frac{1}{(1+|y|)^{3/2+\gamma}} \right) \right) 
\, dy\\
& = & \sum_{n=0}^\infty
\int_{\frac{2\pi\,n}{\xi}}^{\frac{2\pi\,(n+1)}{\xi}}
\left({\cal O}\left(\frac{1}{|\xi|^{\gamma}}\frac{1}{(1+|y|)^{3/2+\gamma}} \right) \right) \, dy\le \frac{C}{ |\xi|^{\gamma}}.
\eean
Finally, properties (\ref{S2EUvedobleUno}) and (\ref{S2EUvedobleDos}) directly follow from (\ref{S2EUvedobleTildeUno})--(\ref{S2cotaderivadauvedobleTilde}).

\qed

We collect now some regularising properties of the semigroups generated by the operators $x_0^{\lambda/2}\, T_1$ and  $S_{\varepsilon, R}$.
\begin{prop}
\label{S2Tsemigrupos} 
For all $\sigma>0$ and $\kappa \in (0, 1]$:
\bear
\label{S2ETsemigrupos1}
\left|\left|\int_0^t S_{\varepsilon, R}(t-s)\, h(s)\, ds\right| \right |^2_{H^{\sigma} (\RR)}\le C\, \int_0^t ||h(s)||^2_{H^{\sigma} (\RR)}\, ds\\
\label{S2ETsemigrupos2}
\int_0^1\left|\left|\int_0^t \kappa \, T_1\, e^{x_0^{\lambda/2}\kappa \, T_1(t-s)}h(s)ds \right|\right|^2_{H^{\sigma}(\RR)}\, dt\le C \int_0^1 ||h(t)||^2_{H^\sigma}(\RR)\, dt\\
\label{S2ETsemigrupos3}
\int_0^1\left|\left|\int_0^t e^{x_0^{\lambda/2}\varepsilon \Lambda(t-s)}h(s)ds \right|\right|^2_{H^{\sigma}(\RR)}\, dt\le 
\frac{C}{\varepsilon ^2} \int_0^1 ||h(t)||^2_{H^{\sigma-1/2}}(\RR)\, dt.
\eear
Moreover, for all $\beta\in (0,1]$ and $\eta$ a $C^\infty$ function of compact support, there exists 
$0<\rho<\min(\sigma, \beta/2)$ such that
\bear
\label{diadic}
||S_{\varepsilon, R}(t)\left[\eta, T_{\varepsilon, R}\right]h||_{H^\sigma (\RR)} \le C\, t^{-\beta}||h||_{H^{\sigma-\rho}(\RR)},
\eear
\bear
\label{diadicdos}
\int_0^1\left|\left| \int_0^t T_1 e^{{x_0^{\lambda/2}}T_1 (t-s)}[\eta,\, T_1]h(s)\, ds\right|\right|^2_{H^\sigma(\RR)}\, ds\le C \int_0^1 ||T_1h||^2_{H^{\sigma-\rho}(\RR)}
\eear
where $C$ denotes a generic positive constant independent of the function $h$ of $R$ and $\varepsilon$ but depending on $\sigma$, $\beta$, $\rho$ and $\eta$.
\end{prop}

In the proof of Proposition \ref{S2Tsemigrupos} we will use the following result.
\begin{lem}
\label{uvedoble}
There exists a positive constant $C$ such that for all $R>1$, $\varepsilon>0$ and $\alpha\ge 0$, $\beta\ge 0$ satisfying $\alpha+\beta=1$:
\bear
\label{S2cotauvedoble}
|W(\xi, \varepsilon, R)-W(z, \varepsilon, R)|\le C\frac{|\xi-z|}{|z|^{\alpha}|\xi|^{\beta}}(1+|W(z, \varepsilon, R)|)^{\alpha}(1+|W(\xi, \varepsilon, R)|)^{\beta}
\eear
for all $z\in \RR$ and $\xi \in \RR$ such that $|z|\ge 1$ and $|\xi|\ge 1$.
\end{lem}
\textbf{Proof of Lemma \ref{uvedoble}.} 
\bean
W(\xi, \varepsilon, R)= -\varepsilon\sqrt{2\, \pi}\,(1+i sign (\xi))|\xi|^{1/2}+(1-\varepsilon) \sqrt R\, \widetilde W(\xi/R).
\eean
The following estimate can be readily obtained studying separately the cases  $sign (\xi)=sign (z)$,  $sign (\xi)=-sign (z)$
\bean
|sign (\xi)|\xi|^{1/2}-sign (z)|z|^{1/2}|\le 2\, \frac{|\xi-z|}{|\xi|^{1/2}+|z|^{1/2}}
\eean
Therefore:
\bean
|\varepsilon\sqrt{2\, \pi}\,(1+i sign (\xi))|\xi|^{1/2}-\varepsilon\sqrt{2\, \pi }\,(1+i sign (z))|z|^{1/2}|
\le C \varepsilon \frac{|\xi-z|}{|\xi|^{1/2}+|z|^{1/2}}.
\eean
Using then $|W(z)|\ge 2\, \sqrt \pi \,\varepsilon\,|z|^{1/2}$ we obtain:
\bean
\frac{|\varepsilon\sqrt{2\, \pi}\,(1+i sign (\xi))|\xi|^{1/2}-\varepsilon\sqrt{2\, \pi }\,(1+i sign (z))|z|^{1/2}|}
{\left(1+|W(z)|\right)^{\alpha}\left(1+|W(\xi)|\right)^{\beta}}\le C\, \frac{|\xi-z|}{|z|^{\alpha}\, |\xi|^{\beta}}.
\eean
In order to prove a similar estimate for $\widetilde W$ we consider the following cases: 
\\

(i) $|\xi|\le 2R$ and $|z|\le 2R$, 
\\

(ii) $|\xi|\ge R/2$ and $|z|\ge R/2$
\\

(iii) $|\xi|\ge 2R$ and $|z|\le R/2$ 
\\

(iv) $|z|\ge 2R$ and $|\xi|\le R/2$.
\\ \\
In the case (i) we have: 
\bean
 C_1 |\xi|^{1/2}\le |\sqrt R\,\, \widetilde W \left(\frac{\xi}{R}\right)|\le C_2  |\xi|^{1/2}\\
  C_1\sqrt |\xi|^{-1/2}\le |\frac{\partial }{\partial \xi} \widetilde W \left(\frac{\xi}{R}\right)|\le C_2|\xi|^{-1/2}
\eean
and similar estimates also for $z$. Defining $g=\sqrt R\,\, \widetilde W \left(\frac{\xi}{R}\right)$ we have by Taylor's theorem:
\bean
|g^2(\xi)-g^2(z)|\le \int_z^{\xi}|g(\eta)||g'(\eta)|\, d\eta\le C\,|\xi-z|.
\eean
Then
\bean
|g(\xi)-g(z)|\le \frac{|\xi-z|}{g(\xi)+g(z)}
\eean
whence:
\bean
\sqrt R \left|\widetilde W \left(\frac{\xi}{R}\right)-\widetilde W \left(\frac{z}{R}\right)\right|\le 
C \frac{|\xi-z|}{|\xi|^{1/2}+|z|^{1/2}}
\eean
and the conclusion follows as above.\\ \\
If condition (ii) holds, suppose first that $sign\, \xi =-sign\, z$. Then, $|\xi-z|=|\xi|+|z|$ and 
\bean
\left|\sqrt R\,\widetilde W \left(\frac{\xi}{R}\right)-\sqrt R\,\widetilde W \left(\frac{z}{R}\right)\right|\le C\, \sqrt R.
\eean
Using Young's inequality:
\bean
\frac{|\xi|+|z|}{|\xi|^{\beta}|z|^{\alpha}}\ge C>0
\eean
for a positive constant $C=C(\alpha, \beta)$, we deduce:
\bean
\left|\sqrt R\,\widetilde W \left(\frac{\xi}{R}\right)-\sqrt R\,\widetilde W \left(\frac{z}{R}\right)\right|\le C\, R^{\alpha/2} R^{\beta/2}\, \frac{|\xi|+|z|}{|\xi|^{\beta}|z|^{\alpha}}\\
\le C\frac{|\xi-z|}{|\xi|^{\beta}|z|^{\alpha}}(1+|W(z)|)^{\alpha}
(1+|W(\xi)|)^{\beta}.
\eean
Suppose now, still under assumption (ii), that $sign\, \xi = sign\, z$. Then, using (\ref{S2cotaderivadauvedobleTilde})  we have
\bean
\left|\frac{\partial }{\partial \xi} \left(\sqrt R\,\,\widetilde W \left(\frac{\eta}{R}\right)\right)\right|\le C\, \frac{ R^{1/2+\alpha}}{|\eta|^{\alpha+1}}\\
\eean
for all $\eta \ge R/2$.
Using Taylor's theorem it then follows that:
\bean
\frac{|\sqrt R\,\widetilde W \left(\frac{\xi}{R}\right)-\sqrt R\,\widetilde W \left(\frac{z}{R}\right)|}
{\sqrt R\, \left(\widetilde W\left(\frac{\xi}{R}\right)\right)^{\alpha}
\left(\widetilde W\left(\frac{z}{R}\right)\right)^{\beta}}
\le R^{\alpha}\,\int_{\xi}^z\frac{d\eta}{\eta^{1+\alpha}}.
\eean
Therefore, if $|\xi|/2 \le |z|\le 2|\xi|$ then,
\bean
 R^{\alpha}\,\int_{\xi}^z\frac{d\eta}{\eta^{1+\alpha}}\le R^{\alpha}\frac{|\xi-z|}{|z|^{\alpha+1}}\le \frac{|\xi-z|}{|z|}\le C\frac{|\xi-z|}{|z|^\alpha |\xi|^{\beta}}.
\eean
Otherwise,
\bean
R^{\alpha}\,\int_{\xi}^z\frac{d\eta}{\eta^{1+\alpha}}\le R^{\alpha}\,\int_{R/2}^\infty\frac{d\eta}{\eta^{1+\alpha}}
=\frac{2^{\alpha}}{\alpha}\le C\frac{|\xi-z|}{|z|}\le C\frac{|\xi-z|}{|z|^\alpha |\xi|^{\beta}}.
\eean
The cases (iii) and (iv) can be treated equivalently. In both cases we have 
\bean
\frac{|\xi-z|}{|z|}\ge C_1>0.
\eean
Moreover, in the case (iv):
\bean
\frac{|\sqrt R\,\widetilde W \left(\frac{\xi}{R}\right)-\sqrt R\,\widetilde W \left(\frac{z}{R}\right)|}
{\sqrt R\, \left(\widetilde W\left(\frac{\xi}{R}\right)\right)^{\alpha}
\left(\widetilde W\left(\frac{z}{R}\right)\right)^{\beta}}\le 
C\,\frac{R^{\alpha/2}}{|\xi|^{\alpha/2}}\le C\,\frac{|z|^{\alpha/2}}{|\xi|^{\alpha/2}}\\
\le C\, \frac{|z|^{\alpha}}{|\xi|^{\alpha}}\le C\frac{|z|}{|z|^{1-\alpha}|\xi|^{\alpha}}\le C \frac{|z-\xi|}{|z|^{1-\alpha}|\xi|^{\alpha}}.
\eean
And this ends the proof of Lemma \ref{uvedoble}.
\qed
\\ \\
\noindent
{\bf Proof of Proposition \ref{S2Tsemigrupos}.}
\bean
&&\left|\left|\int_0^t S_{\varepsilon, R}(t-s)\, h(s)\, ds\right| \right |^2_{H^{\sigma} (\RR)}  
=\\
&&  C\int_{\RR}\int_0^t \int_0^t(1+|\xi|^s)^2\, e^{x_0^{\lambda/2}(t-s_1)T_1(\xi)}e^{x_0^{\lambda/2}(t-s_1)T_2(\xi)}
\widehat h(\xi, s_1)\times\\
&&\hskip 4cm  \times
e^{x_0^{\lambda/2}(t-s_2)T_1(\xi)}e^{-x_0^{\lambda/2}(t-s_2)T_2(\xi)}
\overline{\left(\widehat h(\xi, s_2\right)}\, ds_1\, ds_2\, d\xi\\
&& \le   C\int_{\RR}(1+|\xi|^s)^2\,\left(\int_0^t e^{x_0^{\lambda/2}(t-s_1)T_1(\xi)}\left|\widehat  h(\xi, s_1)\right|ds_1 \right)^2\, d\xi \le  C\int_0^t  ||h||^2_{H^s(\RR)}ds_1,
\eean
which proves (\ref{S2ETsemigrupos1}). \\\\
We prove now (\ref{S2ETsemigrupos2}).  To this end let us define the function
\bean
\varphi(x, t)= \int_0^te^{x_0^{\lambda/2}\kappa \,T_1(t-s)}h(s)ds
\eean
which verifies:
\bean
&&\frac{\partial \varphi}{\partial t}=x_0^{\lambda/2}\kappa\, T_1 (\varphi)+h(x, t),\,\,\,\,\,\,t>0,\,\,x>0,\\
&&\varphi(x, 0)=0.
\eean
Multiplying this equation by $-\kappa \, T_1\, M^{2\sigma}$ in $L^2(\RR)$ where $M$ is the multiplier operator associated to the symbol $|\xi|$ we obtain:
\bean
\frac{\kappa}{2}\frac{\partial}{\partial t} ||M^\sigma (-T_1)^{1/2}\varphi||^2_{L^2(\RR)} & + & \kappa ^2
\, x_0^{\lambda/2}||M^{\sigma}(T_1\varphi)||^2_{L^2(\RR)}\le 
\kappa \, ||M^{\sigma}T_1(\varphi)||_{L^2(\RR)}||M^\sigma h||_{L^2(\RR)}\\
& \le & \frac{1}{2 x_0^{\lambda/2}}||h(s)||^2_{H_x^\sigma(\RR)}+
\frac{\kappa ^2\, x_0^{\lambda/2}}{2}||M^{\sigma}(T_1\varphi)||^2_{L^2(\RR)}
\eean
whence:
\bean
\frac{\kappa }{2}\frac{\partial}{\partial t} ||M^\sigma (-T_1)^{1/2}\varphi||^2_{L^2(\RR)}+
\frac{\kappa ^2\, x_0^{\lambda/2}}{2}||M^{\sigma}(T_1\varphi)||_{L^2(\RR)}
&\le & \frac{1}{2\, x_0^{\lambda/2}}||h(s)||^2_{H_x^\sigma(\RR)}.
\eean
The result follows integrating in time and adding the corresponding inequality for $\sigma=0$. The proof of (\ref{S2ETsemigrupos3})
is similar. We multiply the equation by $-M^{2(\sigma-1/2)}\, \Lambda$ in $L^2(\RR)$ to obtain:
\bean
\frac{1}{2}\frac{d}{dt}||(-\Lambda)^{1/2} \varphi ||^2_{H^{\sigma-1/2}}+\varepsilon||\Lambda \varphi||^2_{H^{\sigma-1/2}}\le
||h||_{H^{\sigma-1/2}}||\Lambda \varphi||_{H^{\sigma-1/2}}.
\eean
Using Young's inequality and integrating in time we obtain (\ref{S2ETsemigrupos3}).

We prove now (\ref{diadic}). By definition:
\bean
\widehat{
\left[ \eta,\, T_{\varepsilon, R}\right]\varphi
} (\xi)=
\int_{\RR }K(\xi-z)\left(W(z, \varepsilon, R)-W(\xi, \varepsilon, R) \right)\hat \varphi(z)dz
\eean
where $K(z)=\hat \eta (z)$. Since the function $\eta$  is $C^\infty$,  for any $m>0$ there is a constant $C_m$ such that:
\bear
\label{S2TCa}
|K(\xi)|\le \frac{C_m}{1+|\xi|^m}\,\,\,\,\hbox{for all}\,\,\,\xi \in \RR.
\eear
 Therefore:
\bean
&&||S_{\varepsilon, R}(t)\left[\eta, T_{\varepsilon, R}\right]h||^2_{H^\sigma (\RR)}=\\
&&\int_{\RR}e^{-2\,t\, |\Re e W(\xi)|}(1+|\xi|^\sigma)^2\left|\int_{\RR}K(\xi-z)\left(W(\xi)-W(z) \right)\hat \varphi (z)\, dz\right|^2d\xi
\eean
We split the integral in two pieces:
\bean
||S_{\varepsilon, R}(t)\left[\eta, T_{\varepsilon, R}\right]h||^2_{H^\sigma (\RR)}=
\int_{|\xi|\le 1}[\cdots]d\, \xi+\int_{|\xi|\ge 1}[\cdots]d\, \xi
\eean
The first term is estimated by:
\bear
\label{S2Tfirsterm}
\left|\int_{|\xi|\le 1}[\cdots]d\, \xi\right|\le C
\int_{|\xi|\le 1}\left(\int_{\RR}\frac{C_m}{1+|\xi-z|^m}(1+|z|^{1/2})|\hat \varphi (z)|\, dz\right)^2d\xi \nonumber \\
\le C\left(\int_{\RR}\frac{C_m}{1+|z|^{m-1/2}}|\hat \varphi (z)|\, dz\right)^2 \le C ||\varphi||_{L^2}^2.
\eear
In the second term we have:
\bean
\int_{|\xi|\ge 1}[\cdots]d\, \xi\le 
2 \int_{|\xi|\ge 1}e^{-2\,t\, |\Re e W(\xi)|}(1+|\xi|^\sigma)^2(\int_{|z|\le 1}[\cdots]dz)^2d\, \xi+\\
+2 \int_{|\xi|\ge 1}e^{-2\,t\, |\Re e W(\xi)|}(1+|\xi|^\sigma)^2(\int_{|z|\ge 1}[\cdots]dz)^2d\, \xi =J_1+J_2.
\eean
We estimate $J_1$ follows:
\bear
J_1\le C\, \int_{|\xi|\ge 1}(1+|\xi|^\sigma)^2(\int_{|z|\le 1}\frac{C_m}{1+|\xi-z|^m}(1+|\xi|^{1/2})|\hat \varphi (z)|\,dz)^2d\, \xi\nonumber\\
\le C \int_{|\xi|\ge 1}\frac{1}{(1+|\xi|^{2m-1-2\sigma})}\left(\int_{|z|\le 1}|\hat \varphi (z)|\,dz\right)^2d\, \xi
\le C ||\varphi||_{L^2}^2.\label{S2Tfirsintegral}
\eear
It only remains to estimate $J_2$.
\bean
J_2 & \le &
\int_{|\xi|\ge 1}e^{-2\,t\, |\Re e W(\xi)|}(1+|\xi|^\sigma)^2\left(\int_{|z|\ge 1}K(\xi-z)\left(W(\xi)-W(z) \right)\hat \varphi (z)\, dz\right)^2d\xi\\
& \le & C_m ||\varphi||^2_{H^{\sigma-\rho}}\times\\
&&\times \int_{|\xi|\ge 1}e^{-2\,t\, |\Re e W(\xi)|}(1+|\xi|^\sigma)^2
\left(\int_{|z|\ge 1}\frac{\left(W(\xi)-W(z)\right)^2}{(1+|\xi-z|^m)^2}\, \frac{dz}{(1+|z|^{\sigma-\rho})^2}\right)
d\xi 
\eean
Using Lemma \ref{uvedoble}:
\bean
J_2&\le &C||\varphi||_{H^{\sigma-\rho}}\int_{|\xi|\ge 1}e^{-2\,t\, |\Re e W(\xi)|}(1+|\xi|^\sigma)^2
\left(\int_{|z|\ge 1}\frac{|\xi-z|^2}{(1+|\xi-z|^m)^2}\times\right.\\
&&\hskip 5cm \left. \times\frac{|W(\xi)|^{2\beta}|W(z)|^{2\alpha}}{|z|^{2\alpha}|\xi|^{2\beta}}\frac{dz}{(1+|z|^{\sigma-\rho})^2}\right)
d\xi \\ \\
&\le &
C\, ||\varphi||_{H^{\sigma-\rho}}t^{-2\beta}\times \\
&&\times\int_{|\xi|\ge 1}|\xi|^{2\sigma}
\left(\int_{|z|\ge 1}\frac{1}{(1+|\xi-z|^{m-1})^2}\frac{|W(z)|^{2\alpha}}{|z|^{2\alpha}|\xi|^{2\beta}}\, \frac{dz}{(1+|z|^{\sigma-\rho})^2}\right)\,d\xi 
\eean
where we have used that, for all $\xi\in \RR$ and all $t>0$: 
\bear
\label{S2Estar}
e^{-2\,t\, |\Re e W(\xi)|}|w(\xi)|^{2\beta}\le \frac{C}{t^{2\, \beta}}.
\eear
Using  now that $|W(z)|/|z| \le |z|^{-1/2}$ we deduce
\bean
J_2
\le C||\varphi||_{H^{\sigma-\rho}}  t^{-2\beta}\int_{|\xi|\ge 1}|\xi|^{2(\sigma-\beta)}
\left(\int_{|z|\ge 1}\frac{1}{(1+|\xi-z|^{m-1})^2}\frac{1}{|z|^{\alpha}}\, \frac{dz}{(1+|z|^{\sigma-\rho})^2}\right)
d\xi .
\eean
We change the order of integration and rewrite the resulting integral as
\bean
\int_{|\xi|\ge 1}\frac{|\xi|^{2(\sigma-\beta)}\, d\xi}{(1+|\xi-z|^{m-1})^2}& = &
\int_{|\xi|\ge 1, \,\,|\xi|\le 8|z|}[\cdots]\, d\xi+\int_{|\xi|\ge 1, \,\,|\xi|\ge 8|z|}[\cdots]\, d\xi\\
& = & I_1+I_2.
\eean
In the second integral we have $|\xi-z|\ge C |\xi|$ and therefore: 
\bean
\int_{|\xi|\ge 1, \,\,|\xi|\ge 8|z|}\frac{|\xi|^{2(\sigma-\beta)}\, d\xi}{(1+|\xi-z|^{m-1})^2}
\le C\int_{|\xi|\ge 1, \,\,|\xi|\ge 8|z|}\frac{|\xi|^{2(\sigma-\beta)}\, d\xi}{|\xi|^{2(m-1)}}\\
\le C  |z|^{2\sigma-2\beta-2m+3}\le C |z|^{2(\sigma-\beta)},
\eean
assuming that $m$ is large. In the first integral 
\bean
\int_{|\xi|\ge 1, \,\,|\xi|\le 8|z|}\frac{|\xi|^{2(\sigma-\beta)}\, d\xi}{(1+|\xi-z|^{m-1})^2}\le
C|z|^{2(\sigma-\beta)}\int_{|\xi|\le 1, \,\,|\xi|\le 8|z|}\frac{d\xi}{(1+|\xi-z|^{m-1})^2}
\eean
Then
\bean
I_1+I_2\le C\, |z|^{2(\sigma-\beta)}
\eean
and
\bean
&&\int_{|\xi|\ge 1}|\xi|^{2(\sigma-\beta)}
\left(\int_{|z|\ge 1}\frac{1}{(1+|\xi-z|^{m-1})^2}\frac{1}{|z|^{\alpha}}\, \frac{dz}{(1+|z|^{\sigma-\rho})^2}\right)
d\xi   \le   C\int_{|z|\ge 1}|z|^{-1-\beta+2\rho}\, dz.
\eean
This integral is bounded as soon as $2\rho<\beta$. This concludes the proof of (\ref{diadic}).

In order to prove (\ref{diadicdos}) we estimate its left hand side as:
\bean
\int_0^1dt \int_0^tds \int_{\RR}e^{-2 (t-s)\Re e W(\xi)}
(1+|\xi|^{2\sigma})|W(\xi)|^2\left|\int_{\RR}K(\xi-z)(W(\xi)-W(z))h(z, s)dz \right|^2 d\xi 
\eean
arguing as in the proof of (\ref{S2Tfirsterm}) and (\ref{S2Tfirsintegral}) we obtain that

\bear
&&\int_0^1dt \int_0^tds \int_{\RR}e^{-2 (t-s)\Re e W(\xi)}
(1+|\xi|^{2\sigma})|W(\xi)|^2\times \nonumber \\
&&\times \left|\int_{\RR} \textbf{1}_{\min(|\xi|, |z|)\le 1}(\xi, z) K(\xi-z)(W(\xi)-W(z))h(z, s) dz\right|^2 d\xi \le C\int_0^1||h(s)||_{L^2(\RR)}^2ds.\nonumber \\
\label{S2horroruno}
\eear
On the other hand,
\bean
&&\int_0^1dt \int_0^tds \int_{|\xi|\ge 1}e^{-2 (t-s)\Re e W(\xi)}
(1+|\xi|^{2\sigma})|W(\xi)|^2\times \nonumber \\
&&\times\left|\int_{|z|\ge 1}K(\xi-z)(W(\xi)-W(z))h(z, s) dz\right|^2 d\xi \\
&&\le  \int_0^1dt \int_0^tds \int_{|\xi|\ge 1}e^{-2 (t-s)\Re e W(\xi)}
(1+|\xi|^{2\sigma})\times \nonumber \\
&&\times\left|\int_{|z|\ge 1}|K(\xi-z)||W(\xi)-W(z)||W(z)||h(z, s)| dz\right|^2 d\xi+ 
\\
&&+\int_0^1dt \int_0^tds \int_{|\xi|\ge 1}e^{-2 (t-s)\Re e W(\xi)}
(1+|\xi|^{2\sigma})\times \nonumber \\
&&\times\left|\int_{|z|\ge 1}|K(\xi-z)||W(\xi)-W(z)|^2|h(z, s)| dz\right|^2 d\xi=I_1+I_2.
\eean
Arguing as in the derivation of (\ref{diadic}) we obtain
\bear
\label{S2horrordos}
I_1\le C\int_0^1dt\int_0^tds \frac{1}{(t-s)^\beta}||T_1 h(s)||_{H^\sigma(\RR)}^2\le C
\int_0^1||T_1 h(s)||_{H^\sigma(\RR)}^2ds.
\eear
On the other hand, in $I_2$ we use (\ref{S2cotauvedoble}) and formula (\ref{S2Estar}),  to obtain:
\bear
&&I_2\le \int_0^1 dt\int_0^t ds\, ||T_1 h(s)||^2_{H_x^{\sigma-\rho}}\int_{|\xi|\ge 1}e^{-2(t-s)\Re e (W(\xi))}|\xi|^{2\sigma} d\xi\times \nonumber \\
&&\times\left( \int_{|z|\ge 1}\frac{|\xi-z|^4}{(1+|\xi-z|^n)^2}\frac{|W(\xi)|^{4\beta}}{|\xi|^{4\beta}}\frac{|W(z)|^{4\alpha-2}}{|z|^{4\alpha}}\frac{dz}{|z|^{2\sigma-2\rho}}
\right)\nonumber \\
&&\le \int_0^1 dt\int_0^t ds\, ||T_1 h(s)||^2_{H_x^{\sigma-\rho}}\int_{|\xi|\ge 1}\frac{|\xi|^{2\sigma}}{(t-s)^{4\beta}} d\xi\times \nonumber \\
&&\times\left( \int_{|z|\ge 1}\frac{1}{(1+|\xi-z|^{n'})^2}\frac{1}{|\xi|^{4\beta}}\frac{|z|^{2\alpha-1}}{|z|^{4\alpha+2\sigma-2\rho}}dz
\right)\nonumber \\
&&\le \int_0^1 dt\int_0^t ds\, \frac{1}{(t-s)^{4\beta}}||T_1 h(s)||^2_{H_x^{\sigma-\rho}}\int_{|\xi|\ge 1} 
\frac{|\xi|^{2\rho}}{|\xi|^{3+2\beta}}d\xi\nonumber \\
&& \le C\int_0^1 ||T_1 h(s)||^2_{H_x^{\sigma-\rho}}ds
\label{S2horrortres}
\eear
Combining (\ref{S2horroruno}), (\ref{S2horrordos}) and (\ref{S2horrortres}), (\ref{diadicdos}) follows and then
Proposition \ref{S2Tsemigrupos}.
\qed
\\ \\
We will also use the following Lemma.
\begin{lem}
\label{S2Tlemasobolev} Let $\alpha \in C^\infty_0(0, +\infty)$ and $\varepsilon_0>0$ such that supp $\alpha \subset (x_0-\varepsilon_0, x_0+\varepsilon_0)$ and 
\bear
\label{S2cotasalpha}
\varepsilon_0^n\left|\frac{d^n \alpha (x)}{d x^n}\right|\le C_n\, \varepsilon_0,\quad \hbox{for all}\,\,x>0 
\eear
for some positive constant $C_n$ independent of $\varepsilon_0$. Then, there exists positive constants $K$ and $C_{\varepsilon_0}$, with $K$ independent of $\varepsilon_0$, such that
\bear
||\alpha\, f||_{H^\sigma(\RR^+)}\le K \varepsilon_0\, ||f||_{H^\sigma(\RR^+)}+C_{\varepsilon_0}||f||_{L^\infty(\RR^+)}
\label{S2TlemasobolevUno}\\
||T_1\alpha\, f||_{H^\sigma(\RR^+)}\le K \varepsilon_0\, ||T_1f||_{H^\sigma(\RR^+)}+C_{\varepsilon_0}||f||_{L^\infty(\RR^+)}.
\label{S2TlemasobolevDos}
\eear

\end{lem}
\textbf{Proof of Lemma \ref{S2Tlemasobolev}} Let us consider the function $m(k)$ that will equal to one in the proof of (\ref{S2TlemasobolevUno}) and $|\Re e(W (k, \varepsilon, R))|$ in the proof of (\ref{S2TlemasobolevDos}) where $W(k, \varepsilon, R)$ is defined in (\ref{S2EUvedobleTres}). Due to the hypothesis (\ref{S2cotasalpha}):
\bear
\label{S2TlemasobolevCotaalphahat}
|\widehat \alpha (k)|\le \frac{C_n\, \varepsilon_0^2}{1+|k\varepsilon_0|^n}\quad \hbox{for all}\,\,\,k\in \RR
\eear
We proceed to estimate

\bear
\label{S2Tlemasobolevjota}
J & = & \int_{\RR}|m(k)|^2 |k|^{2\sigma}\left|\widehat \alpha (k-\xi)\, \widehat f (\xi)\, d\xi\right|^2dk\\
& = & \int_{|k|\le 1} [\cdots]\, dk+\int_{|k|\ge 1} [\cdots]\, dk=J_1+J_2
\eear
The term $J_1$ is estimated as follows
\bear
\label{S2Tlemasobolevjotauno}
|J_1|\le C||f||^2_{L^2(\RR)}||\widehat \alpha||_{L^1}\le C \varepsilon_0\,||f||^2_{L^2(\RR)}
\eear
for some positive constant $C$ independent on $\varepsilon_0$. On the other hand, we split $J_2$ as follows:
\bear
|J_2|& \le & J_{2, 1}+J_{2,2}\\
J_{2, 1} & = & \int_{|k|\ge 1}|m(k)|^2 |k|^{2\sigma}\left|\int_{|\xi|\le 1}\widehat \alpha (k-\xi)\, \widehat f (\xi)\, d\xi\right|^2dk \label{S2Tlemasobolevjotadossplit1}\\
J_{2, 2} & = & \int_{|k|\ge 1}|m(k)|^2 |k|^{2\sigma}\left|\int_{|\xi|\ge 1}\widehat \alpha (k-\xi)\, \widehat f (\xi)\, d\xi\right|^2dk.
\label{S2Tlemasobolevjotadossplit2}
\eear
To estimate $J_{2, 1}$ we use that, for $|k|\ge 1$ and $|\xi|\le 1$, one has $|m(k)-m(\xi)|\le C(1+m(k-\xi))$. We deduce that in the same range of $k$ and $\xi$:
\bean
\left|m(k)|k|^{\sigma}-m(\xi)|\xi|^{\sigma}\right|\le C(1+|k-\xi|^{\sigma})(1+m(k-\xi)).
\eean
Then, since $|m(k-\xi)|\le C(1+\sqrt{|k-\xi|})$ we obtain:
\bear
\label{S2Tlemasobolevjotados1}
J_{2, 1}\le C_{\varepsilon_0, n}\int_{|k|\ge 1}\left(\int_{|\xi|\le 1}\frac{|\widehat f(\xi)|}{1+|k-\xi|^n} d\xi\right)^2dk\le C'_{\varepsilon_0, n}||f||_{L^2(\RR)}
\eear
where $C_{\varepsilon_0, n}$ and $C'_{\varepsilon_0, n}$ are constants depending on $n$ and $\varepsilon$ and using Young's inequality in the last step. Consider finally $J_{2,2}$. To this end we notice that, using (\ref{S2cotauvedoble}) for the case when
 $m(k)=|\Re e(W(k, \varepsilon, R))$:
\bean
&&\left| |k|^\sigma m(k)-|\xi|^\sigma m(\xi)\right|\le |k|^{\sigma}|m(k)-m(\xi)|+\left| |k|^\sigma-|\xi|^\sigma \right|m(\xi)\\
&&\le |k|^\sigma \frac{|\xi -k|}{|\xi|} +|k-\xi|^\sigma m(\xi)
\eean
whence, using once again $|k|\le |k-\xi|+|\xi|$,
\bean
\left| |k|^\sigma m(k)-|\xi|^\sigma m(\xi)\right|\le 
C\, \left(\frac{|k-\xi|^{\sigma+1}}{|\xi|}+\frac{|k-\xi| |\xi|^\sigma}{|\xi|}+|k-\xi|^\sigma\right)m(\xi)
\eean
and then,
\bean
J_{2,2} & \le & \int_{|k|\ge 1}\left|\int_{|\xi|\ge 1}\widehat \alpha (k-\xi)\,m(\xi)\, |\xi|^\sigma\widehat f (\xi)\, d\xi\right|^2dk \nonumber \\
&&+C_{\varepsilon_0, n'}
\int_{|k|\ge 1}\left|\int_{|\xi|\ge 1}\frac{1+|\xi|^{\sigma-1}}{1+|k-\xi|^{n'}}m(\xi)\, |\xi|^\sigma\widehat f (\xi)\, d\xi\right|^2dk .
\eean
Using Young's inequality we obtain
\bear
\label{S2Tlemasobolevjotados2uno}
&&J_{2,2}\le K\varepsilon_0\, ||T_1f||_{H^\sigma(\RR)}+C_{\varepsilon_0}||T_1 f||_{H^{(\sigma-1)_+}(\RR)}\quad \hbox{if}\,\, m(k)=|\Re e W|\\
&&J_{2,2}\le K\varepsilon_0\, ||f||_{H^\sigma(\RR)}+C_{\varepsilon_0}||f||_{H^{(\sigma-1)_+}(\RR)}\quad \hbox{if}\,\, m(k)=1.
\label{S2Tlemasobolevjotados2dos}
\eear
Combining (\ref{S2Tlemasobolevjotauno}), (\ref{S2Tlemasobolevjotados1}), (\ref{S2Tlemasobolevjotados2uno}), (\ref{S2Tlemasobolevjotados2dos}) and a classical interpolation argument to estimate the norm $H^{(\sigma-1)_+}(\RR)$ by the $L^\infty$ and $H^{\sigma}(\RR)$ norms the Lemma follows.
\qed

 \begin{lem}
\label{mediantegral}
Let $\eta$ be a $C^\infty$ compactly supported function in $\RR^+$. Then, for any $\sigma >0$ there exists a positive constant $C$ such that for any $h\in H^\sigma(\RR)$, for any $R>0$ and any $\varepsilon >0$:
\bean
\left|\left|\int_0^{\infty}h(x-y)\left( 
\eta(x)-\eta(x-y)\right)\Phi(y, R, \varepsilon)\, dy\right|\right|_{H^{\sigma+1/2}(\RR)}\le C ||h||_{H^{\sigma}(\RR)}.
\eean
where $\Phi(y, R, \varepsilon)$ is defined by (\ref{S9E10000Phi}).
\end{lem}
\textbf{Proof of Lemma \ref{mediantegral}} We define three functions $M(x, y)$, $P(x, y, R, \varepsilon)$ and $Q(x, R, \varepsilon)$ as follows
\bear
&&Q(y)=y\, \Phi(y, R, \varepsilon) \nonumber \\
&&M(x, y)=\frac{\eta(x)-\eta(x-y)}{y} \nonumber\\
&&P(x, y)=
\left(\eta(x)-\eta(x-y)\right)\Phi(y, R, \varepsilon)=M(x, y)\, Q(y). \label{S2EPxy}
\eear
Where the dependence of $P$ and $Q$ on $R$ and $\varepsilon$ is not explicitely written by shortedness. Notice that $M(x, y)\in C^{\infty}(\RR \times \RR)$. If we suppose that the support of $\eta$ is contained in an interval $I\subset \RR^+$, then the support of $m$ is such that:
\bear
\hbox{supp}\, (M)\subset I\times \RR^+ \cup \left\{(x, y)\in \RR^+\times \RR^+;\,\,x-y \in I \right\}.
\eear
Our goal is then to estimate estimate the the $H^{\sigma+1/2}$ norm of 
\bean
B(h):=\int_0^{\infty}h(x-y, s)\left( 
\eta(x)-\eta(x-y)\right)\Phi(y, R, \varepsilon)\, dy
\eean
 which we write:
\bean
\left|\left|B(h)\right|\right|^2_{H^{\sigma+1/2}_x(\RR)} & = & \int_{\RR}(1+|\xi|^{\sigma+1/2})^2|\widehat {B(h)}(\xi)|^2\, d\xi\\
\widehat{B(h)}(\xi) & = & \frac{1}{2\, \pi}
\int_{\RR}\int_{\RR}\int_0^\infty e^{i\, \eta(x-y)}M(x, y)Q(y)e^{-i\, x\, \xi}\, \hat h(\eta)dy\, d\eta\, d x\\
& = & \int_{\RR} \widehat P (\xi-\eta, \eta) \widehat h(\eta)\, d\eta
\eean
where 
\bear
\label{S2EfourierP}
\widehat P(\zeta_1, \zeta_2)=\frac{1}{2\, \pi}\int_{\RR^2}e^{-i(x\, \zeta_1+y\, \zeta_2)}P(x, y)\, dx\, dy
\eear
is the Fourier  of the function $P$ with respect to the two variables $x$ and $y$.\\
Notice that:
\bean
\left|\left|B(h)\right|\right|^2_{H^{\sigma+1/2}_x(\RR)}=
\int_{\RR}(1+|\xi|^{\sigma+1/2})^2 \left|\int_{\RR}\widehat P (\eta-\xi, -\eta)\hat h(\eta) d\eta\right|^2\, d\xi
\eean
We now proceed to estimate the function $M$. For any $m=0, 1, \cdots$ there is a positive constant $C_m$, independent of $R$ and $\varepsilon$, such that
\bear
\label{S2Tcotaseme}
\left|\frac{\partial^m M(x, y)}{\partial x^m} \right|\le \frac{C_m}{1+|y|}\qquad\hbox{for all}\,\,\,(x, y)\in \,\,\hbox{in supp}(M).
\eear
On the other hand, there exists a positive constant $C$ independent on $R$ and $\varepsilon$ such that for all $y\in \RR^+$:
\bear
\label{S2Tcotasqu}
|Q(y, R, \varepsilon)|\le \frac{C}{\sqrt {|y|}}
\eear
Combining (\ref{S2EPxy}), (\ref{S2EfourierP}), (\ref{S2Tcotaseme}) and (\ref{S2Tcotasqu}) we deduce that $P(x, y)$ is integrable in $\RR^2$ and then $\widehat P$ is a well defined and bounded function on $\RR^2$.

 Moreover, we can also deduce decay estimate for $\widehat P$ for $|\zeta_1|+|\zeta_2|\to +\infty$. To this end we integrate by parts in formula (\ref{S2EfourierP}) 
\bear
\label{S2EPgorro}
\widehat P(\zeta_1, \zeta_2)
& = &\frac{1}{2\pi\, i^n\, \zeta_1^n }\int_{0}^\infty e^{-i\zeta_2\, y}Q(y, R, \varepsilon) S_n(\zeta_1, y) dy\\
\eear
where
\bear
\label{S2Eeseene}
S_n(\zeta_1, y)=\int_{\RR}e^{-i\zeta_1\, x}\frac{\partial^n M(x, y)}{\partial x^n}\, dx.
\eear
Differentiating  (\ref{S2Eeseene}) with respect to $y$ and integrating by parts,  it easily follows  that the function $S_n$ are  such that, for all $m=0, 1, \cdots$, $k=0, 1, \cdots$ there is a positive constant $C_{k, m, n}$, independent on $R$ and $\varepsilon$ satisfying, for all $\zeta_1 \in \RR^+$ and $y\in \RR^+$:
\bear
\label{S2Eeseenederivada}
\left|\frac{\partial^k S_n (\zeta_1, y)}{\partial y^k} \right|\le \frac{C_{k, m, n}}{(1+|y|)\, (1+|\zeta_1|)^m}
\eear

 Let us consider the behaviour of $\widehat P$ with respect to $\zeta_2$. Using (\ref{S2EPgorro})
 \bear
 \hskip -0.5 cm
 \widehat P(\zeta_1, \zeta_2) & = & \frac{\varepsilon}{2\pi\, i^n\, \zeta_1^n }\int_{0}^1 e^{-i\zeta_2\, y}\frac{1}{y^{1/2}} S_n(\zeta_1, y) dy+\nonumber\\
 &+& \frac{(1-\varepsilon)}{2\pi\, i^n\, \zeta_1^n }\int_{0}^1 e^{-i\zeta_2\, y}\, R^{(3+\lambda)/2}\, y^{1+\lambda/2}\, f_0(R y)S_n(\zeta_1, y) dy \nonumber\\
 &+&\frac{1}{2\pi\, i^n\, \zeta_1^n }\int_{1}^\infty e^{-i\zeta_2\, y}Q(y, R, \varepsilon) S_n(\zeta_1, y) dy=
 \frac{1}{2\pi\, i^n\, \zeta_1^n}(J_1+J_2+J_3).
 \label{S2Tlasjotas}
 \eear
 In order to estimate the term $J_1$ we rewrite it as follows:
 \bean
 J_1 & = &
 \int_{0}^1 e^{-i\zeta_2\, y} \frac{\varepsilon}{y^{1/2}}S_n(\zeta_1, 0) dy+
 \int_{0}^1 e^{-i\zeta_2\, y} \frac{\varepsilon}{y^{1/2}}(S_n(\zeta_1, y)- S_n(\zeta_1, 0))dy\\
 &=&\frac{\varepsilon\,S_n(\zeta_1, 0)}{\zeta_2^{1/2}}\int_0^{\zeta_2}e^{-i\, z}\frac{dz}{z^{1/2}}+
  \int_{0}^1 e^{-i\zeta_2\, y} \frac{\varepsilon}{y^{1/2}}(S_n(\zeta_1, y)- S_n(\zeta_1, 0))dy.
 \eean
 The integral $\int_0^{\zeta_2}e^{-i\, z}\frac{dz}{z^{1/2}}$ is uniformly bounded for $\zeta_2 \in \RR$. On the other hand,
 due to (\ref{S2Eeseenederivada}) we have that
 \bean
\left| \frac {\partial}{\partial y} \left( \frac{1}{y^{1/2}}(S_n(\zeta_1, y)- S_n(\zeta_1, 0))\right)\right|\le
\frac{C}{\sqrt y}.
 \eean
Integrating by parts we obtain the existence of a constant $C$ such that for all $\zeta_2\in \RR$:
\bean
\left|\int_{0}^1 e^{-i\zeta_2\, y} \frac{\varepsilon}{y^{1/2}}(S_n(\zeta_1, y)- S_n(\zeta_1, 0))dy \right|
\le \frac{C}{1+|\zeta_2|} \qquad \hbox{for all}\,\,\,\zeta_2\in \RR.
\eean
Therefore:
\bean
|J_1|\le \frac{C}{1+|\zeta_2|^{1/2}.}
\eean

We use similar arguments to estimate $J_2$ that we write as follows
\bean
J_2& = & 
 S_n(\zeta_1, 0)\int_{0}^1 e^{-i\zeta_2\, y}\left( R^{(3+\lambda)/2}\, y^{1+\lambda/2}\, f_0(R y)\right)dy\\
&&+\int_{0}^1 e^{-i\zeta_2\, y}\left( R^{(3+\lambda)/2}\, y^{1+\lambda/2}\, f_0(R y)\right) \left(S_n(\zeta_1, y)-S_n(\zeta_1, 0)  \right)dy\\
& = & I_1+I_2.
\eean
The term $I_2$ may be estimated as above since (\ref{S2Eeseenederivada}) gives:
\bean
\left|\frac{\partial}{\partial y}
\left[\left( R^{(3+\lambda)/2}\, y^{1+\lambda/2}\, f_0(R y)\right) \left(S_n(\zeta_1, y)-S_n(\zeta_1, 0)  \right)\right]
\right|\le C\, y^{-1/2},\,\,y\in [0, 1]
\eean
using that
\bean
\frac{\partial}{\partial y}
\left[\left( R^{(3+\lambda)/2}\, y^{1+\lambda/2}\, f_0(R y)\right) \left(S_n(\zeta_1, y)-S_n(\zeta_1, 0)  \right)\right]\\
=\left(S_n(\zeta_1, y)-S_n(\zeta_1, 0)  \right)\frac{\partial}{\partial y}
\left[\left( R^{(3+\lambda)/2}\, y^{1+\lambda/2}\, f_0(R y)\right)\right]+\\
\left( R^{(3+\lambda)/2}\, y^{1+\lambda/2}\, f_0(R y)\right) \frac{\partial}{\partial y}
\left[\left(S_n(\zeta_1, y)-S_n(\zeta_1, 0)  \right)\right]
\eean 
as well as the bounds on $f_0$ and $f'_0$.
\bean
|I_2|\le \frac{C}{1+|\zeta_2|}
\eean
We use in $I_1$ the change of variables $\eta=\zeta_2 y$ and the auxiliary function $g(y)=y^{1+\lambda/2}\, f_0(y)$ to obtain

\bean
&&I_1=\frac{R^{1/2}}{\zeta_2}\int_0^{\zeta_2} e^{-i \eta}g\left(\frac{R\eta}{\zeta_2}\right)\, d\eta=
\frac{R^{1/2}}{i\, \zeta_2}\int_0^{\zeta_2} (e^{-i \eta}-1)\frac{R}{\zeta_2}g'\left(\frac{R\eta}{\zeta_2}\right)\, d\eta
\eean
after integrating by parts. Then, there is a positive constant $C$ independent on $R$ and $\varepsilon$ such that for all $\zeta_2\in \RR$
\bean
|I_1|\le \frac{C}{1+|\zeta_2|^{1/2}},
\eean
whence, combining the estimates for $I_1$ and $I_2$:
\bean
|J_2|\le \frac{C}{1+|\zeta_2|^{1/2}}.
\eean
We estimate $J_3$ integrating by parts and using (\ref{S2Eeseenederivada}):
\bean
|J_3|\le C \frac{1}{1+|\zeta_2|},
\eean
whence
\bean
|\widehat P(\zeta_1, \zeta_2)|\le \frac{C}{(1+|\zeta_1|^m)(1+|\zeta_2|^{1/2})}.
\eean
To conclude the proof of Lemma \ref{mediantegral} we bound the norm of $B(h)$ in $H^{\sigma +1/2}(\RR)$ as follows:
\bean
&&\left|\left|B(h)\right|\right|^2_{H^{\sigma+1/2}_x(\RR)}\le 
\int_{\RR}(1+|\xi|^{\sigma+1/2})^2 \left|\int_{\RR}\widehat P (\xi-\eta, \eta)\hat h(\eta) d\eta\right|^2\, d\xi\\
&&\le 
\int_{\RR}(1+|\xi|^{\sigma+1/2})^2 \left|\int_{\RR}\frac{\hat h(\eta)}{(1+|\eta-\xi|^m)(1+|\eta|^{1/2})} d\eta\right|^2\, d\xi\\
&&\le 
C\int_{\RR}\left|\int_{\RR}\frac{ (1+|\xi-\eta|^{\sigma+1/2}+|\eta|^{\sigma+1/2}) \hat  h(\eta)}{(1+|\eta-\xi|^m)(1+|\eta|^{1/2})} d\eta\right|^2\, d\xi\\
&&\le 
C\int_{\RR}\left|\int_{\RR}\frac{\hat h(\eta)}{(1+|\eta-\xi|^{m'})} d\eta\right|^2\, d\xi+
C\int_{\RR}\left|\int_{\RR}\frac{ (1+|\eta|^{\sigma}) \hat h(\eta)}{(1+|\eta-\xi|^m)} d\eta\right|^2\, d\xi\\ \\
&&\le C\,|| h||^2_{L^2}\, || {(1+|\cdot|^{-m'})} ||_{L^1(\RR)}^2 +C\,||  h||^2_{H^\sigma}\, || {(1+|\cdot|^{-m})} ||_{L^1(\RR)}^2
\eean
where we have used Young's inequality in the last step.
\qed
\section{Interior regularity theory for the operator $\mathcal L$.}

\setcounter{equation}{0}
\setcounter{theo}{0}

We start with the proof of Theorem \ref{S8T3-101}.
\\
\\
\textbf{Proof of (i) of Theorem \ref{S8T3-101}.} We apply now the classical method of freezing coefficients. To this end let us call $\chi$ a  $C^\infty$ function such that
\bear
\chi(x)=\left\{
\begin{array}{l}
1\quad \hbox{if}\,\,x\in \left(5/8, 11/8\right)\,,\\ \\
0\quad \hbox{if}\,\,\,x\not\in \left(1/2, 2\right)\, .
\end{array}
\right.
\eear
We define $\tilde f(x)= \chi(x)\, f(x).$ Then, for all $x\in \RR$:
\bean
\frac{\partial \tilde f}{\partial t} & = &\kappa\, T_{\varepsilon, R}(M_{\lambda/2}\tilde f)+ \kappa\int_0^{\infty}(x-y)^{\lambda/2}f(x-y)\left( 
\chi(x)-\chi(x-y)\right)\Phi(y, R, \varepsilon))\, dy+\\
&&+\chi(x)\, Q+\chi(x)\, P\\ \\
& = & \kappa T_{\varepsilon, R}(M_{\lambda/2}\tilde f)+\widetilde Q+\widetilde P\\
\widetilde Q & = & \widetilde Q_1+\widetilde Q_2\\
\widetilde Q_1 & = & \kappa\int_0^{\infty}(x-y)^{\lambda/2}f(x-y)\left( 
\chi(x)-\chi(x-y)\right)\Phi(y, R, \varepsilon)\, dy\\
\widetilde Q_2 & = & \chi(x)\, Q\\
\widetilde P&=&\chi(x)\, P.
\eean
It is readily seen that
\bear
\label{S2Epobre}
||\widetilde Q_1||_{L^{\infty}((0, 1);W^{1, \infty}(\RR))} & \le & C\, \kappa\, ||f||_{L^\infty((1/4, 2)\times (0, 1))}
\eear
Equation (\ref{S8T3-101E1}) may be written as
\bear
\label{S8T3-101E2}
&&\frac{\partial \tilde f}{\partial t} = \,  x_0^{\lambda/2 }\, \kappa \,T_{\varepsilon, R}\left(\tilde f\right)+
\kappa \,T_{\varepsilon, R}\,\left((M_{\lambda/2}-M_{\lambda/2, 0})\tilde f \right)+\widetilde Q+\widetilde P
\eear
where $ M_{\lambda/2, 0}\tilde f(x)  = x_0^{\lambda/2}\, \tilde f(x)$. 

Fix now a new cutoff function $\eta$ such that 
\bear
\label{S2E1npcqc0i}
\eta(x)=\left\{
\begin{array}{l}
1\quad \hbox{if}\,\,|x-x_0|\le \delta,\\ \\
0\quad \hbox{if}\,\,\,|x-x_0|\ge 2\delta\, .
\end{array}
\right.
\eear
with $\delta$ such that
\bean
|x^{\lambda/2}-x_0^{\lambda/2}|\le \varepsilon_0,\,\,\,\,\hbox{for}\,\,\,|x-x_0|\le 2 \delta.
\eean
with $\varepsilon_0$ small enough to be chosen later. If we multiply the equation (\ref{S8T3-101E2}) by $\eta$ and denote
$\overline f= \eta \tilde f$ we obtain:
\bear
\label{S8T3-101E3}
&&\frac{\partial \overline f}{\partial t} - \,  x_0^{\lambda/2 }\, \kappa \, T_{\varepsilon, R}\left(\overline f\right)=
\, \kappa\, \eta(x) T_{\varepsilon, R}\left((M_{\lambda/2}-M_{\lambda/2, 0})\tilde f \right)+\eta(x) (\widetilde Q+
\widetilde P)\nonumber\\
&&+\, \kappa \, x_0^{\lambda/2}\int_0^{\infty}\tilde f(x-y)\left( 
\eta(x)-\eta(x-y)\right)\Phi(y, R, \varepsilon)\, dy
\eear
We have the following representation formula for the solution $\overline f$ of (\ref{S8T3-101E3}) in $L^\infty((1/4, 2)\times (0, 1))\cap  L^2(0, 1; H^{1/2}(1/4, 2))\cap\, H^1(0, 1; L^2(1/4, 2))$ is :
\bear
\label{S8T3-101E4}
\overline f(x, t) & = & \int_0^t S_{\varepsilon, R}(\kappa (t-s))\eta(x)\,\left(\widetilde Q (s)+\widetilde P(s)\right)\, ds+\label{S8T3-101E5}\\
&&+\kappa \int_0^tS_{\varepsilon, R}(\kappa (t-s))
\left[\eta(x)T_{\varepsilon, R}\left((M_{\lambda/2}-M_{\lambda/2, 0})\tilde f \right)(s)\right]\, ds \nonumber \\
&&+\,\kappa\, x_0^{\lambda/2}\int_0^tS_{\varepsilon, R}(\kappa (t-s))\int_0^{\infty}\tilde f(x-y, s)\left( 
\eta(x)-\eta(x-y)\right)\Phi(y, R, \varepsilon)\, dy\, ds \nonumber  \\
& = & \overline f_1(x, t)+\overline f_2(x, t)+\overline f_3(x, t) \label{S8T3-101E567}
\eear
for all $x\in \RR$. This follows from the fact that the unique solution 
$f$ in the space  $L^\infty((1/4, 2)\times (0, 1))\cap  L^2(0, 1; H^{1/2}(1/4, 2))\cap\, H^1(0, 1; L^2(1/4, 2))$ of:
\bear
&&\frac{\partial f}{\partial t}-x_0^{\lambda/2}\kappa T_{\varepsilon, R}(f)=G(x, t) \label{S8iqbdvpqiy}\\
&&f(0, x)=0,\nonumber
\eear
where $G\in L^\infty((0, 1)\times (0, +\infty)$ and $f$ and $G$ compactly supported in $(1/4, 2)\times (0, 1)$, is given by Duhamel's formula.  The uniqueness of $f$ can be obtained by taking the difference of two such solutions and taking the scalar product of (\ref{S8iqbdvpqiy}) with that difference in $L^2$. Such computations are possible by the regularity that is assumed on the solutions.
\\ 
\bear
\label{S2Eefeunoestbis}
||\overline f_1||^2_{H^\sigma(\RR)}\le C\int_0^t  ||\eta\, \widetilde Q||^2_{H^\sigma(\RR)}ds_1
+C\left|\left|\int_0^t S_{\varepsilon, R}(\kappa (t-s))\eta(x)\,\widetilde P(s)\, ds\right|\right|^2_{H^\sigma(\RR)}
\eear
Let us estimate the second term in the right hand side of (\ref{S2Eefeunoestbis}). Formulas (\ref{S9E10000}) and (\ref{S2Eoplambda}) imply:
\bean
&&\left|\left|\int_0^t S_{\varepsilon, R}(\kappa (t-s))\eta(x)\,\widetilde P(s)\, ds\right|\right|^2_{H^\sigma(\RR)}
\le\\
&&\left|\left|\int_0^t e^{-\sqrt{2}\Gamma(1/2)\varepsilon \kappa\,|\xi|^{1/2}(t-s)}|\widehat{\eta\,\widetilde P}(\xi, s)|\, ds (1+|\xi|^\sigma)^2  \right|\right|^2_{L^2_\xi(\RR)}=\\
&&\hskip 5cm =\left|\left|\int_0^t e^{-\varepsilon\, \kappa\, \Lambda (t-s)} M\left(\eta\,\widetilde P \right)  ds  \right|\right|^2_{H^\sigma(\RR)}.
\eean
Integration in time and (\ref{S2ETsemigrupos3})  yields:
\bear
\label{S2Eefeunoest}
\int_0^1||\overline f_1(t)||^2_{H^\sigma(\RR)}dt \le C\int_0^1 ||\eta\, \widetilde Q (t)||^2_{H^\sigma(\RR)}dt
+\frac{C}{\varepsilon^2\, \kappa^2}\int_0^1||\eta\, \widetilde P(t)||^2_{H^{\sigma-1/2}}
\eear

In order to estimate the term corresponding to $\overline f_2$ we first write
\bean
\eta(x)T_{\varepsilon, R}\left((M_{\lambda/2}-M_{\lambda/2, 0})\tilde f \right)=
T_{\varepsilon, R}\left(\eta(x)\, (M_{\lambda/2}-M_{\lambda/2, 0})\tilde f \right)+
\left[\eta, T_{\varepsilon, R}\right]\left((M_{\lambda/2}-M_{\lambda/2, 0})\tilde f \right)
\eean
where $\left[\eta, T_{\varepsilon, R}\right]$ is the conmutator of $T_{\varepsilon, R}$ and the multiplication by $\eta$
\bean
\left[\eta, T_{\varepsilon, R}\right](\varphi)(x)=\eta(x)\, T_{\varepsilon, R}(\varphi)(x)-T_{\varepsilon, R}\left(\eta\, \varphi \right)(x)
\eean
Therefore
\bear
\overline f_2&=&\kappa\,\int_0^tS_{\varepsilon, R}(\kappa(t-s))
\left[\eta(x)T_{\varepsilon, R}\left((M_{\lambda/2}-M_{\lambda/2, 0})\tilde f \right)(s)\right]\, ds \nonumber\\
& = & \kappa\, \int_0^tS_{\varepsilon, R}(\kappa(t-s))
\left[T_{\varepsilon, R}\left((M_{\lambda/2}-M_{\lambda/2, 0})\overline f \right)(s)\right]\, ds \nonumber\\
& +&
\kappa\, \int_0^tS_{\varepsilon, R}(\kappa(t-s))
\left[\eta, T_{\varepsilon, R}\right]\left((M_{\lambda/2}-M_{\lambda/2, 0})\tilde f \right)\, ds \nonumber\\
& = & \overline f_{2,1}+\overline f_{2,2}
\label{S2Eefedosdosdescomp}
\eear
where we have used that $\eta \widetilde f=\overline f$. Let us denote 
\bear
\label{S2EPsi}
\Psi (x, s)=(M_{\lambda/2}-M_{\lambda/2, 0})\overline f (s)
\eear
 and define the operator $M$ as:
\bean
M(\varphi)=\frac{1}{\sqrt {2\, \pi}}\int_{\RR}|\widehat{\varphi}(\xi)|\, e^{i\, x\, \xi}\, d\xi.
\eean
Then:
\bean
\left|\widehat {\overline f_{2,1} }(\xi)\right|^2
& \le & C\, \kappa ^2\, \int_0^t \int_0^t\, e^{x_0^{\lambda/2}\kappa (t-s_1)T_1(\xi)}e^{x_0^{\lambda/2}\kappa (t-s_1)T_2(\xi)}\widehat {T(\Psi)}(\xi, s_1)\times\\
&& \times
e^{x_0^{\lambda/2}\kappa (t-s_2)T_1(\xi)}e^{-x_0^{\lambda/2}\kappa (t-s_2)T_2(\xi)}
\overline{\left(\widehat {T(\Psi)}(\xi,s_2)\right)}\, ds_1\, ds_2\\
& \le & C\left(\kappa\, \int_0^t e^{x_0^{\lambda/2}\kappa (t-s)T_1(\xi)}\left|\widehat {T(\Psi)}(\xi, s)\right|ds \right)^2\\
&\le & C\left|\kappa \, \int_0^t e^{x_0^{\lambda/2}\kappa (t-s)T_1(\xi)}\, T_1(M(\Psi))(s)ds\right|^2.
\eean
where we have used that $|W(\xi, \varepsilon, R)|\le C |\Re e W(\xi, \varepsilon, R)|=C\, T_1(\xi)$. Therefore using (\ref{S2ETsemigrupos2}):
\bean
\int_0^1||\overline f_{2, 1}||^2_{H^\sigma(\RR)}\, dt & \le & C \,\int_0^1
||\int_0^t e^{x_0^{\lambda/2}\kappa (t-s)T_1(\xi)}\, \kappa \, T_1(M(\Psi))ds||^2_{H^\sigma(\RR)}dt\\
& \le & C\,\int_0^1 || M(\Psi) ||^2_{H^\sigma(\RR)}ds=C\,\int_0^1 || \Psi ||^2_{H^\sigma(\RR)}ds.
\eean
The function $\Psi$ may be written as $\Psi(x, t)=\alpha(x)\, \overline f(x, t)$ with $\alpha(x)=\widetilde \eta(x)\, (x^{\lambda/2}-x_0^{\lambda/2})$ where $\widetilde \eta$ is a cutoff supported in the interval $|x-x_0|\le \varepsilon_0$ and $\widetilde \eta (x)=1$ in $|x-x_0|\le 2 \delta$ where $\delta$ is given in (\ref{S2E1npcqc0i}). Notice that $\alpha$ may be assumed to satisfy condition (\ref{S2cotasalpha}). Lemma \ref{S2Tlemasobolev} then implies:
\bear
||\Psi||_{H^\sigma_x(\RR)} \le  K\, \varepsilon_0 ||\overline f||_{H^\sigma}+C||\overline f||_{L^{\infty}(\RR \times (0, 1))} \label{S2TPsi}
\eear
where the constant $C$ here and until the end of the Proof of Theorem \ref{S8T3-101} may depend on $\varepsilon_0$ but $K$ independent on it.
We have then obtained:
\bear
\label{S2Eestfdosunoest}
\int_0^1||\overline f_{2, 1}||^2_{H^\sigma(\RR)}\, dt\le K\varepsilon_0\,\int_0^1 ||\overline f(s)||^2_{H^\sigma(\RR)}\, d\, s+C||\overline f||^2_{\infty}
\eear
We consider now $\overline f_{2,2}$.  Using, (\ref{diadic}) we have:
\bean
||\overline f_{2,2}||_{H_x^\sigma(\RR)} 
& \le & C\, \kappa ^{1-\beta}\, \int_0^t (t-s)^{-\beta}||\left(M_{\lambda/2}-M_{0,\lambda/2}\right)\tilde f ||_{H_x^{\sigma-\rho}(\RR)}\, ds\\
 \eean
and by (\ref{S2TPsi}) with $\sigma$ replaced by $\sigma-\rho$ :
\bear
\label{S2Tefetildedosdos}
||\overline f_{2,2}||_{H_x^\sigma(\RR)} & \!\!\!\le \!\!\!& C\, \kappa ^{1-\beta}\, 
\int_0^t(t-s)^{-\beta}||\tilde f ||_{H_x^{\sigma-\rho}(\RR)}\, ds+C\, \kappa ^{1-\beta}\, ||\widetilde f||_{L^\infty(\RR)\times (0, 1)}.
\eear
Squaring and integrating (\ref{S2Tefetildedosdos}) and adding to the results for $\beta $ small estimate, we obtain
\bear
\label{S2Eefedosest}
\int_0^1||\overline f_{2}(s)||^2_{H_x^\sigma(\RR)}ds &\le & \varepsilon_0\,\int_0^1 ||\overline f(s)||^2_{H^\sigma(\RR)}\, d\, s+
C\,\int_0^1 ||\tilde f(s)||^2_{H^{\sigma-\rho}(\RR)}\, d\, s\nonumber 
\\
&&+C||\tilde f||^2_{\infty}
\eear

The last term  $\overline f_3$
is estimated as follows. Using Lemma \ref{mediantegral} we obtain
\bear
\label{S2overlineftrestres}
\left|\left|\int_0^{\infty}\tilde f(x-y, s)\left( 
\eta(x)-\eta(x-y)\right)\Phi(y, R, \varepsilon)\, dy\right|\right|_{H^{\sigma}_x(\RR)}\le C ||\tilde f||_{H^{(\sigma-1/2)_+}(\RR)}.
\eear
Then, (\ref{S2ETsemigrupos1}) in Lemma \ref{S2Tsemigrupos} and an interpolation argument yield:
\bean
||\overline f_3||_{H^\sigma_x(\RR)} & \le & C\, \kappa \, \, \int_0^t||\tilde f(s)||_{H^{(\sigma-1/2)_+}(\RR)}ds,
\eean
whence
\bear
\label{S2Eefetresest}
\int_0^1||\overline f_{3}(s)||^2_{H_x^\sigma(\RR)}ds & \le & 
C\,\kappa ^2\,\int_0^1 ||\tilde f(s)||^2_{H^{(\sigma-1/2)_+}}\, d\, s+C\, \kappa ^2\, ||\tilde f \,||^2_{\infty}.
\eear
Adding  (\ref{S2Eefeunoest}), (\ref{S2Eefedosest}) and (\ref{S2Eefetresest}) and using $\rho<1/2$, we deduce:
\bean
\int_0^1||\overline f (s)||^2_{H^\sigma(\RR)} \le  \varepsilon_0 \int_0^1||\overline f (s)||^2_{H^\sigma(\RR)}
+C \int_0^t||\tilde f(s)||_{H^{\sigma-\rho}(\RR)}ds\\
+C\int_0^1  ||\eta\, \widetilde Q||^2_{H^\sigma(\RR)}ds+C||\widetilde f||^2_{L^{\infty}(\RR \times (0, 1))}+\frac{C}{\varepsilon^2\, \kappa^2}\int_0^1||\eta\, \widetilde P(t)||^2_{H^{\sigma-1/2}}.
\eean
Choosing $\varepsilon_0$ small enough:
\bean
\int_0^1||\overline f (s)||^2_{H^\sigma(\RR)} \le  C \int_0^t||\tilde f(s)||_{H^{\sigma-\rho}(\RR)}ds
+C\int_0^1  ||\eta\, \widetilde Q||^2_{H^\sigma(\RR)}ds+C||\widetilde f||^2_{L^{\infty}(\RR \times (0, 1))}\\
+\frac{C}{\varepsilon^2\, \kappa^2}\int_0^1||\eta\, \widetilde P(t)||^2_{H^{\sigma-1/2}}.
\eean
Using a partition of the unity $(\eta_i)_{i\in \NN}$ of the interval $(1/2, \, 2)$, and adding the contributions of all the terms we obtain:
\bear
\label{S2Eestipart}
&&\int_0^1||\tilde f (s)||^2_{H^\sigma(\RR)} \le  C \int_0^t||\tilde f(s)||_{H^{\sigma-\rho}(\RR)}ds
+C\int_0^1  || \widetilde Q||^2_{H^\sigma(\RR)}ds+C||\widetilde f||^2_{L^{\infty}(\RR \times (0, 1))}+ \nonumber \\
&&\hskip 8cm +\frac{C}{\varepsilon^2\, \kappa^2}\int_0^1||\widetilde P(t)||^2_{H^{\sigma-1/2}}
\eear
where the constants $C$ depend on $\delta$. An interpolation argument then implies:

\bean
\int_0^1||\tilde f (s)||^2_{H^\sigma(\RR)} \le  \varepsilon \int_0^t||\tilde f(s)||_{H^{\sigma}(\RR)}ds
+C\int_0^1  || \widetilde Q||^2_{H^\sigma(\RR)}ds+C||\widetilde f||^2_{L^{\infty}(\RR \times (0, 1))}\\
+
\frac{C}{\varepsilon^2\, \kappa^2}\int_0^1||\widetilde P(s)||^2_{H^{\sigma-1/2}}
\eean
whence part (i) of  Theorem \ref{S8T3-101} follows.
\begin{rem}
\label{S2RefeDosUno}
Notice that in (\ref{S2Eestfdosunoest}), we estimate the $H^\sigma$ norm of $\overline f_{2, 1}$ in terms of the 
$H^\sigma$ norm of $\overline f$, not of $\tilde f$.
\end{rem}

\begin{rem}
\label{S2RefeDosUnoUno}
In the estimates of $\overline f_j$, $j=1, 2, 3$ the term $\overline f_{2, 1}$ is the only one  where we are using the continuity in the freezing coefficients argument to obtain (\ref{S2Eestfdosunoest}).
\end{rem}

{\bf Proof of (ii) of Theorem \ref{S8T3-101}. } In order to prove part (ii) we first notice that the equation satisfied by $\overline f$ is:
\bear
\label{S3eqfbarrra}
&&\frac{\partial \overline f}{\partial t} - \,  x_0^{\lambda/2 }T_{\varepsilon, R}\left(\overline f\right)
+a(t)\, \overline f=
\,\eta(x) T_{\varepsilon, R}\left((M_{\lambda/2}-M_{\lambda/2, 0})\widetilde f \right)+\eta(x) \widetilde Q+\eta(x) \widetilde P\nonumber \\
&&+\, x_0^{\lambda/2}\int_0^{\infty}\tilde f(x-y)\left( 
\eta(x)-\eta(x-y)\right)\Phi(y, R, \varepsilon)\, dy- (a(x, t)-a(t))\, \overline f
\eear
where $x_0$ and $\eta$ have been chosen as before  and where $a(t)=a(x_0, t)$. Then:
\bean
&&\overline f(x, t)  =  \int_0^t \omega(t, s)S_{\varepsilon, R}(t-s)\left(\eta(x)\,\widetilde Q (s)+\eta(x)\,\widetilde P(s)\right)\, ds+\\
&&+\int_0^t \omega(t, s)S_{\varepsilon, R}(t-s)
\left[\eta(x)T_{\varepsilon, R}\left((M_{\lambda/2}-M_{\lambda/2, 0})\widetilde f \right)\right]\, ds +\nonumber \\
&&+\, x_0^{\lambda/2}\int_0^t \omega(t, s)S_{\varepsilon, R}(t-s)\int_0^{\infty}\tilde f(x-y, s)\left( 
\eta(x)-\eta(x-y)\right)\Phi(y, R, \varepsilon)\, dy\, ds \\
&&-\int_0^t \omega(t, s)S_{\varepsilon, R}(t-s)(a(x, t)-a(t))\, \overline f (s)\, ds
\label{S8T3-101E5} \\
& = & \overline f_1(x, t)+ \overline f_2(x, t)+ \overline f_3(x, t)+ \overline f_4(x, t) \label{S8T3-101E567}
\eean
where we have defined:
\bear
\label{S2Eomega}
\omega(t, s) = e^{-\int_s^t a(\lambda)\, d\lambda}
\eear
We estimate first the term with $\overline f _1$.
If $T\le 1$, then the same argument of the proof of point (i) shows that
\bear
\label{cotaefeuno}
&&\left( \int_T^{T+1}||\overline f _1(t)||^2_{H_x^{\sigma}}dt\right)^{1/2}\le \left( \int_0^{2}|| \overline f _1(t)||^2_{H_x^{\sigma}}dt\right)^{1/2}\nonumber \\
&&\hskip 3cm \le C\left( \int_0^{2}|| \widetilde Q(t)||^2_{H_x^{\sigma}}dt\right)^{1/2}
+\frac{C}{\varepsilon^2}\left( \int_0^{2}|| \widetilde P(t)||^2_{H_x^{\sigma}}dt\right)^{1/2}
.
\eear
If $T>1$, using the change of variables: $t=(T-1)+\tau$ we write
\bean
&&\left(\int_T^{T+1}||\int_0^t\omega(t, s) S_{\varepsilon, R}(t-s)\eta(x)\,\left(\widetilde Q (s)+\widetilde P(s)\right)\, ds||^2_{H^{\sigma}}dt\right)^{1/2}\\
 &\le &
\sum_{n=1}^{[T]}\left(\int_T^{T+1}||\int_{n-1}^n \omega(t, s) S_{\varepsilon, R}(t-s)\eta(x)\,\left(\widetilde Q (s)+\widetilde P(s)\right)\, ds||^2_{H^{\sigma}}
dt\right)^{1/2}\\
&&+\left(\int_T^{T+1} ||\int_{[T]}^t  \omega(t, s) S_{\varepsilon, R}(t-s)\eta(x)\,\left(\widetilde Q (s)+\widetilde P(s)\right)\, ds||^2_{H^{\sigma}}dt\right)^{1/2}\\
& = & I_1+I_2.
\eean
The estimate of the term $I_2$ follows as in the proof of point (i) of the Theorem  and gives
\bean
I_2\le C\left(\int_T^{T+1} ||\widetilde Q (s)||^2_{H^{\sigma}}\right)^{1/2}
+\frac{C}{\varepsilon}\left(\int_T^{T+1} ||\widetilde P (s)||^2_{H^{\sigma-1/2}}\right)^{1/2}.
\eean
To estimate $I_1$ we argue as follows. 
Changing  the time variable $t$ as $t=\tau+(T-n)$ and obtain:
\bean
&&I_1=\sum_{n=1}^{[T]}\left(\int_n^{n+1}||\int_{n-1}^n  \omega(\tau+(T-n), s) 
S_{\varepsilon, R}((T-n)+\tau-s)\times \right.\nonumber \\
&&\hskip 7cm \times \left. \eta(x)\,\left(\widetilde Q (s)+\widetilde P(s)\right)\, ds||^2_{H^{\sigma}}
d\tau\right)^{1/2}\\
&&\le \sum_{n=1}^T\left(\int_n^{n+1}||\int_{n-1}^n  \omega(\tau+(T-n), s) 
S_{\varepsilon, R}(\tau-s)\eta(x)\,\left(\widetilde Q (s)+\widetilde P(s)\right)\, ds||^2_{H^{\sigma}}
d\tau\right)^{1/2}
\eean
since $||S_\varepsilon(T-n)\, h||_{H^{\sigma}}\le ||h||_{H^\sigma}$ because $T-N\ge 0$. We use now that for each $n$
\bean
&&\int_n^{n+1}||\int_{n-1}^n \omega(\tau+(T-n), s)\,  S_{\varepsilon, R}(\tau-s)\eta(x)\,\left(\widetilde Q (s)+\widetilde P(s)\right)\, ds||^2_{H^{\sigma}}
d\tau\\
&&\le\int_{n-1}^{n+1}||\int_{n-1}^\tau  \omega(\tau+(T-n), n)\,\omega(n, s)
S_{\varepsilon, R}(\tau-s)\eta(x)\times \nonumber \\
&&\hskip 7cm \times \textbf{1}_{(n-1, n)}(s)\,\left(\widetilde Q (s)+\widetilde P(s)\right)\, ds||^2_{H^{\sigma}}
d\tau\\
&&\le Ce^{-2A(T-n)}\left( \int_{n-1}^{n+1} \,\textbf{1}_{(n-1, n)}(s)\,\omega(n, s)||\widetilde Q||^2_{H^{\sigma}}\, ds+\right.\nonumber \\
&&\left. \hskip 6cm +\frac{1}{\varepsilon^2}\int_{n-1}^{n+1} \,\textbf{1}_{(n-1, n)}(s)\,\omega(n, s)||\widetilde P||^2_{H^{\sigma-1/2}}
\right)
\\
&& \le Ce^{-2A(T-n)}\int_{n-1}^n||\widetilde Q||^2_{H^\sigma}\, ds+\frac{Ce^{-2A(T-n)}}{\varepsilon^2}
\int_{n-1}^n||\widetilde P||^2_{H^{\sigma-1/2}}\, ds,
\eean
whence:
\bean
&&I_1+I_2  \le  C\sum_{n=1}^{[T]} e^{-A(T-n)}\left(\int_{n-1}^n||\widetilde Q||^2_{H^ \sigma}\, ds\right)^{1/2}+C\left(\int_{[T]}^{T+1} ||\widetilde Q (s)||^2_{H^{\sigma}}\right)^{1/2}+ \nonumber \\
& + & \frac{C}{\varepsilon}\sum_{n=1}^{[T]}  e^{-A(T-n)}\left(
\int_{n-1}^n||\widetilde P||^2_{H^{\sigma-1/2}}\, ds\right)^{1/2}+\frac{C}{\varepsilon }\left(\int_{[T]}^{T+1} ||\widetilde P (s)||^2_{H^{\sigma-1/2}}\right)^{1/2}\nonumber\\
&\le & C\, \sup_{0\le T \le T_{max}}\left(\int_{T}^{T+1}||\widetilde Q||^2_{H^\sigma}\, ds\right)^{1/2}
+\frac{C}{\varepsilon}\, \sup_{0\le T \le T_{max}}\left(\int_{T}^{T+1}||\widetilde P||^2_{H^{\sigma-1/2}}\, ds\right)^{1/2}
\eean
and

\bear
\label{S2efeuno}
&&\sup_{0\le T\le T_{max}}\left( \int_T^{\min{(T+1, T_{max})}}||\overline f_1(s)||^2_{H^{\sigma}}ds\right)^{1/2}\le \nonumber \\
&&\hskip 2cm\le  C\, \sup_{0\le T \le T_{max}}\left(\int_{T}^{\min{(T+1, T_{max})}}||\widetilde Q||^2_{H^\sigma}\, ds\right)^{1/2}+\nonumber \\
&&\hskip 3cm+\frac{C}{\varepsilon}\, \sup_{0\le T \le T_{max}}\left(\int_{T}^{\min{(T+1, T_{max})}}||\widetilde P||^2_{H^{\sigma-1/2}}\, ds\right)^{1/2}
\eear
The term $\overline f_2$ is written as $\overline f_2=\overline f_{2, 1}+\overline f_{2,2}$ where $\overline f_{2, 1}$ and $\overline f_{2, 2}$ are defined as in (\ref{S2Eefedosdosdescomp}). We first estimate $\overline f_{2,1}$. Consider then
\bean
&&\left(\int_T^{T+1}||\int_0^t \omega(t, s) S_{\varepsilon, R}(t-s)
\left[T_{\varepsilon, R}\Psi(x, s)\right]\, ds||^2_{H^{\sigma}}
dt\right)^{1/2}\\
 &\le &
\sum_{n=1}^{[T]}\left(\int_T^{T+1}||\int_{n-1}^n  \omega(t, s) S_{\varepsilon, R}(t-s)
\left[T_{\varepsilon, R}\Psi(x, s)\right]\, ds||^2_{H^{\sigma}}
dt\right)^{1/2}\\
&&+\left(\int_T^{T+1} ||\int_{[T]}^t  \omega(t, s) S_{\varepsilon, R}(t-s)
\left[T_{\varepsilon, R}\Psi(x, s)\right]\, ds||^2_{H^{\sigma}}\right)^{1/2}\\
& = & I_1+I_2.
\eean
Arguing as in the derivation of (\ref{S2Eestfdosunoest}) we obtain that there exists a positive constant $\varepsilon_0$ that can be chosen arbitrarily small if $\delta$ is small enough, and such that:
\bear
\label{S2EstIdos29568}
I_2\le \varepsilon_0\left(\int_T^{\min{(T+1, T_{max})}} ||\overline f (s)||^2_{H^{\sigma}}\right)^{1/2}+C\, || \overline f||^2_{L^{\infty}}.\eear
In the first term $I_1$, we change the time variable $t$ as $t=\tau+(T-n)$ and obtain:
\bean
I_1\le\sum_{n=1}^{[T]}\left(\int_n^{n+1}||\int_{n-1}^n \omega(\tau+(T-n), s) S_{\varepsilon, R}(\tau+(T-n)-s)
\left[T_{\varepsilon, R}\Psi(x, s)\right]\, ds||^2_{H^{\sigma}}
d\tau\right)^{1/2}\\
\le \sum_{n=1}^{[T]}\left(\int_{n}^{n+1}||\int_{n-1}^n \omega(\tau+(T-n), s)S_{\varepsilon, R}(\tau-s)
\left[T_{\varepsilon, R}\Psi(x, s)\right]\, ds||^2_{H^{\sigma}}
d\tau\right)^{1/2}.
\eean
Arguing again as in the derivation of (\ref{S2Eestfdosunoest}) we obtain, for $\varepsilon_0$ defined as above:
\bean
&&\int_{n}^{n+1}||\int_{n-1}^n \omega(\tau+(T-n), s)S_{\varepsilon, R}(\tau-s)
\left[T_{\varepsilon, R}\Psi(x, s)\right]\, ds||^2_{H^{\sigma}}
d\tau\\
&&=\int_{n-1}^{n+1}||\int_{n-1}^\tau \omega(\tau+(T-n), s)S_{\varepsilon, R}(\tau-s)
\left[T_{\varepsilon, R}\left(\textbf{1}_{(n-1, n)}(s) \Psi(x, s)\right)\right] ds||^2_{H^{\sigma}}
d\tau
\\
&&\le e^{-2A(T-n)}\left(\varepsilon_0 \int_{n-1}^{n} ||\overline f||^2_{H^\sigma}\, ds+C\,|| \overline f||^2_{L^{\infty}}\right).
\eean
Therefore 
\bean
&&I_1\le \varepsilon_0 \sum_{n=1}^{[T]}e^{-A(T-n)}\left(\int_{n-1}^n||\overline f||^2_{H^ \sigma}\, ds\right)^{1/2}+
C\,\sum_{n=1}^{[T]}e^{-A(T-n)}|| \overline f||_{L^{\infty}}\\
&&\le \varepsilon_0 \sup_{0\le T\le T_{max}}\left(\int_{T}^{\min{(T+1, T_{max})}}||\overline f||^2_{H^ \sigma}\, ds\right)^{1/2}
+C\, || \overline f||_{L^{\infty}},
\eean
whence
\bean
I_1+I_2\le \varepsilon_0\,\sup_{0\le T\le T_{max}}\left(\int_{T}^{\min{(T+1, T_{max})}}||\overline f||^2_{H^ \sigma}\, ds\right)^{1/2}
+C\,|| \overline f||_{L^{\infty}}.
\eean
We then obtain the estimate:

\bear
\label{S2efedosuno}
&&\sup_{0\le T\le T_{max}}\left( \int_T^{\min{(T+1, T_{max})}}||\overline f_{2, 1}(s)||^2_{H^{\sigma}}ds\right)^{1/2} \nonumber \\
&&\hskip 2.2cm \le  \varepsilon_0\sup_{0\le T\le T_{max}}\!\!\!\left(\int_{T}^{\min{(T+1, T_{max})}}||\overline f||^2_{H^ \sigma}\, ds\right)^{1/2} +C\,|| \overline f||_{L^{\infty}}.
\eear

A similar argument using the contractivity of $S_{\varepsilon, R}$ in the spaces $H^\sigma$  gives for $\overline f_{2,2}$ and 
$\overline f_{3}$:
\bear
&&\sup_{0\le T\le T_{max}}\left( \int_T^{\min{(T+1, T_{max})}}||\overline f_{2,2} (t)||^2_{H_x^{\sigma}}dt\right)^{1/2} \nonumber\\
&&\hskip 2cm  \le C\sup_{0\le T\le T_{max}}\left(\int_{T}^{\min{(T+1, T_{max})}}||\widetilde f||^2_{H^{\sigma-\rho}}\, ds\right)^{1/2}
+C|| \widetilde f||_{L^{\infty}} \label{S2efedosdos}\\
&&\sup_{0\le T\le T_{max}}\left( \int_T^{\min{(T+1, T_{max})}}||\overline f_3 (t)||^2_{H_x^{\sigma}}dt\right)^{1/2} \nonumber\\ 
&&\hskip 1.5cm\le C\sup_{0\le T\le T_{max}}\left(\int_{T}^{\min{(T+1, T_{max})}}||\widetilde f||^2_{H^{\sigma-1/2}}\, ds\right)^{1/2}
+C|| \widetilde f||_{L^{\infty}}. \label{S2efetres}
\eear

We now estimate $\overline f_4$ :
\bean
&&\left(\int_T^{T+1}||\overline f_4(s)||^2_{H^{\sigma}}
dt\right)^{1/2}\\
 &\le &
\sum_{n=1}^{[T]}\left(\int_T^{T+1}||\int_{n-1}^n \omega(t, s) S_{\varepsilon, R}(t-s)
(a(x, t)-a(t))\, \overline f (s) ds||^2_{H^{\sigma}}
dt\right)^{1/2}\\
&&+\left(\int_T^{T+1} ||\int_{[T]}^t \omega(t, s) S_{\varepsilon, R}(t-s)
(a(x, t)-a(t))\, \overline f (s) ds||^2_{H^{\sigma}}\right)^{1/2}\\
& = & I_1+I_2.
\eean
We use now the continuity of the semigroup $S_{\varepsilon, R}$ in $H^{\sigma}$, the fact that $a\in H^{\sigma}(\RR)$ and since $\sigma>1/2$, the imbedding of $H^{\sigma}$ into $\textbf C(\RR)$ is continuous to obtain the existence of a positive constant $\varepsilon_0$, which can be made arbitrarily small if $\delta$ is sufficiently small, such that
\bean
||S_{\varepsilon, R}(t-s)
(a(x, t)-a(t))\, \overline f (s)||_{H^{\sigma}} & \le & 
||(a(x, t)-a(t))\, \overline f (s)||_{H^{\sigma}}\\ \\
& \le & \left(\varepsilon_0||\overline f||_{H^\sigma}+C ||\overline f||_{\infty}\right)\, ||a||_{H_x^{\sigma}}.
\eean
Arguing as in the derivation of (\ref{S2efedosuno}) we obtain
\bear
\label{S2efecuatro}
&&\sup_{0\le T\le T_{max}}\left( \int_T^{\min{(T+1, T_{max})}}||\overline f_{4}(s)||^2_{H^{\sigma}}ds\right)^{1/2} \nonumber \\
&&\hskip 2cm \le  \varepsilon_0\sup_{0\le T\le T_{max}}\!\!\!\left(\int_{T}^{\min{(T+1, T_{max})}}||\overline f||^2_{H^ \sigma}\, ds\right)^{1/2} +C\,|| \overline f||_{L^{\infty}}.
\eear

Adding formulas (\ref{S2efeuno}), (\ref{S2efedosuno})--(\ref{S2efecuatro}) we obtain:

\bean
&&\sup_{0\le T \le T_{max}}\left( \int_T^{\min{(T+1, T_{max})}}||\overline f (t)||^2_{H_x^{\sigma}}dt\right)^{1/2}
\le  \varepsilon_0 \!\!\!\sup_{0\le T\le T_{max}}\!\!\left(\int_{T}^{\min{(T+1, T_{max})}}||\overline f||^2_{H^\sigma}\, ds\right)^{1/2}\\
&&+ C\sup_{0\le T\le T_{max}}\left(\int_{T}^{\min{(T+1, T_{max})}}||\widetilde f||^2_{H^{\sigma-\rho}}\, ds\right)^{1/2}
+C|| \widetilde f||_{L^{\infty}}\\
&&+C\, \sup_{0\le T \le T_{max}}\left(\int_{T}^{\min{(T+1, T_{max})}}||\widetilde Q||^2_{H^\sigma}\, ds\right)^{1/2}
+\frac{C}{\varepsilon}\, \sup_{0\le T \le T_{max}}\left(\int_{T}^{T+1}||\widetilde P||^2_{H^{\sigma-1/2}}\, ds\right)^{1/2}
\eean
where we have used that $||\overline f||_{L^\infty}\le ||\widetilde f||_{L^\infty}$ and  $\rho \le  1/2$. Then,
\bear
\label{S2estefe39867}
&&\sup_{0\le T \le T_{max}}\!\!\left( \int_T^{\min{(T+1, T_{max})}}||\overline f (t)||^2_{H_x^{\sigma}}dt\right)^{1/2}\!\!\!
\le  \!C\!\!\!\sup_{0\le T\le T_{max}}\!\!\left(\int_{T}^{\min{(T+1, T_{max})}}||\widetilde f||^2_{H^{\sigma-\rho}}\, ds\right)^{1/2} \nonumber \\
&&+C|| \widetilde f||_{L^{\infty}}+C\, \sup_{0\le T \le T_{max}}\left(\int_{T}^{\min{(T+1, T_{max})}}||\widetilde Q||^2_{H^s}\, ds\right)^{1/2}\nonumber \\
&&\hskip 4cm+\frac{C}{\varepsilon}\, \sup_{0\le T \le T_{max}}\left(\int_{T}^{T+1}||\widetilde P||^2_{H^{\sigma-1/2}}\, ds\right)^{1/2}.
\eear

Using a partition of unity as in the derivation of (\ref{S2Eestipart}), we arrive at
\bear
\label{S2estefe39868}
&&\sup_{0\le T \le T_{max}}\left( \int_T^{\min{(T+1, T_{max})}}||\widetilde f (t)||^2_{H_x^{\sigma}}dt\right)^{1/2}
\le   C\sup_{0\le T\le T_{max}}\left(\int_{T}^{\min{(T+1, T_{max})}}||\widetilde f||^2_{H^{\sigma-\rho}}\, ds\right)^{1/2} \nonumber \\
&&+C|| \widetilde f||_{L^{\infty}}+C\, \sup_{0\le T \le T_{max}}\left(\int_{T}^{\min{(T+1, T_{max})}}||\widetilde Q||^2_{H^\sigma}\, ds\right)^{1/2} \nonumber \\
&&\hskip 4cm+\frac{C}{\varepsilon}\, \sup_{0\le T \le T_{max}}\left(\int_{T}^{T+1}||\widetilde P||^2_{H^{\sigma-1/2}}\, ds\right)^{1/2} 
\eear
where the constants $C>0$ depend on $\delta$. An interpolation argument yields 
\bear
\label{S2ldkvn}
&&\sup_{0\le T\le T_{max}}\left(\int_{T}^{\min(T+1, T_{max})}||\widetilde f(t)||^2_{H^{\sigma}}\, dt\right)^{1/2}\le \nonumber \\
&&\qquad C\sup_{0\le T\le T_{max}}\left(\int_{T}^{\min(T+1, T_{max})}|| \widetilde  Q(t)||^2_{H^{\sigma}}\, dt\right)^{1/2}+C\,|| \widetilde  f||_{L^\infty} \nonumber \\
&&\hskip 4cm +\frac{C}{\varepsilon}\, \sup_{0\le T \le T_{max}}\left(\int_{T}^{T+1}||\widetilde P||^2_{H^{\sigma-1/2}}\, ds\right)^{1/2} 
\eear
Using that $\chi=1$ in the interval $(5/8, 11/8)$ we have:
\bear
\label{S2ldkvn2}
||f||_{H^{\sigma}(3/4, 5/4)}\le ||\widetilde f||_{H^\sigma}.
\eear
Part (ii) of Theorem \ref{S8T3-101} then follows combining (\ref{S2ldkvn}) and (\ref{S2ldkvn2}).
\\

\textbf{Proof of part (iii) of Theorem \ref{S8T3-101}.} The equation satisfied by $\overline f$ is now (\ref{S3eqfbarrra}) with $\widetilde P=0$ and $\varepsilon=0$.  Then:
\bean
\overline f(x, t) & = & \int_0^t \omega(t, s)S_{\varepsilon, R}(t-s)\eta(x)\,\widetilde Q (s)\, ds\\
&&+\int_0^t \omega(t, s)S_{\varepsilon, R}(t-s)
\left[\eta(x)T_{\varepsilon, R}\left((M_{\lambda/2}-M_{\lambda/2, 0})\widetilde f \right)\right]\, ds \nonumber \\
&&+\, x_0^{\lambda/2}\int_0^t \omega(t, s)S_{\varepsilon, R}(t-s)\int_0^{\infty}\tilde f(x-y, s)\left( 
\eta(x)-\eta(x-y)\right)\Phi(y, R, \varepsilon)\, dy\, ds \\
&&-\int_0^t \omega(t, s)S_{\varepsilon, R}(t-s)(a(x, t)-a(t))\, \overline f (s)\, ds
\label{S8T3-101E5} \\
& = & \overline f_1(x, t)+ \overline f_2(x, t)+ \overline f_3(x, t)+ \overline f_4(x, t) \label{S8T3-101E567}
\eean
where $\omega$ is given by (\ref{S2Eomega}). 
The term $\overline f _1$ is estimated using (\ref{S2ETsemigrupos2}) for $T\le 1$. Then,

\bear
\label{cotaefeunoN}
&&\left( \int_T^{T+1}||T_1(\overline f _1)(t)||^2_{H_x^{\sigma}}dt\right)^{1/2}\le \left( \int_0^{2}||T_1( \overline f _1)(t)||^2_{H_x^{\sigma}}dt\right)^{1/2}\nonumber \\
&&\hskip 3cm \le C\left( \int_0^{2}|| \widetilde Q(t)||^2_{H_x^{\sigma}}dt\right)^{1/2}.
\eear
If $T>1$, using the change of variables: $t=(T-1)+\tau$ we write
\bean
&&\left(\int_T^{T+1}||T_1\left(\int_0^t\omega(t, s) S_{\varepsilon, R}(t-s)\eta(x)\,\widetilde Q (s)\, ds\right)||^2_{H^{\sigma}}dt\right)^{1/2}\\
 &\le &
\sum_{n=1}^{[T]}\left(\int_T^{T+1}||\int_{n-1}^n \omega(t, s)T_1\left( S_{\varepsilon, R}(t-s)\eta(x)\,\widetilde Q (s)\right)\, ds||^2_{H^{\sigma}}
dt\right)^{1/2}\\
&&+\left(\int_T^{T+1} ||\int_{[T]}^t  \omega(t, s) T_1\left(S_{\varepsilon, R}(t-s)\eta(x)\,\widetilde Q (s)\right)\, ds||^2_{H^{s}}dt\right)^{1/2}\\
& = & I_1+I_2.
\eean
The estimate of the term $I_2$ can be made as in (\ref{cotaefeunoN}):
\bean
I_2\le C\left(\int_T^{T+1} ||\widetilde Q (s)||^2_{H^{\sigma}}\right)^{1/2}.
\eean
To estimate $I_1$ we argue as follows:
we change the time variable $t$ as $t=\tau+(T-n)$ and obtain:
\bean
I_1=\sum_{n=1}^{[T]}\left(\int_n^{n+1}||\int_{n-1}^n  \omega(\tau+(T-n), s) 
T_1\, S_{\varepsilon, R}((T-n)+\tau-s)\left(\eta(x)\,\widetilde Q (s)\right)\, ds||^2_{H^{\sigma}}
d\tau\right)^{1/2}\\
\le \sum_{n=1}^T\left(\int_n^{n+1}||\int_{n-1}^n  \omega(\tau+(T-n), s) 
T_1\,S_{\varepsilon, R}(\tau-s)\left(\eta(x)\,\widetilde Q (s)\right)\, ds||^2_{H^{\sigma}}
d\tau\right)^{1/2}
\eean
where we have used  $||S_\varepsilon(T-n)\, h||_{H^{\sigma}}\le ||h||_{H^\sigma}$. We notice now that, for each $n$
\bean
&&\int_n^{n+1}||\int_{n-1}^n \omega(\tau+(T-n), s)\,T_1  S_{\varepsilon, R}(\tau-s)(\eta(x)\,\widetilde Q (s))\, ds||^2_{H^{\sigma}}
d\tau\\
&&\le\int_{n-1}^{n+1}||\int_{n-1}^\tau  \omega(\tau+(T-n), s)
T_1\, S_{\varepsilon, R}(\tau-s)(\eta(x)\,\textbf{1}_{(n-1, n)}(s)\,\widetilde Q (s))\, ds||^2_{H^{\sigma}}
d\tau\\
&&=\int_{n-1}^{n+1}||\int_{n-1}^\tau  \omega(\tau+(T-n), n)\,\omega(n, s)T_1\, 
S_{\varepsilon, R}(\tau-s)(\eta(x)\,\textbf{1}_{(n-1, n)}(s)\,\widetilde Q (s))\, ds||^2_{H^{\sigma}}
d\tau\\
&&\le Ce^{-2A(T-n)}\left( \int_{n-1}^{n+1} \,\textbf{1}_{(n-1, n)}(s)\,\omega(n, s)||\widetilde Q||^2_{H^{\sigma}}\, ds
\right)
\\
&& \le Ce^{-2A(T-n)}\int_{n-1}^n||\widetilde Q||^2_{H^\sigma}\, ds,
\eean
whence:
\bean
I_1+I_2 & \le & C\sum_{n=1}^{[T]} e^{-A(T-n)}\left(\int_{n-1}^n||\widetilde Q||^2_{H^ \sigma}\, ds\right)^{1/2}+C\left(\int_{[T]}^{T+1} ||\widetilde Q (s)||^2_{H^{\sigma}}\right)^{1/2}\nonumber \\
&\le &  C\, \sup_{0\le T \le T_{max}}\left(\int_{T}^{T+1}||\widetilde Q||^2_{H^\sigma}\, ds\right)^{1/2}
\eean
and

\bear
\label{S2efeunoN}
&&\sup_{0\le T\le T_{max}}\left( \int_T^{\min{(T+1, T_{max})}}||T_1(\overline f_1)(s)||^2_{H^{\sigma}}ds\right)^{1/2} \nonumber \\
& \le &  C\, \sup_{0\le T \le T_{max}}\left(\int_{T}^{\min{(T+1, T_{max})}}||\widetilde Q||^2_{H^\sigma}\, ds\right)^{1/2}.
\eear

The term $\overline f_2$ is written as $\overline f_2=\overline f_{2, 1}+\overline f_{2,2}$ where $\overline f_{2, 1}$ and $\overline f_{2, 2}$ are defined as in (\ref{S2Eefedosdosdescomp}). We first estimate $\overline f_{2,1}$ . Consider then
\bean
&&\left(\int_T^{T+1}||T_1\int_0^t \omega(t, s) S_{\varepsilon, R}(t-s)
\left[T_{\varepsilon, R}\Psi(x, s)\right]\, ds||^2_{H^{\sigma}}
dt\right)^{1/2}\\
 &\le &
\sum_{n=1}^{[T]}\left(\int_T^{T+1}||\int_{n-1}^n  \omega(t, s) T_1\, S_{\varepsilon, R}(t-s)
\left[T_{\varepsilon, R}\Psi(x, s)\right]\, ds||^2_{H^{\sigma}}
dt\right)^{1/2}\\
&&+\left(\int_T^{T+1} ||\int_{[T]}^t  \omega(t, s) T_1\, S_{\varepsilon, R}(t-s)
\left[T_{\varepsilon, R}\Psi(x, s)\right]\, ds||^2_{H^{\sigma}}\right)^{1/2}\\
& = & I_1+I_2.
\eean
Arguing as in the derivation of (\ref{S2Eestfdosunoest}), but using (\ref{S2TlemasobolevDos}) instead of (\ref{S2TlemasobolevUno})
we obtain that there exists a positive constant $\varepsilon_0$ that can be chosen arbitrarily small such that:
\bear
\label{S2EstIdos29568N}
I_2\le \varepsilon_0\left(\int_T^{\min{(T+1, T_{max})}} ||T_1\,\overline f (s)||^2_{H^{\sigma}}\right)^{1/2}+C\, || \overline f||^2_{L^{\infty}}.\eear
In the first term $I_1$, we change the time variable $t$ as $t=\tau+(T-n)$ and obtain:
\bean
I_1\le\sum_{n=1}^{[T]}\left(\int_n^{n+1}||\int_{n-1}^n \omega(\tau+(T-n), s) T_1\,S_{\varepsilon, R}(\tau+(T-n)-s)
\left[T_{\varepsilon, R}\Psi(x, s)\right]\, ds||^2_{H^{\sigma}}
d\tau\right)^{1/2}\\
\le \sum_{n=1}^{[T]}\left(\int_{n}^{n+1}||\int_{n-1}^n \omega(\tau+(T-n), s) T_1\,S_{\varepsilon, R}(\tau-s)
\left[T_{\varepsilon, R}\Psi(x, s)\right]\, ds||^2_{H^{\sigma}}
d\tau\right)^{1/2}.
\eean
Arguing again as in the derivation of (\ref{S2Eestfdosunoest}) we obtain, for $\varepsilon_0$ defined as above,
\bean
&&\int_{n}^{n+1}||\int_{n-1}^n \omega(\tau+(T-n), s) T_1\,S_{\varepsilon, R}(\tau-s)
\left[T_{\varepsilon, R}\Psi(x, s)\right]\, ds||^2_{H^{\sigma}}
d\tau\\
&&\le e^{-2A(T-n)}\left(\varepsilon_0 \int_{n-1}^{n+1} \,\textbf{1}_{(n-1, n)}(s)||T_1\,\overline f||^2_{H^s}\, ds+C\,|| \overline f||^2_{L^{\infty}}\right)\\
&&= e^{-2A(T-n)}\left(\varepsilon_0 \int_{n-1}^{n} ||T_1\,\overline f||^2_{H^s}\, ds+C\,|| \overline f||^2_{L^{\infty}}\right).
\eean
Therefore 
\bean
&&I_1\le \varepsilon_0 \sum_{n=1}^{[T]}e^{-A(T-n)}\left(\int_{n-1}^n||T_1\,\overline f||^2_{H^ \sigma}\, ds\right)^{1/2}+
C\,\sum_{n=1}^{[T]}e^{-A(T-n)}|| \overline f||_{L^{\infty}}\\
&&\le \varepsilon_0 \sup_{0\le T\le T_{max}}\left(\int_{T}^{\min{(T+1, T_{max})}}||T_1\,\overline f||^2_{H^ \sigma}\, ds\right)^{1/2}
+C\, || \overline f||_{L^{\infty}},
\eean
whence
\bean
I_1+I_2\le \varepsilon_0\,\sup_{0\le T\le T_{max}}\left(\int_{T}^{\min{(T+1, T_{max})}}||T_1\,\overline f||^2_{H^ \sigma}\, ds\right)^{1/2}
+C\,|| \overline f||_{L^{\infty}}
\eean
and therefore

\bear
\label{S2efedosunoN}
&&\sup_{0\le T\le T_{max}}\left( \int_T^{\min{(T+1, T_{max})}}||T_1\, \overline f_{2, 1}(s)||^2_{H^{\sigma}}ds\right)^{1/2} \nonumber \\
&&\hskip 1.5cm \le  \varepsilon_0\sup_{0\le T\le T_{max}}\!\!\!\left(\int_{T}^{\min{(T+1, T_{max})}}||T_1\,\overline f||^2_{H^ \sigma}\, ds\right)^{1/2} +C\,|| \overline f||_{L^{\infty}}.
\eear

A similar argument using the contractivity of $S_{\varepsilon, R}$ in the spaces $H^\sigma$ and formula (\ref{diadicdos}) gives 
\bear
&&\sup_{0\le T\le T_{max}}\left( \int_T^{\min{(T+1, T_{max})}}||T_1\,\overline f_{2,2} (t)||^2_{H_x^{\sigma}}dt\right)^{1/2} \nonumber\\
&&\hskip 1.5cm  \le C\sup_{0\le T\le T_{max}}\left(\int_{T}^{\min{(T+1, T_{max})}}||T_1\widetilde f||^2_{H^{\sigma-\rho}}\, ds\right)^{1/2}
+C|| \widetilde f||_{L^{\infty}} \label{S2efedosdosN}\eear
To estimate  $\overline f_3$ we combine (\ref{S2ETsemigrupos2}) and (\ref{S2overlineftrestres}) to obtain
\bear
&&\sup_{0\le T\le T_{max}}\left( \int_T^{\min{(T+1, T_{max})}}||T_1\overline f_3 (t)||^2_{H_x^{\sigma}}dt\right)^{1/2} \nonumber\\ 
&&\hskip 1.7cm\le C\sup_{0\le T\le T_{max}}\left(\int_{T}^{\min{(T+1, T_{max})}}||\widetilde f||^2_{H^{\sigma-1/2}}\, ds\right)^{1/2}
+C|| \widetilde f||_{L^{\infty}}. \label{S2efetresN}
\eear

We now estimate $\overline f_4$ :
\bean
&&\left(\int_T^{T+1}||T_1\overline f_4(s)||^2_{H^{\sigma}}
dt\right)^{1/2}\\
 &\le &
\sum_{n=1}^{[T]}\left(\int_T^{T+1}||\int_{n-1}^n \omega(t, s) T_1\, S_{\varepsilon, R}(t-s)
(a(x, t)-a(t))\, \overline f (s) ds||^2_{H^{\sigma}}
dt\right)^{1/2}\\
&&+\left(\int_T^{T+1} ||\int_{[T]}^t \omega(t, s) T_1\,S_{\varepsilon, R}(t-s)
(a(x, t)-a(t))\, \overline f (s) ds||^2_{H^{\sigma}}\right)^{1/2}\\
& = & I_1+I_2.
\eean
Using (\ref{S2ETsemigrupos2}) we get
\bean
I_2\le \left(\int_T^{T+1}||(a(x, t)-a(t))\overline f||_{H^\sigma(\RR)}^2\right)^{1/2}
\eean
Since $a\in H^{\sigma+1}\subset C^{1, \alpha}$ for some $\alpha > 0$, we obtain,for $\delta$  sufficiently small
\bean
I_2\le C\left(\int_T^{T+1}\left(\varepsilon_0\, ||\overline f (t)||_{H^\sigma(\RR)}^2+C||\overline f(t)||^2_{L^\infty}\right) dt\right)^{1/2}
\eean
The term $I_1$ can be estimated similarly using  the exponential decay of  $\omega (t, s)$ as in the previous cases. Then
\bear
\label{S2efecuatroN}
&&\sup_{0\le T\le T_{max}}\left( \int_T^{\min{(T+1, T_{max})}}||T_1\, \overline f_{4}(s)||^2_{H^{\sigma}}ds\right)^{1/2} \nonumber \\
&&\hskip 2cm \le  \varepsilon_0\sup_{0\le T\le T_{max}}\!\!\!\left(\int_{T}^{\min{(T+1, T_{max})}}||\overline f||^2_{H^ \sigma}\, ds\right)^{1/2} +C\,|| \overline f||_{L^{\infty}}.
\eear

Adding formulas (\ref{S2efeuno}), (\ref{S2efedosuno})--(\ref{S2efecuatro}) we obtain:

\bean
&&\sup_{0\le T \le T_{max}}\left( \int_T^{\min{(T+1, T_{max})}}||T_1\,\overline f (t)||^2_{H_x^{\sigma}}dt\right)^{1/2}
\le  \varepsilon_0 \sup_{0\le T\le T_{max}}\left(\int_{T}^{\min{(T+1, T_{max})}}||T_1\,\overline f||^2_{H^\sigma}\, ds\right)^{1/2}\\
&&+ C\sup_{0\le T\le T_{max}}\left(\int_{T}^{\min{(T+1, T_{max})}}||\widetilde f||^2_{H^{\sigma-\rho}}\, ds\right)^{1/2}
+C|| \widetilde f||_{L^{\infty}}\\
&&+C\, \sup_{0\le T \le T_{max}}\left(\int_{T}^{\min{(T+1, T_{max})}}||\widetilde Q||^2_{H^\sigma}\, ds\right)^{1/2}
\eean
where we have used that $||\overline f||_{L^\infty}\le ||\widetilde f||_{L^\infty}$ and  $\rho \le  1/2$. Then,
\bear
\label{S2estefe39867N}
&&\sup_{0\le T \le T_{max}}\left( \int_T^{\min{(T+1, T_{max})}}||T_1\,\overline f (t)||^2_{H_x^{\sigma}}dt\right)^{1/2}
\le   C\sup_{0\le T\le T_{max}}\left(\int_{T}^{\min{(T+1, T_{max})}}||\widetilde f||^2_{H^{\sigma-\rho}}\, ds\right)^{1/2} \nonumber \\
&&+C|| \widetilde f||_{L^{\infty}}+C\, \sup_{0\le T \le T_{max}}\left(\int_{T}^{\min{(T+1, T_{max})}}||\widetilde Q||^2_{H^s}\, ds\right)^{1/2}.
\eear

Using a partition of unity as for the derivation of (\ref{S2Eestipart}), we arrive at
\bear
\label{S2estefe39868N}
&&\sup_{0\le T \le T_{max}}\left( \int_T^{\min{(T+1, T_{max})}}||T_1\widetilde f (t)||^2_{H_x^{\sigma}}dt\right)^{1/2}
\le   C\sup_{0\le T\le T_{max}}\left(\int_{T}^{\min{(T+1, T_{max})}}||\widetilde f||^2_{H^{\sigma-\rho}}\, ds\right)^{1/2} \nonumber \\
&&+C|| \widetilde f||_{L^{\infty}}+C\, \sup_{0\le T \le T_{max}}\left(\int_{T}^{\min{(T+1, T_{max})}}||\widetilde Q||^2_{H^s}\, ds\right)^{1/2}
\eear
where the constants $C>0$ depend on $\delta$. An interpolation argument yields 
\bean
\label{S2estefe39868NN}
&&\sup_{0\le T\le T_{max}}\left(\int_{T}^{\min(T+1, T_{max})}||T_1\,\widetilde f(t)||^2_{H^{\sigma}}\, dt\right)^{1/2}\le \\
&&\qquad C\sup_{0\le T\le T_{max}}\left(\int_{T}^{\min(T+1, T_{max})}|| \widetilde  Q(t)||^2_{H^{\sigma}}\, dt\right)^{1/2}+C\,|| \widetilde  f||_{L^\infty}  
\eean
Using that $\chi=1$ in the interval $(5/8, 11/8)$, we obtain
\bear
||T_1f||_{H^{\sigma}(3/4, 5/4)}\le C||T_1\widetilde f||_{H^\sigma}.
\eear
On the other hand,
\bean
|W(k, R, 0)|\le C \min\{|k|, R\}
\eean
Therefore estimate (\ref{S8T3-101E143bisbis}) holds. 
This concludes the proof of part (iii) of Theorem \ref{S8T3-101}.

\qed

\noindent
We state now the main result of this Section.

\begin{theo}
\label{S8T3-100}
Suppose that $\sigma \in (1/2,\, 2)$, $\nu \in L^2_t(0, 1;H_x^\sigma(1/4, 4))$, 
$\varepsilon \in [0, 1]$, $K\in L^\infty((1/4, 4)\times (0, 1))\cap L^2_t(0, 1;H_x^\sigma(1/4, 4))$,  and $h\in L^\infty((1/8, 4)\times (0, 1))\cap  L^2(0, 1; H^{1/2}(1/4, 2))\cap H^1(0, 1; L^2(1/4, 2))$, $W\in L^2_t(0, 1;H_x^{\sigma-1/2}(1/4, 4))$ satisfies :
\bean
\frac{\partial h}{\partial t} & = & \varepsilon \int_0^{x/2}\frac{(x-y)^{\lambda/2}h(x-y)-x^{\lambda/2}h(x)}{y^{3/2}}+\\
&&+(1-\varepsilon) R^{3/2}\int_0^{x/2}\left((x-y)^{\lambda/2}h(x-y)-x^{\lambda/2}h(x)\right){(Ry)^{\lambda/2}}f_0(R\, y)\, dy+\\
&&+
K(x, t)\, h(x, t)+\nu(x, t)+W(x, t),
\eean
for all $x \in (1/4, 4)$ and $R>1$ and $h(x, 0)=0$. Then for any $T\in [0, 1]$:
\bean
&&||h||_{L^2_t(0, T;H^\sigma_x(7/8, 9/8))}\le \\
&& \hskip 2.5cm C\left(||\nu||_{ L^2_t(0, 1;H_x^\sigma(1/4, 4))}+||h||_{L^\infty((1/8, 4)\times (0, 1))}
+ \frac{1}{\varepsilon}||W||_{ L^2_t(0, 1;H_x^{\sigma-1/2}(1/4, 4))}\right)
\eean
where the constant $C$ is independent of $\varepsilon$ and $R$ but depends on 
$||K||_{L^\infty((1/2, 2)\times (0, 1))}$ and
$||K||_{L^2_t(0, 1;H_x^\sigma(1/4, 4))}$.
\end{theo}

\textbf{Proof of Theorem \ref{S8T3-100}.}
Let  $\chi$ be a $C^{\infty}$ function such that  
\bear
\chi(x)=\left\{
\begin{array}{l}
1\quad \hbox{if}\,\,x\in \left(1/2, 2\right)\,,\\ \\
0\quad \hbox{if}\,\,\,x\not\in \left(1/4, 4\right)\, .
\end{array}
\right.
\eear
We define $\tilde h(x, t)= \chi(x)\, h(x).$ Then, for all $x\in \RR$ the function $\tilde h$ satisfies
\bear
\frac{\partial \tilde h}{\partial t} & = & \int_0^{x/2}\left((x-y)^{\lambda /2}\tilde h(x-y)-x^{\lambda /2}\tilde h(x)
\right)\Phi(y, R, \varepsilon)dy  \nonumber \\
&&+ \int_0^{x/2}(x-y)^{\lambda /2} h(x-y)\left(\chi(x)-\chi (x-y) \right)\, \Phi(y, R, \varepsilon)\, dy \nonumber \\
&&+K(x, t)\, \tilde h(x, t)+\chi (x)\, \nu +\chi (x)\, W\nonumber \\
& = & \int_0^{+\infty}\left((x-y)^{\lambda /2}\tilde h(x-y)-x^{\lambda /2}\tilde h(x)
\right)\Phi(y, R, \varepsilon)dy \nonumber \\
&& - \int_{x/2}^{+\infty}\left((x-y)^{\lambda /2}\tilde h(x-y)-x^{\lambda /2}\tilde h(x)
\right)\Phi(y, R, \varepsilon)dy\nonumber \\
&&+ \int_0^{x/2}(x-y)^{\lambda /2} h(x-y)\left(\chi(x)-\chi (x-y) \right)\, \Phi(y, R, \varepsilon)\, dy \nonumber \\
&&+K(x, t)\, \tilde h(x, t)+\chi (x)\, \nu +\chi (x)\, W
.\label{S9E1}
\eear
where the function $ \Phi(y, R, \varepsilon)$ has been defined in (\ref{S9E10000Phi}).\\Ê\\
We write the equation (\ref{S9E1}) in terms of the new function $\overline h$ defined as:
\bean
\tilde h(x, t)=e^{\left(\int_0^tK(x, s)\, ds+c_0(\varepsilon, R, x) t\right)}\, \overline{h }(x, t)
\eean
with,
\bean
c_0(\varepsilon, R, x)=x^{\lambda/2}\int_{x/2}^{\infty}\Phi(y, R, \varepsilon)\, dy
\eean
\bean
\frac{\partial \overline h}{\partial t}& = &  T_{\varepsilon, R}\left(M_{\lambda/2}\, \overline h \right)+Q_1+Q_2+Q_3+Q_4
\eean
where $T_{\varepsilon, R}$ has been defined in (\ref{S9E10000}) and
\bean
Q_1 & = &-\int_{x/2}^{+\infty}(x-y)^{\lambda /2}\overline h(x-y)
\Phi(y, R, \varepsilon)dy\\
Q_2 & = &  e^{-\left(\int_0^tK(x, s)\, ds+2\sqrt 2 \varepsilon t x^{(\lambda-1)/2}\right)}\,\int_0^{x/2}(x-y)^{\lambda /2} h(x-y)\left(\chi(x)-\chi (x-y) \right)\, \Phi(y, R, \varepsilon)\, dy \\
Q_3 & = & e^{-\left(\int_0^tK(x, s)\, ds+c_0(\varepsilon, R, x)t\right)}\,\chi (x)\,\nu\\
Q_4 & = & e^{-\left(\int_0^tK(x, s)\, ds+c_0(\varepsilon, R, x)t\right)}\,\chi (x)\,W.
\eean
These terms are estimated as follows:
\bean
||Q_1||_{L^{\infty}(0, 1; W^{1, \infty}((1/4, 4)))}+
||Q_2||_{L^{\infty}(0, 1; H_x^{\tilde \sigma}((1/4, 4)))}
 & \le & C\, ||h||_{L^{\infty}((1/8, 4)\times (0, 1))}\\
||Q_3||_{ L^2_t(0, 1;H_x^\sigma(1/4, 4))} & \le & C\, ||\nu||_{ L^2_t(0, 1;H_x^\sigma(1/4, 4))},
\\
||Q_4||_{ L^2_t(0, 1;H_x^{\sigma-1/2}(1/4, 4))} & \le & C\, ||W||_{ L^2_t(0, 1;H_x^{\sigma-1/2}(1/4, 4))},
\eean
where $\tilde \sigma=\min\{\sigma, 1\}$.

If $1/2<\sigma\le 1$ Theorem \ref{S8T3-101} immediately yields:
\bean
||\overline h||_{L^2_t(0, 1;H^\sigma_x(3/4, 5/4))} & \le & C\left(||h||_{L^{\infty}((1/8, 4)\times (0, 1))}+\, ||\nu||_{ L^2_t(0, 1;H_x^\sigma(1/4, 4))}
+||h||_{L^{\infty}((1/4, 2)\times (0, 1))}
 \right) +\nonumber \\
 &&+\frac{C}{\varepsilon}||W||_{ L^2_t(0, 1;H_x^{\sigma-1/2}(1/4, 4))}\nonumber \\
 & \le & C\left(||\nu||_{ L^2_t(0, 1;H_x^\sigma(1/4, 4))}
+||h||_{L^{\infty}((1/8, 4)\times (0, 1))}\right)+\nonumber \\
&&+\frac{C}{\varepsilon}||W||_{ L^2_t(0, 1;H_x^{\sigma-1/2}(1/4, 4))} .
\eean
If $\sigma > 1$ we apply Theorem \ref{S8T3-101} with $\sigma=1$ to obtain:
\bear
\label{S8T3-101E14339865}
||\overline h||_{L^2_t(0, 1;H^1_x(3/4, 5/4))} & \le & C\left(||h||_{L^{\infty}((1/8, 4)\times (0, 1))}+\, ||\nu||_{ L^2_t(0, 1;H_x^\sigma(1/4, 4))}
+||h||_{L^{\infty}((1/4, 2)\times (0, 1))}
 \right) +\nonumber \\
 &&+\frac{C}{\varepsilon}||W||_{ L^2_t(0, 1;H_x^{\sigma-1/2}(1/4, 4))}\nonumber \\
 & \le & C\left(||\nu||_{ L^2_t(0, 1;H_x^\sigma(1/4, 4))}
+||h||_{L^{\infty}((1/8, 4)\times (0, 1))}
 \right)+\nonumber \\
 && +\frac{C}{\varepsilon}||W||_{ L^2_t(0, 1;H_x^{\sigma-1/2}(1/4, 4))}.
\eear
Since $Q_1$ and $Q_2$ involve integrals of the function $\overline h$, 
(\ref{S8T3-101E14339865}) provides better estimates on $Q_1$ and $Q_2$ although on the smaller interval $(3/4, 5/4)$:
\bean
||Q_1||_{L^{\infty}(0, 1; H^{\sigma}((3/4, 5/4)))}& +&||Q_2||_{L^{\infty}(0, 1; H^{\sigma}((3/4, 5/4)))} \le   C\, ||\overline h||_{L^2_t(0, 1;H^1_x(3/4, 5/4))}\\
& \le & C\left(||\nu||_{ L^2_t(0, 1;H_x^\sigma(1/4, 4))}
+||h||_{L^{\infty}((1/8, 4)\times (0, 1))}\right).
\eean
Using again the Theorem \ref{S8T3-101} with $\sigma<2$:
\bean
||\overline h||_{L^2_t(0, 1;H^\sigma_x(7/8, 9/8))} & \le & C\left(||h||_{L^{\infty}((1/8, 4)\times (0, 1))}+\, ||\nu||_{ L^2_t(0, 1;H_x^s(1/4, 4))}
+\right.\nonumber \\
&&\left. +||h||_{L^{\infty}((1/4, 2)\times (0, 1))}
 \right)+\frac{C}{\varepsilon}||W||_{ L^2_t(0, 1;H_x^{\sigma-1/2}(1/4, 4))} \nonumber \\
 & \le & C\left(||\nu||_{ L^2_t(0, 1;H_x^s(1/4, 4))}
+||h||_{L^{\infty}((1/8, 4)\times (0, 1))}
 \right)+ \nonumber \\
 &&+\frac{C}{\varepsilon}||W||_{ L^2_t(0, 1;H_x^{\sigma-1/2}(1/4, 4))}.
\eean
This ends the proof of Theorem \ref{S8T3-100}.
\qed
\\ \\
We end this Section with the following property of the operator $\mathcal L -L$.
\begin{lem}
\label{S2Tpseudo}
Consider the operators $\mathcal W_R$ and $\mathcal W_\infty$ defined as:
\bear
&&\mathcal W_R(h)=R^{(3+\lambda)/2}\int_0^{x/2}\left((x-y)^{\lambda/2}h(x-y)-x^{\lambda/2}h(x) \right)y^{\lambda/2}f_0(Ry) dy \label{S2Eopelredondaerre}\\
&&\mathcal W_\infty(h)=\int_0^{x/2}\left((x-y)^{\lambda/2}h(x-y)-x^{\lambda/2}h(x) \right)y^{-3/2}dy \label{S2Eopelredondainfty}\\
&&\mathcal W_{\infty,\, \varepsilon}(h)=\int_0^{x/2}\left((x-y)^{\lambda/2}h(x-y)-x^{\lambda/2}h(x) \right)\frac{dy}{y^{3/2}+\varepsilon^{3/2}x^{3/2}} \label{S2Eopelredondaepsilon}
\eear
Then, for any $\eta \in {\bf C}^{\infty}(\RR) $ of compact support contained in $(1/2, 3/2)$ such that $\eta=1$ on $(3/4, 5/4)$, and for all $\sigma \ge 1/2$, there exists a positive constant $C$, depending only on the function $\eta$ and its derivatives, such that for all $\psi \in H^\sigma(\RR)$:
\bear
\label{S2Ederivadasigmamedios}
||\mathcal W_\infty\,(\eta\, \psi)||_{H^{\sigma-1/2}(\RR)}+||\mathcal W_R\,(\eta\, \psi)||_{H^{\sigma-1/2}(\RR)}
+||\mathcal W_{\infty,\,\varepsilon} \,(\eta\, \psi)||_{H^{\sigma-1/2}(\RR)}\le C ||\eta\,  \psi||_{H^{\sigma}(\RR)}.\nonumber\\
\eear
Moreover, for all $h\in H^\sigma (\RR)$ fixed:
\bear
\label{S2Elimites}
\lim_{\varepsilon \to 0}||\eta\,\cdot\left(\mathcal W_{\infty,\,\varepsilon} -\mathcal W_{\infty}\right)(\eta\, h) ||_{H^{\sigma-1/2}(\RR)}=0.
\eear
\end{lem}
{\bf Proof of Lemma \ref{S2Tpseudo}.} The function $\mathcal W_R\,(\eta\, \psi)$ can  be written as follows
\bean
R^{(3+\lambda)/2}\int_0^{x/2}\left((x-y)^{\lambda/2}\eta(x-y)\, \psi(x-y)-x^{\lambda/2}\eta (x)\psi(x) \right)y^{\lambda/2}f_0(Ry) dy\\
=
T_{0, R}\circ M_{\lambda/2}(\eta\, \psi)+{\cal Z}
\eean
with
\bean
||{\cal Z}||_{H^\sigma}\le C ||\eta\, \psi||_{H^\sigma}.
\eean
Using now the fact that the operator $T_{0, R}$ is the  multiplier by a function bounded by $|\xi|^{1/2}$ and  $M_{\lambda/2}\, h$ is the product of $h$  by $x^{\lambda/2}$ which is a smooth function in the interval $(1/2, /2)$ the result follows. The same argument yields the estimate for $\mathcal W_\infty$:
\bear
\label{S3EstiZeta25}
||\mathcal W_\infty\,(\eta\, \psi)||_{H^{\sigma-1/2}(\RR)}\le C  ||\eta\, \psi||_{H^\sigma}.
\eear
The third operator 
$\mathcal W_{\infty,\,\varepsilon}$  may be written as a pseudo differential operator with symbol 
\bear
\label{S3EstiZeta35}
P_{\varepsilon}(x, k)=\int_0^\infty
\frac{(e^{-iky}-1)}{y^{3/2}+\varepsilon^{3/2}x^{3/2}}dy.
\eear
Therefore,
\bean
||\eta\cdot\left(\mathcal W_{\infty,\,\varepsilon} -\mathcal W_{\infty}\right)(\eta\, h)||^2_{H^{\sigma-1/2}(\RR)}=
\int_{\RR}  d\tilde k (1+|\tilde k|^2)^{\sigma-1/2}\int_{\RR} dk_1\, \overline{\widehat \psi (k_1)}\times\\
\times \int_{\RR} dk_2\, \widehat \psi (k_2)\, \mathcal Z_\varepsilon (k_1, k_2, \tilde k)
\eean
where
\bean
&&\mathcal Z_\varepsilon (k_1, k_2, \tilde k)=\int_{\RR}dx_1\int_{\RR}dx_2\overline{\left[P_\varepsilon (x_1, k_1)-P_0(x_1, k_1)\right]}\times \\
&&\hskip 3cm \times \left[P_\varepsilon (x_2, k_2)-P_0(x_2, k_2)\right]e^{-i\, (k_1-\tilde k)\, x_1}e^{-i\, (k_2-\tilde k)\, x_2}\eta (x_1)\eta (x_2)
\eean
We now show:
\bear
\label{S3EstiZeta}
\left| \mathcal Z_\varepsilon (k_1, k_2, \tilde k)\right|\le C_m\frac{|k_1|^{1/2}|k_2|^{1/2}}{(1+|\tilde k-k_1|^m)(1+|\tilde k-k_1|^m)}.
\eear
To this end we notice that we may write:
\bean
P_\varepsilon (x, k)-P_0(x, k) & = & \int_0^\infty dy\left(e^{-i k y}-1 \right) \frac{\varepsilon ^{3/2}x^{3/2}}{y^{3/2}(y^{3/2}+\varepsilon ^{3/2}x^{3/2})}
\eean
whence,
\bear
\label{S3EstiUni}
\int_{\RR}e^{-i (k-\tilde k) x}\left( P_\varepsilon (x, k)-P_0(x, k)\right)\eta (x)dx & = & 
\int_{\RR}e^{-i (k-\tilde k) x}\int_0^\infty \frac{e^{-i k y}-1}{y^{3/2}}R\left(\frac{y}{\varepsilon x} \right)\eta (x)dx\, dy\nonumber \\\\
R(\xi)& = & \frac{1}{\xi^{3/2}+1}\nonumber
\eear
For $|k-\tilde k|\le 1$ we immediately obtain from (\ref{S3EstiUni}) that, for some positive constant $C$ independent of $\varepsilon$:
\bear
\label{S3EstiDos}
\left|\int_{\RR}e^{-i (k-\tilde k) x}\left( P_\varepsilon (x, k)-P_0(x, k)\right)\eta (x)dx\right|\le C
\eear
On the other hand, using:
\bean
e^{-i(k-\tilde k) x}=\frac{i}{k-\tilde k}\frac{\partial}{\partial x}\left(e^{-i(k-\tilde k) x}\right)
\eean
and integrating by parts $m$ times in the right hand side of (\ref{S3EstiUni}) we obtain that for any $m\in \NN$ there exists a positive constant $C_m$ such that
\bear
\label{S3EstiTres}
\left|\int_0^\infty e^{-i (k-\tilde k) x}\, R\left(\frac{y}{\varepsilon x} \right)\eta (x)dx\right|\le \frac{C_m}{1+|k-\tilde k|^m}.
\eear
In the derivation of (\ref{S3EstiTres}) we have used:
\bean
\frac{\partial}{\partial x}R\left(\frac{y}{\varepsilon x} \right)=-\frac{1}{x}\, \xi\, R'\left(\xi\right),
\qquad \xi=\left(\frac{y}{\varepsilon x} \right)
\eean
the function $\xi\, R'\left(\xi\right)$ has the same structure than $R(\xi)$: it is a rational function of $\xi^{3/2}$ decreasing as $\xi \to \infty$ like $\xi^{-3/2}$. This is also true for all the derivatives of higher order.  Moreover, since
 $\hbox{supp}(\eta) \subset (1/2, 2)$, the term $\eta(x)/x$ is uniformly bounded in $\RR$. 
 
 Define now the function 
 \bean
 M(y, k-\tilde k)=\frac{1}{y^{3/2}}\int_0^\infty e^{-i (k-\tilde k) x}\, R\left(\frac{y}{\varepsilon x} \right)\eta (x)dx
 \eean
An integration by parts yields:
\bear
\int_0^\infty(e^{-i k y}-1) M(y, k-\tilde k)dy=-ik \int_0^\infty e^{-iky}\int_y^\infty M(\sigma, k-\tilde k)\, d\sigma dy.
\eear
This identity still holds in the straight lines $\Gamma$ of the complex plane defined by
\bean
|\Im m (y)=\varepsilon_0 |\Re e (y)|,\quad sign (\Im m(y))=-sign (k)
\eean
Using then (\ref{S3EstiTres}) we obtain:
\bear
\label{S3Esti4}
\left|ik \int_\Gamma e^{-iky}\int_y^\infty M(\sigma, k-\tilde k)\, d\sigma dy\right|\le \frac{C_m |k|}{1+|k-\tilde k|^m}\int_\Gamma \frac{e^{-ky}}{|y|^{1/2}}dy\le \frac{C'_m |k|^{1/2}}{1+|k-\tilde k|^m}.
\eear
Using (\ref{S3Esti4}) twice, estimate (\ref{S3EstiZeta}) follows. Therefore,
\bear
\label{S3Esti5}
&&\left|\int_{\RR}  d\tilde k (1+|\tilde k|^2)^{\sigma-1/2}\int_{\RR} dk_1\, \overline{\widehat \psi (k_1)}\int_{\RR} dk_2\, \widehat \psi (k_2)\, \mathcal Z_\varepsilon (k_1, k_2, \tilde k)\right|\\
&&\le \int_{\RR}  d\tilde k (1+|\tilde k|^2)^{\sigma-1/2}\int_{\RR} dk_1\, \overline{\widehat \psi (k_1)} \int_{\RR} dk_2\, \widehat \psi (k_2)\, \frac{|k_1|^{1/2}|k_2|^{1/2}}{(1+|k_1-\tilde k|)^m\, (1+|k_2-\tilde k|)^m}.\nonumber 
\eear
Using that $|\tilde k|\le |k_1|+|\tilde k-k_1|$ we have:
\bear
\label{S3Esti6}
\int_{\RR}  \frac{(1+|\tilde k|^2)^{\sigma-1/2} d\tilde k}{(1+|k_1-\tilde k|)^m\, (1+|k_2-\tilde k|)^m}
\le  C\frac{(1+|k_1|)^{2\sigma-1}}{(1+|k_2-k_1|)^{m'}}
\eear
for some $m'<m$. Using (\ref{S3Esti6}) in (\ref{S3Esti5}) and Cauchy-Schwartz's inequality we obtain:
\bear
\label{S3Esti7}
\left|\int_{\RR}  d\tilde k (1+|\tilde k|^2)^{\sigma-1/2}\int_{\RR} dk_1\, \overline{\widehat \psi (k_1)}\int_{\RR} dk_2\, \widehat \psi (k_2)\, \mathcal Z_\varepsilon (k_1, k_2, \tilde k)\right|\le \nonumber \\
\le ||\psi ||_{H^\sigma(\RR)}
\int _{\RR}dk_1\left|\int_{\RR}dk_2 \frac{|k_2|^{1/2}(1+|k_1|)^{\sigma-1/2}|\widehat\psi(k_2)|}{(1+|k_1-k_2|)^{m'}} \right|^2 \nonumber\\
\le ||\psi ||_{H^\sigma(\RR)}
\int _{\RR}dk_1\left|\int_{\RR}dk_2 \frac{(1+|k_2|)^{\sigma}|\widehat\psi(k_2)|}{(1+|k_1-k_2|)^{m''}} \right|^2
\eear
for some $m''<m'$. Young's inequality then emplies:
\bear
\label{S3Esti8}
\left|\int_{\RR}  d\tilde k (1+|\tilde k|^2)^{\sigma-1/2}\int_{\RR} dk_1\, \overline{\widehat \psi (k_1)}\int_{\RR} dk_2\, \widehat \psi (k_2)\, \mathcal Z_\varepsilon (k_1, k_2, \tilde k)\right|\le \nonumber\\
\le C
 ||\psi ||^2_{H^\sigma(\RR)}.
\eear
and therefore
\bear
\label{S3Esti9}
||\eta\cdot\left(\mathcal W_{\infty,\,\varepsilon} -\mathcal W_{\infty}\right)(\eta\, h)||_{H^{\sigma-1/2}(\RR)}
\le C
 ||\psi ||_{H^\sigma(\RR)}.
\eear
Combining (\ref{S3EstiZeta25}) and (\ref{S3Esti9}) we obtain the estimate for $\mathcal W_{\infty, \, \varepsilon}(\eta \, h)$ in (\ref{S2Ederivadasigmamedios}).

It remains to prove that (\ref{S2Elimites}) holds true.
By the estimate (\ref{S3EstiZeta}) in $\mathcal Z_\varepsilon (k_1, k_2, \tilde k)$ this is reduced to prove that for any $k_1$, $k_2$ and $\tilde k$, $\mathcal Z_\varepsilon (k_1, k_2, \tilde k)\to 0$ as $\varepsilon \to 0$. This follows from the fact that the support of $\eta$ is compact and that $P_\varepsilon(x, k)\to P_0(x, k)$ as $\varepsilon \to 0$ as it follows from the explicit expressions (\ref{S3EstiZeta35}) .

If $y=\varepsilon\, t$, we obtain,
\bean
P_{\varepsilon}(x, k)=\frac{1}{\sqrt \varepsilon}\int_0^\infty
\frac{(e^{-ik\varepsilon t}-1)}{t^{3/2}+x^{3/2}}dt.
\eean
Therefore it follows that, for some positive constant $C$ independent of $\varepsilon>0$ and $x>0$:
\bean
|P_{\varepsilon}(x, k)|\le C|k|^{1/2},\,\,\,\forall \varepsilon>0,\,\,\forall x>0
\eean
and (\ref{S2Eopelredondaepsilon}) follows.
\qed
\section{Estimating the difference between $\mathcal L$ and $L$.}
\setcounter{equation}{0}
\setcounter{theo}{0}
In this Section we estimate the operator $\mathcal L -L$ which appear in the equation (\ref{S8E10-2}). 
\bear
(\mathcal{L}-L)(\varphi)(x, t) & = & A_1+A_2,\nonumber\\
A_1(x) & = & \int_0^{x/2}\left(H(x-y)-H(x)\right)
y^{\lambda /2} \varphi(y, t)dy \nonumber \\
&&- H(x)\int_{x/2}^{\infty }y^{\lambda /2} \varphi(y, t)dy-x^{\lambda/2}\varphi(x, t)\left(\int_{x/2}^{\infty }H(y)\, dy\right)
 \label{S8E15} \\
A_2(x) & = & \int_0^{x/2}\left((x-y)^{\lambda /2} \varphi(x-y, t)-x^{\lambda /2} \varphi(x, t) \right)H(y) dy \label{S8E15-4}\\
H(y) & = & y^{\lambda/2}\, f_0(y)-y^{-3/2}.\label{S3EHdef}
\eear

Since it will be needed in the Section 6, we shall actually  estimate more general operators where the function $A_2$ has the more general form:
\bear
A_{2,\varepsilon}(x) & = & \int_0^{x/2}\left((x-y)^{\lambda /2} \varphi(x-y, t)-x^{\lambda /2} \varphi(x, t) \right)H_\varepsilon(x, y) dy \label{S8E15-4bis}\\
H_\varepsilon(x, y) & = & y^{\lambda/2}\, f_0(y)-\frac{1}{y^{3/2}+\varepsilon^{3/2}x^{3/2}}.\label{S3EHdefbis}
\eear
Notice that $\varepsilon=0$ corresponds to the functions $A_2$ and $H$ defined in (\ref{S8E15-4}) and (\ref{S3EHdef}).
 
In the two following Lemmas we  estimate the two terms $A_1$ and $A_{2, \varepsilon}$ assuming some conditions of the function $f_0$.
\begin{lem}
\label{S8T3}
Suppose that $f_0$ satisfies conditions (\ref{defhachecero}), (\ref{S2Ecotasregdato}) and $|||\varphi|||_{3/2, (3+\lambda)/2}<\infty$. Then
\bean
|||A_1|||_{3/2,\, 2+\delta} & \le & C\, |||\varphi|||_{3/2,\, (3+\lambda)/2}.
\eean
\end{lem}
\textbf{Proof of Lemma \ref{S8T3}.} 
The estimate on $A_1(x)$ for $0<x<1$ is immediate:
\bear
\left|\int_0^{x/2}(H(x-y)-H(x))
y^{\lambda /2} \varphi(y)dy\right| & \le & |||\varphi|||_{3/2, (3+\lambda)/2}\int_0^{x/2}\left|H(x-y)+H(x)\right|
y^{(-3+\lambda) /2} dy \nonumber\\
& \le & C\, |||\varphi|||_{3/2, (3+\lambda)/2}\, x^{-3/2} \label{S8T3E1}\\
\left|H(x)\int_{x/2}^{\infty }y^{\lambda /2} \varphi(y)dy\right| & \le & C |||\varphi|||_{3/2, (3+\lambda)/2}\, x^{-3/2},
\label{S8T3E2}\\
\left|x^{\lambda/2}\left(\int_{x/2}^{\infty }H(y)\, dy\right) \varphi(x)\right| & \le &
 C\, |||\varphi|||_{3/2, (3+\lambda)/2}\, x^{\lambda/2-2}\le C\, |||\varphi|||_{3/2, (3+\lambda)/2}\, x^{-3/2}.\nonumber \\\label{S8T3E3}
\eear

 Let us consider the case when $x>1$.
In order to estimate the first term in the right hand side of (\ref{S8E15}) we write:
\bean
H(x-y)-H(x)=y\int_0^1H'(x-\theta y)d\theta
\eean
where
\bean
H'(z)=\frac{\lambda}{2}z^{(\lambda-2)/2}(f_0(z)-z^{-(3+\lambda)/2})+z^{\lambda/2}(f'_0(z)+\frac{3+\lambda}{2}z^{-(3+\lambda)/2-1})
\eean
By assumptions (\ref{defhachecero}) (\ref{S2Ecotasregdato}), for all $z>1$:
\bean
|H'(z)|\le\left(1+\frac{\lambda}{2}\right)z^{-5/2 -\delta}
\eean
In particular, for all $y<x/2$ and $0<\theta<1$ we have $x-\theta y >x/2$ and so, if $x>2$:
\bean
|H(x-y)-H(x)|=\left|y\int_0^1H'(x-\theta y)d\theta\right|\le C\, y\,x^{-5/2 -\delta}
\eean
and
\bear
\label{S8E16} 
\left|\int_0^{x/2}\left(H(x-y)-H(x)\right)
y^{\lambda /2} \varphi(y)dy\right| &\le &C\,x^{-5/2 -\delta}\!\!\!
\int_0^{x/2} y^{1+\lambda /2} |\varphi(y)|dy \nonumber \\
& \le & C|||\varphi|||_{3/2,\, (3+\lambda)/2} \,x^{-2 -\delta}
\eear
In order to estimate the second term in (\ref{S8E15}) we use:
\bean
|H(x)|=x^{\lambda/2}\left|f_0(x)-G(x) \right|\le C\, x^{\lambda/2}x^{-(3+\lambda)/2-\delta}\quad \hbox{for}\,\,\,x>1
\eean
whence, for $x>1$:
\bear
\left|H(x)\int_{x/2}^{\infty }y^{\lambda /2} \varphi(y)dy\right|  \le 
x^{-3/2-\delta}\!\!\int_{x/2}^{\infty }y^{\lambda /2} \varphi(y)dy 
\le  C\, |||\varphi|||_{3/2,\, (3+\lambda)/2}\,x^{-2-\delta}.
\label{S8E17} 
\eear
The third term  of (\ref{S8E15}) is bounded by
\bear
x^{\lambda/2}|\varphi(x)| \int_{x/2}^\infty y^{-3/2-\delta}\, dy & = & C\, |||\varphi|||_{3/2,\, (3+\lambda)/2}\, x^{-2-\delta}
\quad \hbox{for}\,\,\,x>1. \label{S8E18} 
\eear
Lemma \ref{S8T3} then follows combining (\ref{S8T3E1})-(\ref{S8E18}).
\qed
\begin{lem}
\label{S3estA1reesc}
Let $0\le T \le 1$. Then, there exists a constant $C>0$ such that, for any $\varphi \in  \mathcal E_{T;\sigma}$ and  for all $t_0\in (0, T)$ :
\bean
R^2\, N_{2;\, \sigma}\left(A_1;\, R,\, t_0 \right)\le C\, ||| \varphi|||,\,\,\,\forall R>1,\\
R^{2-\lambda/2}\, M_{2;\, \sigma}\left(A_1;\, R \right)\le C\, |||\varphi|||,\,\,\,\forall\,\,\, 0< R<1.
\eean
\end{lem}
\textbf{Proof of Lemma \ref{S3estA1reesc}.}
For $R>1$ we write
\bean
\varphi(x, t)=\sum_{n=0}^{\infty}\chi(x/2^n)\, \varphi(x, t)
\eean
where $\chi\in C_0^\infty$,  $\hbox{supp}\, \chi \subset (1/2, 2)$. Let us consider $R=2^{n_0}$, $x\in (R/2, 2R)$ and rescale $x=RX$,  $y=R\, Y$, $\tau=(t-t_0)R^{(\lambda-1)/2}$,  $\varphi(x, t)=R^{-(3+\lambda)/2)}\psi(X, \tau)$, 
${\cal A}_{1}(X, \tau)=A_{1}(x, t)$ to obtain:
\bear
\label{S3estA1reescE1}
R^2\, |\mathcal A _1(X, \tau )| & = & \int_0^{X/2}\left[ H_R(X-Y)-H_R(X)\right]\, Y^{\lambda/2}\, \psi(Y, \tau)\, dY-\\
&&-H_R(X)\int_{X/2}^\infty
Y^{\lambda/2}\psi(Y, \tau)\, dY-X^{\lambda/2}\, \psi(X, \tau )\int_{X/2}^\infty H_R(Y)\, dY \nonumber
\eear
where the function $H_R$ is defined as follows:
\bear
\label{S3HacheErre}
H_R(X)=R^{(3+\lambda)/2} X^{\lambda/2}\, f_0(R\, X)-X^{-3/2}.
\eear
Since $|||\varphi|||_{3/2, (3+\lambda)/2}<\infty$, we have the following bound on $\Psi(X, \tau)$ 
\bear
\label{S3Eestipsi}
|\Psi(X, \tau)|\le C \min\left\{\frac{R^{\lambda/2}}{X^{3/2}}, \frac{1}{X^{(3+\lambda)/2}}  \right\}|||\varphi (t)|||_{3/2, (3+\lambda)/2}
\eear
 for all $X\ge 0$ and $\tau \in (0, T\,R^{{(\lambda-1)/2}})$. Using this estimate it then follows that the integrals in the right hand side of (\ref{S3estA1reescE1}) are convergent. Moreover, using conditions (\ref{defhachecero}) and (\ref{S2Ecotasregdato}) we obtain:
\bear
\label{S3estA1reescE2}
R^2\left(\int_0^1d\tau \int_{1/2}^2|D_x^\sigma \mathcal A _1(X, \tau)|^2dX d\tau \right) ^{1/2}
\le C |||\varphi|||, \quad \forall R>1.
\eear
For $R\in (0, 1)$ we scale the variables $x\in (R/2, 2R)$ and $\varphi$ as   $x=RX$,  $y=R\, Y$,  $\varphi(x, t)=R^{-3/2}\psi(X, t)$, 
${\cal A}_{1} (X, t)=A_{1} (x, t)$ to obtain in this case:
\bear
\label{S3estA1reescE3}
R^{2-\lambda/2}\, |\mathcal A _1(X, t )| & = & \int_0^{X/2}\left[ H_R(X-Y)-H_R(X)\right]\, Y^{\lambda/2}\, \psi(Y, t)\, dY-\\
&&-H_R(X)\int_{X/2}^\infty
Y^{\lambda/2}\psi(Y, \tau)\, dY-X^{\lambda/2}\, \psi(X, t )\int_{X/2}^\infty H_R(Y)\, dY \nonumber
\eear
Using again (\ref{defhachecero}), (\ref{S2Ecotasregdato}) and (\ref{S3Eestipsi}) we deduce
\bear
\label{S3estA1reescE4}
R^{2-\lambda/2}\left(\int_0^1d\tau \int_{1/2}^2|D_x^\sigma \mathcal A _1(X, \tau)|^2dX d\tau \right) ^{1/2}
\le C |||\varphi|||, \quad \forall R\in (0, 1).
\eear
Lemma \ref{S3estA1reesc} follows from (\ref{S3estA1reescE2}) and (\ref{S3estA1reescE4}). \qed
\vskip 0.5cm
\noindent
The following technical Lemma will be needed in order to estimate $A_{2, \varepsilon}$.
\begin{lem}
\label{lematech}
For any given function $h\in H^\sigma (\RR)$ supported in $(1/2, 2)$, and \hfill \break$\delta\in [0, \min(1/2, 1-\sigma))$ there holds:
\bean
\int_0^{5/8}
\left| h(X-Y)-h(X) \right|Y^{-3/2-\delta} dY\le C||h||_{H^\sigma}
\eean
\end{lem}
{\bf Proof of Lemma \ref{lematech}.}  Using 
\bean
h(X)=\frac{1}{\sqrt {2\pi}}\int_{\RR}\widehat h(\xi)e^{ix\xi} d\xi
\eean
we obtain
\bean
\int_0^{5/8}
\left| h(X-Y)-h(X) \right|Y^{-3/2-\delta} dY \le C
\int_0^{1/2}\int_{\RR}|\widehat h(\xi)|
\left| e^{-i\xi Y}-1 \right|Y^{-3/2-\delta}d\xi dY\\
\le C \left(\int_{\RR}| \widehat h(\xi)|^2(1+|\xi|^{\sigma})^2d\xi\right)^{1/2}
\left(\int_{\RR}
\left|\int_0^{5/8} \left| e^{-i\xi Y}-1 \right| Y^{-3/2-\delta} dY\right|^2\frac{d\xi}{(1+|\xi|^{2 \sigma})} \right)^{1/2}.
\eean
Using the change of variables $\xi y=z$ we arrive at
\bean
\int_0^{5/8} \left| e^{-i\xi Y)}-1 \right| Y^{-3/2-\delta} dY\le 
C\xi^{1/2+\delta}\int_0^{5\, \xi/8}|e^{-iz}-1|\frac{dz}{z^{3/2+\delta}}\le C \xi^{1/2+\delta},
\eean
and
\bean
\int_{\RR}
\left|\int_0^{5/8} \left| e^{-i\xi Y)}-1 \right| Y^{-3/2-\delta} dY\right|^2\frac{d\xi}{1+|\xi|^{2\sigma}} \le \int_{\RR}|\xi|^{1+2\delta}
\frac{d\xi}{(1+|\xi|^{2 \sigma})}<\infty
\eean
since $\sigma >1+\delta$. 
\qed

\vskip 0.5cm
\noindent
We have the following estimate  for $A_{2,\, \varepsilon}$ in (\ref{S8E15-4bis}).

\begin{lem}
\label{S8Tprov}
Suppose that $f_0$ satisfies conditions (\ref{defhachecero})(\ref{S2Ecotasregdato}) and $\sigma> 1+\delta$, then
\bear
\sup_{t_0\in (0, T)}\sup_{R>1}\,R^{2+\delta}N_{\infty}(A_{2,\varepsilon};\, t_0, R)
\le C |||\varphi|||, \label{S8TprovE1}\\
\sup_{t_0\in (0, T)}\sup_{R>1}\, R^2\,N_{2;\, \sigma-\frac{1}{2}}(A_{2,\varepsilon};\, t_0, R)
\le C |||\varphi|||,\label{S8TprovE2}\\
\sup_{0<R<1}\,R^{2-\lambda/2}M_\infty (A_{2,\varepsilon};\, R)
\le C |||\varphi|||, \label{S8TprovE3}\\
\sup_{0<R<1}\,R^{2-\lambda/2}M_{2;\, \sigma-\frac{1}{2}} (A_{2,\varepsilon};\, R)
\le C |||\varphi|||, \label{S8TprovE4}
\eear
where the functions $N_{\infty}(\cdot\,;\, t_0, R)$, $N_{2;\, \sigma}(\cdot\,;\, t_0, R)$, $M_\infty (\cdot\,;\, R)$ and $M_{2; \sigma} (\cdot\, ;\, R)$ are defined in 
(\ref{S3TnormaNinfty}) - (\ref{S3TnormaMsigma}).

\end{lem}
\textbf{Proof of Lemma \ref{S8Tprov}. }
For $R>1$ we write
\bean
\varphi(x, t)=\sum_{n=0}^{\infty}\chi(x/2^n)\, \varphi(x, t)
\eean
where $\chi\in C_0^\infty$,  $\hbox{supp} \chi \subset (1/2, 2)$
\bean
|A_{2, \, \varepsilon}(x, t)| \le \!\!\! \sum_{n=n_0-2}^{n_0+1}\int_0^{x/2}
\left|(x-y)^{\lambda /2} \varphi(x-y, t)\chi \left(\frac{x-y}{2^n}\right)-x^{\lambda /2} \varphi(x, t) 
\chi \left(\frac{x}{2^n}\right)\right|y^{-3/2-\delta} dy.
\eean
Let us consider $R=2^{n_0}$, $x\in (R/2, 2R)$ and rescale $x=RX$,  $y=R\, Y$, $\tau=(t-t_0)R^{(\lambda-1)/2}$,  $\varphi(x, t)=R^{-(3+\lambda)/2)}\psi(X, \tau)$, 
${\cal A}_{2, \varepsilon}(X, \tau)=A_{2, \varepsilon}(x, t)$ to obtain:
\bean
&&|{\cal A}_{2, \varepsilon}(X,\tau)|  \le   R^{-2-\delta}\times\\
&& 
\times\sum_{\ell=-2}^{1}\int_0^{X/2}\left|(X-Y)^{\lambda /2} \psi(X-Y, \tau)\chi \left(\frac{X-Y}{2^\ell}\right)-X^{\lambda/2}\psi(X, \tau)\chi \left(\frac{X}{2^\ell}\right) \right|Y^{-3/2-\delta} dY.
\eean
\noindent
Using Lemma \ref{lematech}, we deduce, for $X\in (3/4, 5/4)$:
\bean
|{\cal A}_{2, \varepsilon} (X, \tau)|\le R^{-(2+\delta)}\, C\, \left( ||\psi(\tau)||_{L^{\infty}(1/8, 8)}+ ||\psi(\tau)||_{H^{\sigma}(1/8, 8)}\right)
\eean
whence:
\bean
&&R^{2+\delta}\left(\int_{0}^{\min(1,\, R^{(\lambda-1)/2}(T-t_0))}||{\cal A}_{2, \varepsilon} (t)||^2_{L^\infty(3/4, 5/4)}d\tau\right)^{1/2} \\
&&\le 
C\, \sup_{0\le \tau \le \min(1,\, R^{(\lambda-1)/2}(T-t_0))}||\psi(\tau)||_{L^{\infty}(1/8, 8)}+\nonumber \\
&&+ C\,
\left( \int_{0}^{\min(1,\, R^{(\lambda-1)/2}(T-t_0))}
 ||\psi(\tau)||^2_{H^{\sigma}(1/8, 8)}d\tau\right)^{1/2}.
\eean
Therefore,
\bean
R^{2+\delta}N_\infty(A_{2, \varepsilon} ; t_0, R) & \le &
C\, R^{(3+\lambda)/2}\left[ \sup_{t_0\le t \le \min(t_0+R^{-(\lambda-1)/2}, T)}||\varphi(t)||_{L^{\infty}(R/8, 8 R)}+\right.\nonumber \\
&&+  \left. \sum_{\ell =-3}^3N_{2; \sigma}(\varphi; t_0, 2^\ell R)
\right] \le C|||\varphi|||,
\eean
and (\ref{S8TprovE1}) follows.

We now prove (\ref{S8TprovE2}). To this end notice that:
\bean
&&{\cal A}_{2, \varepsilon} (X,\tau)  =   R^{-2}\sum_{\ell=-2}^{1}\int_0^{X/2}\\
&&\left((X-Y)^{\lambda /2} \psi(X-Y, \tau)\chi \left(\frac{X-Y}{2^\ell}\right)-X^{\lambda/2}\psi(X, \tau)\chi \left(\frac{X}{2^\ell}\right) \right)
\mathcal H_\varepsilon(X, Y) dY
\eean
where $\mathcal H_\varepsilon(X, Y)=R^{3/2}H_\varepsilon(x, y)$, using Lemma \ref{S2Tpseudo} we obtain:
\bean
&&R^{2}\left(\int_0^{\min(1,\, R^{(\lambda-1)/2}(T-t_0))} ||{\cal A}_{2, \varepsilon} (\tau) ||^2_{H^{\sigma-1/2} (3/4, 5/4)}d\tau\right)^{1/2}\le \\
&&\hskip 2cm\le
C\left(\int_0^{\min(1,\, R^{(\lambda-1)/2}(T-t_0))}||\psi(\tau)||^2_{H^{\sigma}(1/8, 8)}d\tau\right)^{1/2}+\\
&&\hskip 4cm +C
\sup _{0\le \tau \le \min(1,\, R^{(\lambda-1)/2}(T-t_0))}||\psi(\tau)||_{L^\infty(1/8, 8)}.
\eean
Therefore
\bean
 R^{\,\sigma+1}N_{2;\, \sigma-\frac{1}{2}}(\mathcal A_{2, \varepsilon} ; t_0, R) & \le & 
C\, R^{(3+\lambda)/2} \left[\sum_{\ell =-3}^3N_{2; \sigma}(\varphi; t_0, 2^\ell R)+\right.\\
&&\hskip 1cm+\left.
\sup _{t_0\le t \le \min(t_0+R^{-(\lambda-1)/2}, T)}|| \varphi(t)||_{L^\infty(R/8, 8 R)}\right].
\eean
whence (\ref{S8TprovE2}) follows.

We consider now the case where $0<R\le 1$. The arguments are very similar to those used in the previous case. In order to prove (\ref{S8TprovE2}) we write 
\bean
\varphi(x, t)=\sum_{n=0}^{\infty}\chi(2^n\, x)\, \varphi(x, t)
\eean
where $\chi\in C_0^\infty$,  $\hbox{supp} \chi \subset (1/2, 2)$
\bean
|\mathcal A_{2, \varepsilon} (x, t)| & \le  & C\, \sum_{n=n_0-2}^{n_0+1}\int_0^{x/2}
\left|(x-y)^{\lambda /2} \varphi(x-y, t)\chi \left(2^n(x-y)\right)-x^{\lambda /2} \varphi(x, t) 
\chi \left(2^n\, x\right)\right|y^{-3/2} dy.
\eean
Let us consider $R=2^{n_0}$, $x\in (R/2, 2R)$ and rescale $x=RX$,  $y=R\, Y$,  $\varphi(x, t)=R^{-3/2}\psi(X, t)$, 
${\cal A}_{2, \varepsilon} (X, t)=A_{2, \varepsilon} (x, t)$ to obtain:
\bean
|{\cal A}_{2, \varepsilon} (X,t)| & \le  & R^{\lambda/2-2}\sum_{\ell=-2}^{1}\int_0^{X/2}\\
&& 
\left|(X-Y)^{\lambda /2} \psi(X-Y, t)\chi \left(2^\ell(X-Y)\right)-X^{\lambda/2}\psi(X, t)\chi \left(2^\ell\, X\right) \right|Y^{-3/2} dY.
\eean
Using Lemma \ref{lematech} we deduce that for $X\in (3/4, 5/4)$:

\bean
|{\cal A}_{2, \varepsilon} (X, t)|\le R^{-2+\frac{\lambda}{2}}\, C\, \left( ||\psi(t)||_{L^{\infty}(1/8, 8)}+ ||\psi(t)||_{H^{\sigma}(1/8, 8)}\right)
\eean
whence,
\bean
R^{2-\lambda/2}\left(\int_{0}^{T}||{\cal A}_{2,\, \varepsilon}(t)||^2_{L^\infty(3/4, 5/4)}dt\right)^{1/2}\le 
C\, \sup_{0\le t \le T}||\psi(t)||_{L^{\infty}(1/8, 8)}+\\ 
+C\,
\left(\int_{0}^{T}
 ||\psi(t)||^2_{H^{\sigma}(1/8, 8)}dt\right)^{1/2}.
\eean
Therefore
\bean
R^{2-\lambda/2}M_\infty (\mathcal A_{2, \varepsilon} , R) & \le &
C\, R^{3/2}\left[ \sup_{0\le t \le T}||\varphi(t)||_{L^{\infty}(R/8, 8 R)}+
\sum_{\ell=-3}^3M_{2;\, \sigma}(\varphi, 2^\ell\, R)\right]\\
&\le &  C|||\varphi|||,
\eean
and (\ref{S8TprovE3}) follows.
\\ \\
We now prove (\ref{S8TprovE4}). Since:
\bean
&&{\cal A}_{2, \varepsilon} (X,t)  =   R^{-2+\lambda/2}\sum_{\ell=-2}^{1}\int_0^{X/2}\\
&&\left((X-Y)^{\lambda /2} \psi(X-Y, t)\chi \left(2^\ell (X-Y)\right)-X^{\lambda/2}\psi(X, t)\chi \left(2^\ell\, X\right) \right)
\mathcal H_\varepsilon(X, Y) dY
\eean
where $\mathcal H_\varepsilon(X, Y)=R^{3/2}H_\varepsilon(x, y)$, using Lemma \ref{S2Tpseudo} we obtain:
\bean
R^{2-\lambda/2}\left(\int_0^{T} ||{\cal A}_{2, \varepsilon} (t) ||^2_{H^{\sigma-1/2} (3/4, 5/4)}dt\right)^{1/2} & \le &
C\left(\int_0^{T}||\psi(t)||^2_{H^{\sigma}(1/8, 8)}dt\right)^{1/2}\\
&&+C
\sup _{0\le t \le T}||\psi(t)||_{L^\infty(1/8, 8)}.
\eean
Therefore
\bean
&& R^{2-\lambda/2}M_{2; \, \sigma-\frac{1}{2}}(\mathcal A_{2, \varepsilon} ; R) \le
C\, R^{3/2} \left[\sum_{\ell=-3}^3M_{2; \, \sigma}(\varphi; 2^\ell \, R)+
\sup _{0\le t \le T}|| \varphi(t)||_{L^\infty(R/8, 8 R)}\right].
\eean
whence (\ref{S8TprovE4}) follows.
\qed
\\ \\ 
The following result has been proved in \cite{EV}:
\begin{prop}
\label{S2Tsolfund}
 The fundamental solution $g(t, x, x_0)$ of the operator $L$ defined in (\ref{S1Eopele}) such that $g(0, x, x_0)=\delta (x-x_0)$ satisfies:
\bear
&&g(t, x, x_0)=\frac{1}{x_0}g\, \left(t\, x_0^{(\lambda-1)/2}, \frac{x}{x_0},1\right) \label{S2TsolfundE1}\\
&&|g(t, x, 1)|\le C\, t\, x^{-3/2}, \quad\hbox{for all}\,\,\,0\le t\le 1,\, 0<x\le 1/2,\label{S2TsolfundE2}\\
&&|g(t, x, 1)|\le C\, t\, x^{-(3+\lambda)/2}, \quad\hbox{for all}\,\,\,0\le t\le 1,\, x\ge  3/2,\label{S2TsolfundE3}\\
&&|g(t, x, 1)|\le C\, t^{-2}\,\Phi\left(\frac{x-1}{t^2} \right) x^{-3/2}, \quad\hbox{for all}\,\,\,0\le t\le 1,\, 1/2\le x\le 3/2,\label{S2TsolfundE4}
\eear
where,
\bear
\label{S2TsolfundE5}
\Phi(\xi)=\frac{1}{1+|\xi|^{3/2-\delta}}.
\eear
Moreover
\bear
&&g(t, x, x_0)\le Ct^{2/(\lambda-1)}\, \sigma^{-3/2},
\quad\hbox{for all}\,\,\, t\ge 1,\, 0<\sigma \le 1,\label{S2TsolfundE6}\\
&&|g(t, x, 1)|\le C\, t^{2/(\lambda-1)}\, \sigma^{-(3+\lambda)/2}, \quad\hbox{for all}\,\,\, t\ge 1,\, \sigma \ge 1,\label{S2TsolfundE7}
\eear
with
\bear
\label{S2TsolfundE8}
\sigma=t^{2/(\lambda-1)}\, x.
\eear
\end{prop}

\begin{lem}
\label{S8T4}Ê
For $T\in (0, 1]$ there is a constant $C>0$ such that, for all $||\nu||_{X_{3/2, \, 2+\delta}(T)}<\infty$:
\bean
&&\sup_{0\le t \le T}\left|\left| \left |\int_0^tG(t-s)\,\nu (s)ds\right|\right|\right|_{3/2, \, \frac{3+\lambda}{2}}\le 
CT^{\beta}||\nu||_{X_{3/2, \, 2+\delta}(T)}
\eean
where
\bear
\label{S3Ebeta}
\beta=\min \left(1, \frac{2\delta}{\lambda-1}\right).
\eear

\end{lem}
\textbf{Proof of Lemma \ref{S8T4}.}  We assume first that 
\bear
R^{-(\lambda-1)/2}\le t. \label{S3Ezona2}
\eear
Let us suppose that 
\bear
x \in \left(\frac{3R}{4}, \frac{5R}{4}\right).\label{S3Ezona1}
\eear
Using Proposition \ref{S2Tsolfund}:
\bear
\label{S8E22}
&&\int_0^t G(t-s)\nu (s, y)\, ds  =\int_0^t ds\int_0^\infty dy \nu (s, y) g\left((t-s)  y^{\frac{\lambda-1}{2}}, \frac{x}{y}\right)\frac {dy}{y} \nonumber \\
&&
\le 
\int_{t-R^{-\frac{\lambda-1}{2}}}^t ds\int_{|x-y|\le R/2}\nu(s, y)g\left((t-s) y^{\frac{\lambda-1}{2}}, \frac{x}{y}\right)\frac {dy}{y} \nonumber \\
&&
+\int_0^{t-R^{-\frac{\lambda-1}{2}}} ds\int_{|x-y|\le R/2}\nu(s, y) g\left((t-s) y^{\frac{\lambda-1}{2}}, \frac{x}{y}\right)\frac {dy}{y} \nonumber \\
&&+\int_0^t ds\int_{|y|\le R/2} \nu (s, y)g\left((t-s) y^{\frac{\lambda-1}{2}}, \frac{x}{y}\right)\frac {dy}{y} \nonumber \\
&&+\int_0^t ds\int_{|y| \ge 2 R}\nu(s, y)g\left((t-s) y^{\frac{\lambda-1}{2}}, \frac{x}{y}\right)\frac {dy}{y} \nonumber \\
&&={\mathcal I}_1+{\mathcal I}_2+{\mathcal I}_3+{\mathcal I}_4.
\eear
To estimate $\mathcal I_1$ we use the fact that (\ref{S2TsolfundE4}) implies:
\bear
\label{S2TsolfundE4bis}
g(s, z)\le \frac{C}{s^2}\Phi\left(\frac{z-1}{s^2}\right)
\eear
for $0\le s \le 1$, $z\in (1/2, 3/2)$. 
\bean
|{\mathcal I}_1|& \le &  
C\, \int_{t-R^{-\frac{\lambda-1}{2}}}^t ds\int_{|x-y|\le R/2}
\frac{\nu(s, y)}{((t-s)\, y^{\frac{\lambda-1}{2}})^2}
\Phi\left(\frac{\frac{x}{y}-1}{(t-s)^2\, y^{\lambda-1}}\right)\frac{d\, y}{y} \nonumber\\
& \le & C\, \int_{t-R^{-\frac{\lambda-1}{2}}}^t ||\nu (s)||_{L^\infty (R/2,\, 2R)}ds\int_{|x-y|\le R/2}
\frac{1}{(t-s)^2\, y^\lambda}
\Phi\left(\frac{x-y}{(t-s)^2\, y^{\lambda}}\right)\, dy \nonumber\\
& \le & C\, \int_{t-R^{-\frac{\lambda-1}{2}}}^t ||\nu (s)||_{L^\infty (R/2,\, 2R)}ds.
\eean
where we have used (\ref{S3Ezona2})  in the last inequality. Using H\"older's inequality we deduce:
\bean
|{\mathcal I}_1|& \le &  C\, R^{-(\lambda-1)/2}N_\infty(\nu; t_0, R)
\eean
whence:
\bear
|{\mathcal I}_1|\le   C\, R^{-(3+\lambda)/2}\, R^{-\delta}\left[R^{2+\delta}\, N_\infty(\nu; t_0, R)\right]\le 
C\, R^{-(3+\lambda)/2}\, t^{2\delta/(\lambda-1)}|||\nu|||_{X_{3/2,\, 2+\delta}}.\label{S3estIuno}
\eear
We consider now the term ${\mathcal I}_2$:

\bean
&&{\mathcal I}_2=\int_0^{t-R^{-\frac{\lambda-1}{2}}} ds\int_{|x-y|\le R/2}\nu(y) g\left((t-s) y^{\frac{\lambda-1}{2}}, \frac{x}{y}\right)\frac {dy}{y}.
\eean
In the region of integration we have $(t-s) y^{\frac{\lambda-1}{2}}\ge 1$. Using then (\ref{S2TsolfundE7}) we deduce
\bear
\label{S2TsolfundE78}
\left|g\left((t-s) y^{(\lambda-1)/2}, \frac{x}{y}\right)\right|\le C\, (t-s)^{-\frac{\lambda+1}{\lambda-1}}\frac{y}{x^{(3+\lambda)/2}}
\eear
for $s\ge 1$ and $1/7 \le |z|\le 7$. Therefore:
\bear
|{\mathcal I}_2| & \le & R^{-(3+\lambda)/2}\, \int_{R^{-\frac{\lambda-1}{2}}}^{t}||\nu(t-s)||_{L^\infty(R/2, 2R)}\, s^{-\frac{\lambda+1}{\lambda-1}} ds \nonumber\\
& = & R^{-(3+\lambda)/2}\,\sum_{n=1}^{\left[t\, R^{(\lambda-1)/2} \right]}\int_{n\, R^{-(\lambda-1)/2}}^{\min\{ (n+1)R^{-(\lambda-1)/2},\, t\}}||\nu(t-s)||_{L^\infty(R/2, 2R)}\, s^{-\frac{\lambda+1}{\lambda-1}} ds
 \nonumber\\
& \le & C R^{-(3+\lambda)/2}\,  \sum_{n=1}^{\left[t\, R^{(\lambda-1)/2} \right]} R^{(\lambda+1)/2}n^{-(\lambda+1)/(\lambda-1)}R^{-(\lambda-1)/2}
N_\infty (\nu; nR^{-(\lambda-1)/2}, R)\nonumber \\
&\le & C R^{-(3+\lambda)/2}\, R^{-1-\delta}|||\nu|||_{X_{3/2,\, 2+\delta}}\le C R^{-(3+\lambda)/2}\, t^{2(1+\delta)/(\lambda-1)}|||\nu|||_{X_{3/2,\, 2+\delta}}, 
\label{S3estIdos}
\eear
where we have used (\ref{S3Ezona2}) in the last step.

We next consider the term ${\mathcal I}_3$.
\bear
\mathcal I _3 & = & \int_0^t ds\int_{|y|\le R/2} \nu (y)g\left((t-s) y^{\frac{\lambda-1}{2}}, \frac{x}{y}\right)\frac {dy}{y}=\\
&&=\int_0^t ds\int_{0}^{t^{-2/(\lambda-1)}} \nu (y)g\left((t-s) y^{\frac{\lambda-1}{2}}, \frac{x}{y}\right)\frac {dy}{y}+\nonumber \\
&&\hskip 1.5cm +
\int_0^t ds\int_{t^{-2/(\lambda-1)}}^{R/2} \nu (y)g\left((t-s) y^{\frac{\lambda-1}{2}}, \frac{x}{y}\right)\frac {dy}{y}=\mathcal I_{3,1}
+\mathcal I_{3,2}. 
\label{S3EsplitItres}
\eear
We can use (\ref{S2TsolfundE3}) in the region of integration of $\mathcal I_{3,1}$. Therefore:
\bear
\label{S2TsolfundE3bis}
\left|g\left((t-s) y^{\frac{\lambda-1}{2}}, \frac{x}{y}\right)\right|\le C (t-s)\, x^{-(3+\lambda)/2}\,y^{\lambda+1}.
\eear
Then:
\bear
|\mathcal I_{3,1}| &\le &
C x^{-(3+\lambda)/2}\int_{0}^t ds (t-s)\int_0^{t^{-2/(\lambda-1)}} |\nu (y) |y^{\lambda}dy \nonumber\\
& = & C\, x^{-(3+\lambda)/2}\int_{0}^t ds (t-s)\left(\int_0^{1} |\nu (y) |y^{\lambda}dy+
\int_1^{t^{-2/(\lambda-1)}} |\nu (y) |y^{\lambda}dy \right) \nonumber \\
& = & \mathcal I_{3, 1, 1}+\mathcal I_{3, 1, 2}.
\label{S3estItresuno}
\eear
\bear
\mathcal I_{3, 1, 1} & \le & C\, x^{-(3+\lambda)/2}\sum_{n=0}^{\infty}\, 2^{-n(\lambda+1)}\int_{0}^t ds (t-s)  ||\nu (s)||_{L^\infty (2^{-(n+1)},\, 2^{-n})} \nonumber\\
& \le & C x^{-(3+\lambda)/2}t^{3/2}\sum_{n=0}^{\infty}2^{-n(\lambda+1)}M_\infty(\nu; 2^{-n})\nonumber \\
& \le & C x^{-(3+\lambda)/2}t^{3/2}|||\nu|||_{X_{3/2,\, 2+\delta}}\sum_{n=0}^{\infty}2^{-n(\lambda-1/2)}
\label{S3estItresunouno}
\eear

\bear
\mathcal I_{3, 1, 2} & \le & C\, x^{-(3+\lambda)/2}\, t  \!\!\!\!\!\!\!\!  \sum_{0\le 2^n \le t^{-2/(\lambda-1)}} 
 \!\!\! \!\!\!\sum_{\ell=1}^{\left[t\, (2^n)^{(\lambda-1)/2} \right]} 
\int_{ 2^{-n(\lambda-1)/2}\, \ell}^{\min\{2^{-n(\lambda-1)/2}\, (\ell+1),  t\}} ||\nu (s)||_{L^\infty(2^n,\, 2^{n+1})}dsÊ \nonumber \\
&&\hskip 7cm \times\int _{2^n}^{2^{n+1}} y^{\lambda}\, dy\nonumber \\
& \le & C\, x^{-(3+\lambda)/2}\, t \!\! \!\!\!\!\!\!\!\!  \sum_{0\le 2^n \le t^{-2/(\lambda-1)}} \!\!\!\!\!\!\!\!\sum_{\ell=0}^{\left[t\, (2^n)^{(\lambda-1)/2} \right]}
\int_{ 2^{-n(\lambda-1)/2}\, \ell}^{\min\{2^{-n(\lambda-1)/2}\, (\ell+1),  t\}} ||\nu (s)||_{L^\infty(2^n,\, 2^{n+1})}ds \timesÊ \nonumber \\
&&\hskip 7cm \times 2^{n (\lambda+1)} \nonumber \\
& \le & C\, x^{-(3+\lambda)/2}\, t  \!\!\!\!\!\!\!\!  \sum_{0\le 2^n \le t^{-2/(\lambda-1)}} \!\!\!\!\!\!\!\!\!\! \sum_{\ell=0}^{\left[t\, (2^n)^{(\lambda-1)/2} \right]}\!\!\!
 2^{-n(\lambda-1)/2} \times N_\infty (\nu; 2^{-n(\lambda-1)/2}\ell, 2^n) ds \,  2^{n (\lambda+1)}\nonumber \\
& \le &  \!\!\! C\, x^{-(3+\lambda)/2}\, t^2\, |||\nu|||_{X_{3/2, 2+\delta}} \sum_{0\le 2^n \le t^{-2/(\lambda-1)}}(2^n)^{\lambda-1-\delta}\nonumber \\
& \le & \!\!\!C\, x^{-(3+\lambda)/2}\, t^2\, |||\nu|||_{X_{3/2, 2+\delta}} (t^{-2/(\lambda-1)})^{\lambda-1-\delta}=
C\, \frac{t^{2\delta/(\lambda-1)}}{x^{(3+\lambda)/2}\, }\, |||\nu|||_{X_{3/2, 2+\delta}}.
\label{S3estItresunodos}
\eear

On the other hand: 
\bear
&&{\mathcal I}_{3, 2}=\int_0^tds\int_{t^{-2/(\lambda-1)}}^{R/2}\nu(y, s)g\left((t-s) y^{\frac{\lambda-1}{2}}, \frac{x}{y}\right)\frac {dy}{y} \nonumber\\
&& = 
\int_{t^{-2/(\lambda-1)}}^{R/2}\frac {dy}{y}\int_0^{t-y^{-(\lambda-1)/2}}\nu(y, s) g\left((t-s) y^{\frac{\lambda-1}{2}}, \frac{x}{y}\right)ds+\nonumber\\
& + & \,\int_{t^{-2/(\lambda-1)}}^{R/2}\frac {dy}{y}\int_{t-y^{-(\lambda-1)/2}}^t\nu(y, s) g\left((t-s) y^{\frac{\lambda-1}{2}}, \frac{x}{y}\right)ds \nonumber\\
&&={\mathcal I}_{3, 2,1}+{\mathcal I}_{3, 2, 2}. \label{S8E24}
\eear
In the term ${\mathcal I}_{3, 2,1}$ we use (\ref{S2TsolfundE78}) that gives

\bear
\label{S3estItresdosuno}
|{\mathcal I}_{3, 2,1}|& \le & x^{-(3+\lambda)/2}\int_{t^{-2/(\lambda-1)}}^{R/2}dy\int_0^{t-y^{-(\lambda-1)/2}}(t-s)^{-(\lambda+1)/(\lambda-1)}|\nu(y, s)| ds \nonumber\\
& \le & C\, x^{-(3+\lambda)/2}\,   \sum_{0\le 2^n \le t^{-2/(\lambda-1)}} \sum_{\ell=1}^{\left[t\, (2^n)^{(\lambda-1)/2} \right]} 
\int_{2^n}^{2^{n+1}}dy \int_{2^{-n (\lambda-1)/2}\, \ell}^{2^{-n (\lambda-1)/2}\, (\ell+1) }ds \nonumber \\
&& \hskip 7cm s^{-(\lambda+1)/(\lambda-1)} |\nu(y, t-s)|\nonumber \\
& \le & C\, x^{-(3+\lambda)/2}\,   \sum_{0\le 2^n \le t^{-2/(\lambda-1)}} \sum_{\ell=1}^{\left[t\, (2^n)^{(\lambda-1)/2} \right]} 
\int_{2^n}^{2^{n+1}}dy \nonumber \\
&&(2^{-n(\lambda-1)/2}\ell)^{-(\lambda+1)/(\lambda-1)}\, 2^{-n(\lambda-1)/2}
\left(2^{n(\lambda-1)/2}\int_{2^{-n (\lambda-1)/2}\, \ell}^{2^{-n (\lambda-1)/2}\, (\ell+1) }ds|\nu(y, t-s)| \right)\nonumber \\
 &\le & C\, x^{-(3+\lambda)/2}\,   \sum_{0\le 2^n \le t^{-2/(\lambda-1)}} \sum_{\ell=1}^{\left[t\, (2^n)^{(\lambda-1)/2} \right]} 
\int_{2^n}^{2^{n+1}}dy\nonumber \\
&&(2^{-n(\lambda-1)/2}\ell)^{-(\lambda+1)/(\lambda-1)}\, 2^{-n(\lambda-1)/2} N_\infty(\nu; t-2^{-n(\lambda-1)/2}\, \ell, 2^n)\nonumber \\
& \le  & C\, x^{-(3+\lambda)/2}\,|||\nu|||_{X_{3/2, 2+\delta}}   \sum_{0\le 2^n \le t^{-2/(\lambda-1)}}
(2^n)^{-\delta}\sum_{\ell=1}^{\left[t\, (2^n)^{(\lambda-1)/2} \right]} 
\ell^{-(\lambda+1)/(\lambda-1)}\nonumber \\
& = & C\, x^{-(3+\lambda)/2}\,|||\nu|||_{X_{3/2, 2+\delta}} t^{2\delta/(\lambda-1)}
\eear
In the term ${\mathcal I}_{3, 2,2}$, we use (\ref{S2TsolfundE3bis}) which gives:
\bear
|{\mathcal I}_{3, 2,2}|& \le & x^{-(3+\lambda)/2}\int_{t^{-2/(\lambda-1)}}^{R/2}y^{\lambda+1}\frac {dy}{y}\int_{t-y^{-(\lambda-1)/2}}^t (t-s) |\nu(y, s)|ds \nonumber\\
 & \le  & C x^{-(3+\lambda)/2}\sum_{t^{-2/(\lambda-1)}\le 2^n\le R/2}\int_{2^n}^{2^{n+1}}y^{\lambda+1}\frac {dy}{y}
 \int_{t-2^{-n(\lambda-1)/2}}^t (t-s) |\nu(y, s)|ds \nonumber \\
 & \le  & C x^{-(3+\lambda)/2}\sum_{t^{-2/(\lambda-1)}\le 2^n\le R/2}\int_{2^n}^{2^{n+1}} y^\lambda dy\, 2^{-n(\lambda-1)/2}2^{-n(\lambda-1)/2}\times\nonumber \\
&&\times2^{n(\lambda-1)/2} \int_{t-2^{-n(\lambda-1)/2}}^t |\nu(y, s)|ds \nonumber \\
& \le & C x^{-(3+\lambda)/2}\sum_{t^{-2/(\lambda-1)}\le 2^n\le R/2}\int_{2^n}^{2^{n+1}}dy\, 2^n
N_\infty(\nu; t-2^{-n(\lambda-1)/2}, 2^n)\nonumber \\
& \le & C x^{-(3+\lambda)/2}|||\nu|||_{X_{3/2, 2+\delta}}\sum_{t^{-2/(\lambda-1)}\le 2^n\le R/2}2^{-n\delta}\nonumber \\
&\le & 
C x^{-(3+\lambda)/2}|||\nu|||_{X_{3/2, 2+\delta}}t^{2\delta/(\lambda-1)}
.\label{S3estItresdosdos}
\eear
Estimates (\ref{S3estItresdosuno}) and (\ref{S3estItresdosdos}) yield
\bear
\label{S3estItresdos}
|\mathcal I_{3, 2}|\le C x^{-(3+\lambda)/2}\, t^{\frac{2\delta}{\lambda-1}}\, |||\nu|||_{X_{3/2, \,2+\delta}}.
\eear
Then, using also (\ref{S3EsplitItres}) and (\ref{S3estItresuno}), we deduce that
\bear
\label{S3estItres}
|\mathcal I_{3}|\le C x^{-(3+\lambda)/2}\, t^{\frac{2\delta}{\lambda-1}}\, |||\nu|||_{X_{3/2, \,2+\delta}} .
\eear

We estimate now the term ${\mathcal I}_4$. To this end we have:
\bean
\int_0^t ds\int_{y \ge 2 R}\nu(y)g\left((t-s) y^{\frac{\lambda-1}{2}}, \frac{x}{y}\right)\frac {dy}{y}\le
\int_{y \ge 2 R}\frac{dy}{y}\int_0^{t-y^{-(\lambda-1)/2}}[\cdots] ds+\\
+\int_{y \ge 2 R}\frac{dy}{y}\int_{t-y^{-(\lambda-1)/2}}^t[\cdots] ds=\mathcal I_{4,1}+\mathcal I_{4,2}.
\eean

We split ${\mathcal I}_{4,1}$ in two pieces as follows:
\bear
\label{S3decIcuatrouno}
\mathcal I_{4, 1} & = & \int_{y \ge 2 R}\frac{dy}{y}\int_0^{t-x^{-(\lambda-1)/2}}[\cdots] ds+
\int_{y \ge 2 R}\frac{dy}{y}\int_{t-x^{-(\lambda-1)/2}}^{t-y^{-(\lambda-1)/2}}[\cdots] ds\nonumber \\
& = & \mathcal I_{4, 1, 1}+\mathcal I_{4, 1, 2}.
\eear
In the term $\mathcal I_{4, 1,1}$ we are in the region where (\ref{S2TsolfundE7}) holds. Then, we use (\ref{S2TsolfundE78}) to obtain:
\bear
|\mathcal I_{4, 1, 1}| 
& \le &  Cx^{-(3+\lambda)/2}\int_{y \ge 2 R}dy\int_{x^{-(\lambda-1)/2}}^ts^{-(\lambda+1)/(\lambda-1)}|\nu(y, (t-s))| ds
\nonumber\\
 & \le  & C x^{-(3+\lambda)/2}\sum_{2^n\ge 2R}\int_{2^n}^{2^{n+1}}dy\sum_{x^{-(\lambda-1)/2}\le 2^{-n(\lambda-1)/2}\ell\le t}
 \int_{2^{-n(\lambda-1)/2}\ell}^{2^{-n(\lambda-1)/2}(\ell+1)} s^{-(\lambda+1)/(\lambda-1)}|\nu(y, (t-s))|ds \nonumber \\
 & \le  & C x^{-(3+\lambda)/2}\sum_{2^n\ge 2R} 2^{n(\lambda+1)/2}2^n \sum_{x^{-(\lambda-1)/2}\le 2^{-n(\lambda-1)/2}\ell\le t}
  \ell^{-(\lambda+1)/(\lambda-1)}\times \nonumber \\
 && \hskip 4cm \times 2^{-n(\lambda-1)/2}  N_\infty (\nu; t-2^{-n(\lambda-1)/2}\ell, 2^n) \nonumber \\
  & \le  &
  C x^{-(3+\lambda)/2}|||\nu|||_{X_{3/2, 2+\delta}}\sum_{2^n\ge 2R} 2^{-n\delta}\nonumber\\
  & \le & C x^{-(3+\lambda)/2}|||\nu|||_{X_{3/2, 2+\delta}}R^{-2\delta}\le C x^{-(3+\lambda)/2}|||\nu|||_{X_{3/2, 2+\delta}}t^{2\delta/(\lambda-1)}
  \label{S3estIcuatrounouno}
\eear

In the integral $\mathcal I_{4, 1, 2}$ we use (\ref{S2TsolfundE6}) to obtain:
\bear
\label{S2TsolfundE6bis}
|g\left((t-s) y^{\frac{\lambda-1}{2}}, \frac{x}{y}\right)|\le C (t-s)^{-1/(\lambda-1)}\, y\, x^{-3/2}.
\eear
This yields,
\bear
|\mathcal I_{4, 1, 2}| & \le & C x^{-3/2}\int_{y \ge 2 R}dy
\int_{t-x^{-(\lambda-1)/2}}^{t-y^{-(\lambda-1)/2}}(t-s)^{-1/(\lambda-1)}|\nu(y, s)|ds \nonumber\\
& = & C x^{-3/2}\int_{y \ge 2 R}dy
\int_{y^{-(\lambda-1)/2}}^{x^{-(\lambda-1)/2}}s^{-1/(\lambda-1)}|\nu(y, (t-s))|ds \nonumber\\
& \le  & C x^{-3/2}\sum_{2^n\ge 2R}\int_{2^n}^{2^{n+1}}dy\times \nonumber \\
&&\hskip 2cm \times \sum_{\substack{1\le \ell\le 2^{-n(\lambda-1)/2} \\ \ell\le x^{-(\lambda-1)/2}}}
 \int_{2^{-n(\lambda-1)/2}\ell}^{2^{-n(\lambda-1)/2}(\ell+1)} s^{-1/(\lambda-1)}||\nu(t-s)||_{L^\infty(2^n, 2^{n+1})}ds \nonumber 
 \eear
 \bear
 \label{S3estIcuatrounodos}
& \le  & C x^{-3/2}\sum_{2^n\ge 2R}2^{2n} 2^{-n\lambda/2}
N_\infty(\nu, t- 2^{-n(\lambda-1)/2}\ell , 2^n)\nonumber \\
& \le & C x^{-3/2}R^{-\lambda/2}|||\nu|||_{X_{3/2, 2+\delta}}\sum_{2^n\ge 2R}2^{-n\delta}
\le Cx^{-(3+\lambda)/2}|||\nu|||_{X_{3/2, 2+\delta}}t^{2\delta/(\lambda-1)}
\eear
using (\ref{S3Ezona2})  in the last step.

 In the term $\mathcal I_{4, 2}$ we use (\ref{S2TsolfundE2}) to obtain,
\bear
\label{S2TsolfundE2bis}
|g\left((t-s) y^{\frac{\lambda-1}{2}}, \frac{x}{y}\right)|\le C (t-s)\, y^{(\lambda+2)/2}\, x^{-3/2}
\eear
and then
\bear
\label{S3estIcuatrodos}
|I_{4, 2}| & \le & C x^{-3/2}\int_{y \ge 2 R}y^{(\lambda+2)/2}\frac{dy}{y}\int_{t-y^{-(\lambda-1)/2}}^t(t-s)|\nu(y, s)| ds \nonumber\\
& \le & C x^{-3/2}\sum_{2^n\ge 2R}\int_{2^n}^{2^{n+1}}y^{\lambda/2}\, dy \int_{t-y^{-(\lambda-1)/2}}^t(t-s)|\nu(y, s)| ds \nonumber\\
& \le & C x^{-3/2}\sum_{2^n\ge 2R}2^{-n(\lambda-1)}2^{n\lambda/2}2^n
N_\infty(\nu, t- 2^{-n(\lambda-1)/2}, 2^n)
 \nonumber\\
&\le &  C x^{-(3+\lambda)/2}\, t^{2\delta/(\lambda-1)}\, |||\nu|||_{X_{3/2, \,2+\delta}}.
\eear
where we have used  (\ref{S3Ezona2}) in the last inequality.\\ \\
Estimates (\ref{S3estIcuatrounouno}), (\ref{S3estIcuatrounodos}) and (\ref{S3estIcuatrodos}) give
\bear
\label{S3estIcuatro}
|I_{4}|\le C  R^{-(3+\lambda)/2}\, t^{2\delta/(\lambda-1)}\, |||\nu|||_{X_{3/2, \,2+\delta}},
\eear
which, combined with (\ref{S3estIuno}),  (\ref{S3estIdos}) and   (\ref{S3estItres}) yields,
\bear
\label{S3estmrgrande}
R^{(3+\lambda)/2}\left|\left|\int_0^t G(t-s)\, \nu(s)\, ds  \right|\right|_{L^{\infty}(R/2,\, 2R)}\le C t^{2\delta/(\lambda-1)}|||\nu|||_{X_{3/2, \,2+\delta}}
\eear
for $R\ge t^{-2/(\lambda-1)}$.
\\

We assume now:
\bear
\label{S3Ezona3}
1\le R\le t^{-2/(\lambda-1)}
\eear
 Then,
\bear
\label{S8E22bis}
&&\int_0^t G(t-s)\nu (s, y)\, ds  =\int_0^t ds\int_0^\infty dy \nu (s, y) g\left((t-s)  y^{\frac{\lambda-1}{2}}, \frac{x}{y}\right)\frac {dy}{y} \nonumber \\
&&
\le\int_0^{t} ds\int_{|x-y|\le R/2}\nu(s, y) g\left((t-s) y^{\frac{\lambda-1}{2}}, \frac{x}{y}\right)\frac {dy}{y} \nonumber \\
&&+\int_0^t ds\int_{y\le 1} \nu (s, y)g\left((t-s) y^{\frac{\lambda-1}{2}}, \frac{x}{y}\right)\frac {dy}{y} \nonumber \\
&&+\int_0^t ds\int_{1\le y \le 5 R/4}\nu(s, y)g\left((t-s) y^{\frac{\lambda-1}{2}}, \frac{x}{y}\right)\frac {dy}{y} \nonumber \\
&&+\int_0^t ds\int_{|y| \ge 5 R/4}\nu(s, y)g\left((t-s) y^{\frac{\lambda-1}{2}}, \frac{x}{y}\right)\frac {dy}{y} \nonumber \\
&&={\mathcal J}_1+{\mathcal J}_2+{\mathcal J}_3+{\mathcal J}_4.
\eear
In the term $\mathcal J _1$ we use again $(\ref{S2TsolfundE4bis})$ to obtain
\bear
\label{S3estjoyauno}
|{\mathcal J}_1|& \le &  
C\, \int_{0}^t ds\int_{|x-y|\le R/2}
\frac{\nu(s, y)}{((t-s)\, y^{\frac{\lambda-1}{2}})^2}
\Phi\left(\frac{\frac{x}{y}-1}{(t-s)^2\, y^{\lambda-1}}\right)\frac{d\, y}{y} \nonumber\\
& \le &  
C\, \int_{0}^t ds||\nu (s) ||_{L^\infty(R/2, 2R)}\le C \sqrt t \left( \int_{0}^t ds||\nu (s) ||^2_{L^\infty(R/2, 2R)}\right)^{1/2}\nonumber \\
& \le & C \sqrt t R^{-(\lambda-1)/4}\left( R^{(\lambda-1)/2}\int_{0}^{R^{-(\lambda-1)/2}} ds||\nu (s) ||^2_{L^\infty(R/2, 2R)}\right)^{1/2}\nonumber \\
& \le & 
 C R^{-(3+\lambda)/2}\left[\sqrt t R^{(\lambda-1)/4}R^{-\delta}\right]|||\nu|||_{X_{3/2, 2+\delta}}\nonumber \\
 & \le &  
 C R^{-(3+\lambda)/2} t^{2\delta(\lambda-1)}|||\nu|||_{X_{3/2, 2+\delta}}.
 \eear
In the term $\mathcal J _2$ we have again (\ref{S2TsolfundE3bis}) and therefore:
\bear
\label{S3estjoyados}
|\mathcal J _2|& \le & C\, x^{-(3+\lambda)/2}
\int_0^t ds \sum_{n=0}^\infty||\nu(s)||_{L^\infty(2^{-(n+1)}, 2^{-n})}(t-s)\int_{2^{-(n+1)}}^{2^{-n}} y^{\lambda }dy \nonumber\\
& \le & C\, x^{-(3+\lambda)/2}t
 \sum_{n=0}^\infty\int_0^t ds||\nu(s)||_{L^\infty(2^{-(n+1)}, 2^{-n})}2^{-n(\lambda+1)} \nonumber\\
 & \le & C\, x^{-(3+\lambda)/2}t
 \sum_{n=0}^\infty 2^{-n(\lambda+1)}  \sqrt t M_\infty (\nu; 0, 2^{-n}), \nonumber\\
  & \le & C\, x^{-(3+\lambda)/2}t^{3/2}|||\nu|||_{X_{3/2, 2+\delta}}
 \sum_{n=0}^\infty 2^{-n(\lambda-1/2)}.
\eear
We consider now $\mathcal J _3$ where we still have (\ref{S2TsolfundE3bis}) and then,
\bear
|\mathcal J _3|& \le & C\, x^{-(3+\lambda)/2}
\int_0^t ds (t-s)\int_{1}^{5R/4} y^{\lambda} |\nu(y, s)|dy \nonumber\\
& \le & C\, x^{-(3+\lambda)/2}\sum_{1\le 2^n \le 5R/4}
\int_0^t ds (t-s)||\nu(s)||_{L^\infty(2^n, 2^{n+1})}\int_{2^n}^{2^{n+1}} y^{\lambda} dy \nonumber\\
& \le & C\, x^{-(3+\lambda)/2}\sum_{1\le 2^n \le 5R/4}2^{n(\lambda+1)}t
\sqrt t \left(\int_0^{R^{-(\lambda-1)/2}} ds||\nu(s)||^2_{L^\infty(2^n, 2^{n+1})}\right)^{1/2} \nonumber\\
 & \le & C\, x^{-(3+\lambda)/2}t^{3/2}\sum_{1\le 2^n \le 5R/4}2^{n(\lambda+1)}
 2^{-n(\lambda-1)/4}N_\infty(\nu, 0, 2^n) \nonumber\\
 & \le & C\, x^{-(3+\lambda)/2}t^{3/2}|||\nu|||_{X_{3/2, 2+\delta}}\sum_{1\le 2^n \le 5R/4}(2^n)^{\frac{3}{4}(\lambda-1)-\delta}
  \nonumber\\
   & \le & C\, x^{-(3+\lambda)/2}t^{3/2}|||\nu|||_{X_{3/2, 2+\delta}}R^{\frac{3}{4}(\lambda-1)-\delta}
  \le C\,x^{-(3+\lambda)/2}|||\nu|||_{X_{3/2, 2+\delta}}t^{\frac{2\delta}{\lambda-1}}.\label{S3estjoyatres}
\eear
In the term $\mathcal J_4$,
\bear
\label{S3descjoyatres}
\mathcal J _4=\int_{|y| \ge 5 R/4}\frac {dy}{y}\int_0^{\left(t-y^{-(\lambda-1)/2}\right)_+} ds\, \nu(s, y)g\left((t-s) y^{\frac{\lambda-1}{2}}, \frac{x}{y}\right)+\nonumber \\
\int_{|y| \ge 5 R/4}\frac {dy}{y}\int_{\left(t-y^{-(\lambda-1)/2}\right)_+}^t ds\, \nu(s, y)g\left((t-s) y^{\frac{\lambda-1}{2}}, \frac{x}{y}\right)=
\mathcal J _{4,1}+\mathcal J _{4,2}.
\eear
In the first term at the right hand side of (\ref{S3descjoyatres}) we are in the region where (\ref{S2TsolfundE6}) holds and then we have (\ref{S2TsolfundE6bis}) to obtain:
\bean
|\mathcal J _{4,1}|\le  C\, x^{-3/2} \int_{|y| \ge 5 R/4}dy
\int_0^{\left(t-y^{-(\lambda-1)/2}\right)_+}(t-s)^{-1/(\lambda-1)}|\nu(y, s)| ds.
\eean
Notice that this integral is nonzero if and only if $y\ge t^{-2/(\lambda-1)}$. In that case:
 
\bear
\label{S3estjoyacuatrouno}
|\mathcal J _{4,1}| &\le & C\, x^{-3/2} \sum_{2^n\ge t^{-2/(\lambda-1)}\ge R}
\int_{2^n}^{2^{n+1}}dy
\int_{y^{-(\lambda-1)/2}}^{t}s^{-1/(\lambda-1)}||\nu(t-s)||_{L^\infty(2^n, 2^{n+1})} ds \nonumber \\
&\le & C\, x^{-3/2} \sum_{2^n\ge t^{-2/(\lambda-1)}\ge R}\times \nonumber \\
&&\times \sum_{\ell=1, \, 2^{-n(\lambda-1)/2}\, \ell\le t}
\int_{2^n}^{2^{n+1}}dy
\int_{2^{-n(\lambda-1)/2}\, \ell}^{2^{-n(\lambda-1)/2}\, (\ell+1)}s^{-1/(\lambda-1)}||\nu(t-s)||_{L^\infty(2^n, 2^{n+1})} ds \nonumber \\
&\le & C\, x^{-3/2} \sum_{2^n\ge t^{-2/(\lambda-1)}\ge R}\,\,\sum_{\substack{1\,\le\, \ell\le 2^{-n(\lambda-1)/2} \\ \ell\le t}} 2^n2^{-n(\lambda-1)}\ell^{-1/(\lambda-1)}2^{n/2}\times \nonumber \\
&&\times 
N_\infty(\nu; t-2^{-n(\lambda-1)/2}\ell, 2^n)
\eear
\bear
&\le& C\, x^{-3/2} |||\nu|||_{X_{3/2, 2+\delta}}\sum_{2^n\ge t^{-2/(\lambda-1)}\ge R}
2^{-n\lambda/2}2^{-n\delta}\nonumber \\
&\le& C\, x^{-3/2} |||\nu|||_{X_{3/2, 2+\delta}}t^{2\delta/(\lambda-1)}R^{-\lambda/2}\, \le \, C\, |||\nu|||_{X_{3/2, 2+\delta}}t^{2\delta/(\lambda-1)} R^{-(3+\lambda)/2}.
\eear 
In the integral  $\mathcal J _{4,2}$, we are in a region where  (\ref{S2TsolfundE2}) holds true. Then we may use 
(\ref{S2TsolfundE2bis}) $(t-s)\, y^{(\lambda+2)/2}\, x^{-3/2}$ to get:
\bean
|\mathcal J _{4,2}| & \le & C x^{-3/2}\,\int_{|y| \ge 5 R/4}y^{\lambda/2}dy\int_{\left(t-y^{-(\lambda-1)/2}\right)_+}^t ds\, (t-s) |\nu(y, s)|.
\eean
The last integral is bounded as follows:
\bean
|\mathcal J _{4,2}|
&\le & C x^{-3/2}\,\int_{5 R/4}^{t^{-2/(\lambda-1)}}y^{\lambda/2}dy\int_{0}^t ds\, (t-s) |\nu(y, s)|+\nonumber \\
&&+C x^{-3/2}\,\int_{t^{-2/(\lambda-1)}}^{\infty}y^{\lambda/2}dy\int_{\left(t-y^{-(\lambda-1)/2}\right)_+}^t ds\, (t-s) |\nu(y, s)|
\nonumber \\
&\le & C x^{-3/2}t^{3/2}\sum_{R\le 2^n\le t^{-2/(\lambda-1)}}(2^n)^{\lambda/2+1}2^{-n(\lambda-1)/4} (2^{n(\lambda-1)/2}\int_{0}^{2^{-n(\lambda-1)/2}} ds\,||\nu( s)||^2_{L^\infty(2^n, 2^{n+1})})^{1/2}+\nonumber \\
&&+C x^{-3/2}\sum_{2^n\ge t^{-2/(\lambda-1)}}2^{2n}2^{-n\lambda/2}N_\infty(\nu; t-2^{-n(\lambda-1)/2}, 2^n)
\nonumber \\
&\le & C x^{-3/2}t^{3/2}\sum_{R\le 2^n\le t^{-2/(\lambda-1)}}(2^n)^{\lambda/2+1}2^{-n(\lambda-1)/4} 
N_\infty (\nu; 0, 2^n)+\nonumber \\
&&+C x^{-3/2}\sum_{2^n\ge t^{-2/(\lambda-1)}}2^{2n}2^{-n\lambda/2}N_\infty(\nu; t-2^{-n(\lambda-1)/2}, 2^n)
\nonumber \\
&\le & C x^{-3/2}R^{-\lambda/2}t^{3/2}|||\nu|||_{X_{3/2, 2+\delta}}t^{-\frac{2}{\lambda-1} \frac{3(\lambda-1)}{4}}
+\nonumber \\
&+ & C x^{-3/2}|||\nu|||_{X_{3/2, 2+\delta}}R^{-\lambda/2}\sum_{2^n\ge t^{-2/(\lambda-1)}}2^{-n\delta}
=C R^{-(3+\lambda)/2}|||\nu|||_{X_{3/2, 2+\delta}}t^{2\delta/(\lambda-1)}.
\eean
This yields,
\bear
\label{S3estjoyacuatrodos}
|\mathcal J _{4,2}|&\le &C\, R^{-(3+\lambda)/2} t^{2\delta/(\lambda-1)}|||\nu|||_{X_{3/2, \, 2+\delta}},
\eear
which, combined with (\ref{S3estjoyacuatrodos}) gives
\bear
\label{S3estjoyacuatro}
|\mathcal J _{4}|&\le &C\, R^{-(3+\lambda)/2} t^{2\delta/(\lambda-1)}|||\nu|||_{X_{3/2, \, 2+\delta}}.
\eear
Adding (\ref{S3estjoyauno}), (\ref{S3estjoyados}), (\ref{S3estjoyatres}) and (\ref{S3estjoyacuatro}):
\bear
\label{S3estmrpeq}
R^{(3+\lambda)/2}\left| \left| \right| \right|_{L^\infty(R/2, 2\, R)} \le C t^{2\delta/(\lambda-1)} |||\nu|||_{X_{3/2, \, 2+\delta}}
\eear
for all $R\ge t^{-2/(\lambda-1)}$.
Adding (\ref{S3estmrgrande}) and (\ref{S3estmrpeq}) yields, for all $R>1$:
\bear
\label{S3estmrmayuno}
R^{(3+\lambda)/2}\left|\left|\int_0^t G(t-s)\, \nu(s)\, ds  \right|\right|_{L^{\infty}(R/2,\, 2R)}\le C t^{2\delta/(\lambda-1)}|||\nu|||_{X_{3/2, \, 2+\delta}}.
\eear

We now consider the region where $0<R<1$.
Then,  for $|x-R|\le R/8$:
\bear
&&\int_0^t G(t-s)\nu (s, y)\, ds  \le\int_0^t ds\int_{y\le 3R/4} |\nu (s, y)| g\left((t-s)  y^{\frac{\lambda-1}{2}}, \frac{x}{y}\right)\frac {dy}{y} \nonumber \\
&&+ \int_0^t ds\int_{y\ge 5R/4} |\nu (s, y)| g\left((t-s)  y^{\frac{\lambda-1}{2}}, \frac{x}{y}\right)\frac {dy}{y}\nonumber \\
&&+\int_0^t ds\int_{|x-y|\le R/2} \nu (s, y) g\left((t-s)  y^{\frac{\lambda-1}{2}}, \frac{x}{y}\right)\frac {dy}{y}\nonumber\\
&&=\mathcal K_1+\mathcal K_2+\mathcal K_3\label{S3desc}
\eear
The last integral in the right hand side of (\ref{S3desc}) is estimated as follows. Since in that term (\ref{S2TsolfundE4}) holds we still have (\ref{S2TsolfundE4bis}) and then

\bear
\label{S3estcatres}
|\mathcal K_3|& \le &  
C\, \int_{0}^t ds\int_{|x-y|\le R/2}
\frac{\nu(s, y)}{((t-s)\, y^{\frac{\lambda-1}{2}})^2}
\Phi\left(\frac{\frac{x}{y}-1}{(t-s)^2\, y^{\lambda-1}}\right)\frac{d\, y}{y} \nonumber\\
& \le & C\, \int_{0}^t ||\nu(s)||_{L^\infty(R/2, 2R)}ds\int_{|x-y|\le R/2}
\frac{1}{(t-s)^2\, y^\lambda}
\Phi\left(\frac{x-y}{(t-s)^2\, y^{\lambda}}\right)\, dy \nonumber\\
& \le & \!\!C\, \int_{0}^t ||\nu(s)||_{L^\infty(R/2, 2R)}\le C \sqrt t \, M_\infty(\nu; R)\le C \sqrt t\, R^{-3/2}\, |||\nu|||_{X_{3/2, 2+\delta}}.
\eear
Using (\ref{S2TsolfundE3}) we deduce that, in the integral $\mathcal K_1$ the following estimate holds:
\bean
g\left((t-s)y^{(\lambda-1)/2}, \frac{x}{y}\right)\le C(t-s)y^{\lambda+1}x^{-(3+\lambda)/2}.
\eean
Using this estimate we deduce:

\bear
\label{S3estcauno}
|\mathcal K_1|
& \le & C\, x^{-(3+\lambda)/2}\sum_{2^{-n}\le R}
\int_0^t ds\int_{2^{-(n+1)}}^{2^{-n}} y^{\lambda} ||\nu (s)||_{L^\infty(2^{-(n+1), 2^{-n}})}(t-s) dy \nonumber\\
& \le & C\, x^{-(3+\lambda)/2}t\sum_{2^{-n}\le R} 2^{-n(\lambda+1)}
\sqrt t\,\, M_\infty(\nu, 2^{-n}) \nonumber \\
& \le & C\, x^{-(3+\lambda)/2}t^{3/2}|||\nu|||_{X_{3/2, 2+\delta}}\sum_{2^{-n}\le R} 2^{-n(\lambda+1-3/2)}\nonumber \\
& \le & C\, x^{-(3+\lambda)/2}t^{3/2}|||\nu|||_{X_{3/2, 2+\delta}}\,R^{\lambda-1/2}
 \le\,\,  C\, x^{-3/2}t^{3/2}|||\nu|||_{X_{3/2, 2+\delta}}
\eear
for $x\in (R/2, 2R)$.
We are then left with the term $|\mathcal K_2|$.
\bear
\label{S3desccados}
\mathcal K_2 & = & \int_{5R/4\le y \le 2} \frac {dy}{y} \int_0^{t} ds\, \nu (s, y) g\left((t-s)  y^{\frac{\lambda-1}{2}}, \frac{x}{y}\right) \nonumber\\
& + & \int_{y\ge 2} \frac {dy}{y} \int_0^{(t-y^{-(\lambda-1)/2})_+} ds\, \nu (s, y) g\left((t-s)  y^{\frac{\lambda-1}{2}}, \frac{x}{y}\right) \nonumber
\\
& + & \int_{y\ge 2} \frac {dy}{y} \int_{(t-y^{-(\lambda-1)/2})_+}^t ds\, \nu (s, y) g\left((t-s)  y^{\frac{\lambda-1}{2}}, \frac{x}{y}\right) \nonumber\\
& = & \mathcal K_{2, 1} +\mathcal K_{2, 2} +\mathcal K_{2, 3}.
\eear
In the term $\mathcal K_{2, 1}$, we may use (\ref{S2TsolfundE2bis}) to obtain:
\bear
\label{S3estcadosuno}
|\mathcal K_{2, 1}| & \le &  C\, x^{-3/2}
\int_{5R/4\le y\le 2}dy y^{\lambda/2} \int_0^t ds|\nu (s, y)|(t-s) \nonumber\\
& \le &  C\, x^{-3/2}t \sum_{n=0,\, R\le 2^{-n}}
2^{-n(1+\lambda/2)}\sqrt t\,\,\left( \int_0^t ds||\nu (s)||^2_{L^\infty(2^{-(n+1)}, 2^{-n})}\right)^{1/2} \nonumber \\
& \le &  C\, x^{-3/2}t^{3/2} \sum_{n=0,\, R\le 2^{-n}}
2^{-n(1+\lambda/2)}\,M_\infty(\nu, 2^{-n}) \nonumber \\
& \le &  C\, x^{-3/2}t^{3/2} |||\nu|||_{X_{3/2, 2+\delta}}\sum_{n=0,\, R\le 2^{-n}}
2^{-n(\lambda-1)/2}
\nonumber \\
& \le &  C\, x^{-3/2}t^{3/2} |||\nu|||_{X_{3/2, 2+\delta}}.
\eear

In
$\mathcal K_{2, 2}$, (\ref{S2TsolfundE6bis}) holds and then,
\bear
\label{S3estcadosdoster}
|\mathcal K_{2, 2}| & \le & C\, x^{-3/2}\int_{y\ge 2} dy \int_0^{(t-y^{-(\lambda-1)/2})_+} ds\, |\nu (s, y) |
(t-s)^{-1/(\lambda-1)}
\eear
We notice also here that the last integral in the right hand side of (\ref{S3estcadosdoster}) is nonzero only if $y \ge t^{-2/(\lambda-1)}$. Therefore
\bear
\label{S3estcadosdos}
|\mathcal K_{2, 2}| 
& \le & C\, x^{-3/2}\sum_{n=1,\, 2^n\ge t^{-2/(\lambda-1)}}2^n 
\int_{y^{-(\lambda-1)/2}}^{t} ds\, s^{-1/(\lambda-1)}||\nu (s) ||_{L^\infty(2^n, 2^{n+1})}
\nonumber\\
& \le & C\, x^{-3/2}\sum_{n=1,\, 2^n\ge t^{-2/(\lambda-1)}}2^{3n/2} 
\sum_{\ell =1,\, 2^{-n(\lambda-1)/2}\, \ell\le t}\ell^{-1/(\lambda-1)}2^{-n(\lambda-1)/2}\times\nonumber \\
&&\hskip 8cm \times N_\infty (\nu; t-2^{-n(\lambda-1)/2}\, \ell, 2^n)
\nonumber\\ \nonumber \\
& \le & C\, x^{-3/2}|||\nu|||_{X_{3/2, 2+\delta}}\sum_{n=1,\, 2^n\ge t^{-2/(\lambda-1)}}2^{-n(\lambda/2+\delta)} \nonumber \\ 
& \le & C\, x^{-3/2}|||\nu|||_{X_{3/2, 2+\delta}}t^{\frac{2}{\lambda-1}(\lambda/2 +\delta)}.
\eear

In $\mathcal K_{2, 3}$, we may use (\ref{S2TsolfundE2bis}), whence
\bear
\label{S3estcadostres}
|\mathcal K_{2, 3}| & \le & C\, x^{-3/2}
\int_{y\ge 2}y^{\lambda/2}dy \int_{(t-y^{-(\lambda-1)/2})_+}^t |\nu(y, s)| (t-s)ds \nonumber \\
& \le & C\, x^{-3/2}t\sum_{n=1}^\infty
2^{n(\lambda/2+1)}2^{-n(\lambda-1)/2}(2^{n(\lambda-1)/2}  \int_{(t-2^{-n(\lambda-1)/2})_+}^t ||\nu( s)||_{L^\infty(2^n, 2^{n+1})}ds)\nonumber \\
& \le & C\, x^{-3/2}t\sum_{n=1}^\infty
2^{n(\lambda/2+1)}2^{-n(\lambda-1)/2}N_\infty(\nu; t-2^{-n(\lambda-1)/2}, 2^n)\nonumber \\
& \le & C\, x^{-3/2}t|||\nu|||_{X_{3/2, 2+\delta}}\sum_{n=1}^\infty
2^{-n(1/2+\delta)}\le  C\, x^{-3/2}t|||\nu|||_{X_{3/2, 2+\delta}}.
\eear
By (\ref{S3desccados}), (\ref{S3estcadosuno}), (\ref{S3estcadosdos}) and (\ref{S3estcadostres}) we have
\bear
\label{S3estcados}
|\mathcal K_{2}| & \le & C\, x^{-3/2}\left(t+ t^{(2\delta+\lambda)/(\lambda-1)}\right)|||\nu|||_{X_{3/2, 2+\delta}}.
\eear
Adding (\ref{S3estcatres}), (\ref{S3estcauno}) and (\ref{S3estcados})  we obtain the following estimate for
$0< R \le 1$:
\bear
\label{S3estmracero}
R^{3/2}\left|\left|\int_0^t G(t-s)\, \nu(s)\, ds  \right|\right|_{L^{\infty}(R/2,\, 2R)}\le C t\, |||\nu|||_{X_{3/2, 2+\delta}}.
\eear
The Lemma follows combining (\ref{S3estmrmayuno}) and (\ref{S3estmracero}).\qed

\begin{lem}
\label{lemcor} For all $\varphi \in Y^{\sigma}_{3/2, (3+\lambda)/2}(T)$ with $\sigma >1+\delta$, $\varepsilon \ge 0$  and $0<T<1$:
\bean
\sup_{0\le t \le T}\left|\left|\left|\int_0^tG(t-s)(\mathcal{L}-L_\varepsilon)\varphi(s)ds\right|\right|\right|_{{3/2,(3+\lambda)/2}}\le C T^\beta|||\varphi |||\eean
for some constant $C>0$ independent of $T$, of $\varepsilon$ and $\varphi$.
\end{lem}
\textbf{Proof of Lemma \ref{lemcor}}.
By Lemma \ref{S8T4}:
\bean
\sup_{0\le t \le T}||\int_0^tG(t-s)A_1(s)ds||_{3/2, (3+\lambda)/2}\le CT^\beta ||A_1||_{X_{3/2, (3+\lambda)/2}(T)}.
\eean
Moreover, for all $h(t, x), q, p$:
\bean
||h||_{X_{q, p}(T)}\le C \sup_{0\le t \le T} |||h(t)|||_{q, p}
\eean
Then, using also Lemma \ref{S8T3} and the definition of the norm $|||.|||$
\bean
T^\beta ||A_1||_{X_{3/2, (3+\lambda)/2}(T)}\le T^\beta |||\varphi|||_{3/2, (3+\lambda)/2}\le C|||\varphi|||
\eean
And then
\bear
\label{S3Elemaauno}
\sup_{0\le t \le T}||\int_0^tG(t-s)A_1(s)ds||_{3/2, (3+\lambda)/2}\le  C\, T^\beta\,|||\varphi|||.
\eear
A similar argument is used for the term $A_{2,\, \varepsilon}$. First, by Lemma \ref{S8T4},
\bean
\sup_{0\le t \le T}||\int_0^tG(t-s)A_{2,\, \varepsilon}(s)ds||_{3/2, (3+\lambda)/2}\le CT^\beta ||A_{2,\, \varepsilon}||_{X_{3/2, (3+\lambda)/2}(T)}.
\eean
Then, by Lemma \ref{S8Tprov},
\bean
||A_{2,\, \varepsilon}||_{X_{3/2, (3+\lambda)/2}(T)}\le C |||\varphi|||
\eean
whence, using the definition of the norm $|||.|||$:
\bear
\label{S3Elemados}
\sup_{0\le t \le T}||\int_0^tG(t-s)A_{2,\, \varepsilon}(s)ds||_{3/2, (3+\lambda)/2}\le CT^\beta |||\varphi|||
\eear
and Lemma \ref{lemcor} follows from (\ref{S3Elemaauno}) and (\ref{S3Elemados}).
\qed
\\ \\

\begin{lem} 
\label{regularity} 
There exists a positive constant $C$ such that, for all $0<T^*<1$, for all $\theta \in [0, 1]$, for all $\nu\in Y^\sigma_{3/2, \, 2+\delta}(T)$  and  all $\varphi$ satisfying $|||\varphi |||<+\infty$ and solving:
\bear
\label{S3Elemregeq}
\frac{\partial \varphi}{\partial t}=L(\varphi)+\theta\,\left(\mathcal {L}-L\right)(\varphi)+\nu,\,\,\,x>0, \,\,t\in(0, T^*)
\eear
we have:
\bean
|||\varphi|||\le C\, ||\nu||_{Y^\sigma_{3/2, \, 2+\delta}(T^*)}.
\eean
\end{lem}
\begin{rem}
The result of Lemma \ref{regularity} remains true if the space $Y^\sigma_{3/2, \, 2+\delta}(T^*)$ is replaced by 
$Y^\sigma_{3/2, \, 2}(T^*)$. However, a solution of (\ref{S3Elemregeq}) satisfying $|||\varphi |||<+\infty$ does not exists in general if $\nu \in Y^\sigma_{3/2, \, 2}(T^*)$.
\end{rem}
\textbf{Proof of Lemma \ref{regularity}.} We first rewrite the equation (\ref{S3Elemregeq}) as follows:

\bean
\frac{\partial \varphi}{\partial t}=(1-\theta)L(\varphi)+\theta\,\mathcal {L}(\varphi)+\nu
\eean
Then, for $x\in (3R/4, 5R/5)$ and $R>1$ we define the new variables:
$x=XR$, $y=YR$, $t=(\tau/ R^{(\lambda-1)/2})$ and $\varphi(x, t)=R^{-(3+\lambda)/2}\, \Psi(X, \tau)$. Since $t\in (0, T_*)$, $\tau\in (0, T_*\,R^{(\lambda-1)/2})$.
\bear
\frac{\partial \Psi}{\partial \tau}&  = & (1-\theta)\,L(\Psi)+\theta\left[
 R^{3/2}\int_0^{X/2}\left((X-Y)^{\lambda/2}\Psi(X-Y)-X^{\lambda/2}\Psi(X)\right){(Ry)^{\lambda/2}}f_0(R\, y)\, dY\right]\nonumber\\
&&-\theta \,X^{\lambda/2}\Psi(X)\int_{X/2}^{\infty}Y^{\lambda/2}f_0(R Y)dY+\tilde \nu_1 \nonumber\\
\tilde \nu_1 & = & R^2\, \nu(R\, X, \tau\, R^{-(\lambda-1)/2})+
\theta R^{(3+\lambda)/2}\times \nonumber \\
&&\times\int_0^{X/2}\left((X-Y)^{\lambda/2}f_0(R(X-Y))-X^{\lambda/2}f_0(R X) \right)Y^{\lambda/2}\Psi(Y)dY\nonumber \\
&&
- \theta \,X^{\lambda/2}R^{(3+\lambda)/2}f_0(RX)\int_{X/2}^{\infty}Y^{\lambda/2}\Psi(Y)dY \label{S3nutildeuno} 
\eear
Using the expression of the operator $L$ given in (\ref{S1Eopele})
\bear
\frac{\partial \Psi}{\partial \tau}&  = & (1-\theta)\int_0^{X/2}\left((X-Y)^{\lambda /2}\Psi(X-Y)-X^{\lambda /2}\Psi(x)
\right)Y^{\lambda /2}Y^{-3/2}dy \nonumber\\
&&+[R^{(3+\lambda)/2}???]\left[
 R^{3/2}\int_0^{X/2}\left((X-Y)^{\lambda/2}\Psi(X-Y)-X^{\lambda/2}\Psi(X)\right){(Ry)^{\lambda/2}}f_0(R\, y)\, dY\right] \nonumber\\
&&-2(1-\theta)\sqrt 2 X^{(\lambda -1)/2}\Psi(X)-\theta\, 
R^{(3+\lambda)/2}\,X^{\lambda/2}\Psi(X)\int_{X/2}^{\infty}Y^{\lambda/2}f_0(R Y)dY\nonumber \\
&& +\tilde \nu_1+\tilde \nu_2 \nonumber\\
\tilde \nu_2 & = & (1-\theta) \int_0^{X/2}\left((X-Y)^{-3 /2}-X^{-3/2}\right)Y^{\lambda /2}\Psi(Y)dY -\nonumber\\
&&\hskip 6cm -(1-\theta) X^{-3 /2}\int_{X/2}^{\infty }Y^{\lambda /2}\Psi(Y)dY \label{S3nutildedos} 
\eear
We can rewrite the equation as
\bear
&&\frac{\partial \Psi}{\partial \tau}  =  T_{1-\theta,\, R}\left(M_{\lambda/2} \Psi\right)-a(X, t)\Psi+Q \label{S3EpsiUno}\\
&&a(X, t)=2(1-\theta)\sqrt 2 X^{(\lambda -1)/2}+\theta\,R^{(3+\lambda)/2} \,X^{\lambda/2}\int_{X/2}^{\infty}Y^{\lambda/2}f_0(R Y)dY \label{S3EpsiDos}\\
&& Q=\tilde \nu_1+\tilde \nu_2.\label{S3EpsiTres}
\eear
Since $|||\varphi|||_{3/2, (3+\lambda)/2}<\infty$, we can combine (\ref{S8T3-101E143}) in  Theorem \ref{S8T3-101} with (\ref{S3Eestipsi}) to obtain: 

\bean
\sup_{0\le T\le R^{{(\lambda-1)/2}}}\left(\int_T^{\min{(T+1, T^*\,R^{{(\lambda-1)/2}} )}}
||\Psi(s)||^2_{H^\sigma_x(3/4, 5/4)}ds \right)^{1/2}\le C\, \sup_{0\le t \le T^*}|||\varphi (t)|||_{3/2, (3+\lambda)/2}\\
+\sup_{0\le T\le R^{{(\lambda-1)/2}}}\left(\int_T^{\min{(T+1, T^*\,R^{{(\lambda-1)/2}} )}}
||Q(s)||^2_{H^\sigma_x(1/2, 2)}ds \right)^{1/2}
\eean
Moreover, in order to estimate the norm of $Q(s)$ we first notice that, using (\ref{S3Eestipsi}):
\bean
||\theta X^{-3 /2}\int_{X/2}^{\infty }Y^{\lambda /2}\Psi(Y, s)dY||_{H^\sigma_x(1/2, 2)}  \le  C|||\varphi(t)|||_{3/2, (3+\lambda)/2}+
C||\Psi(s)||_{H^{(\sigma-1)_+}(1/2, 2)}.
\eean
The same estimate holds trivially for the term $\theta \,X^{\lambda/2}R^{(3+\lambda)/2}f_0(RX)\int_{X/2}^{\infty}Y^{\lambda/2}\Psi(Y)dY $ in $\widetilde \nu_1$. We are then left with the term \\$\int_0^{X/2}\left((X-Y)^{-3 /2}-X^{-3/2}\right)Y^{\lambda /2}\Psi(Y)dY$. Using that
\bean
|\Psi(Y)|\le \frac{1}{Y^{(3+\lambda)/2}}
\eean 
we deduce:
\bean
||\int_0^{X/2}\left((X-Y)^{-3 /2}-X^{-3/2}\right)Y^{\lambda /2}\Psi(Y)dY||_{H^\sigma(1/2, 2)}
\le 
C|||\varphi(t)|||_{3/2, (3+\lambda)/2}\\+C||\Psi(s)||_{H^{(\sigma-1)_+}(1/2, 2)}.
\eean
 This gives
\bean
&&\sup_{0\le T\le T^*\,R^{{(\lambda-1)/2}}}\left(\int_T^{\min{(T+1, T_*\,R^{{(\lambda-1)/2}} )}}
||\Psi(s)||_{H^\sigma_X(3/4, 5/4)}ds \right)^{1/2}\le C\sup_{0\le t \le T^*}|||\varphi|||_{3/2, (3+\lambda)/2}\\
&&+C\,\sup_{0\le T\le T^*\,R^{{(\lambda-1)/2}}}\left(\int_T^{\min{(T+1, T_*\,R^{{(\lambda-1)/2}} )}}
||\Psi(s)||_{H^{(\sigma-1)_+}_X(1/2, 2)}ds \right)^{1/2}+C ||\nu||_{Y^\sigma_{3/2, \, 2+\delta}(T)}
\eean
A bootstrap argument then yields:
\bear
\label{S3estboots}
\sup_{0\le T\le T_*\,R^{{(\lambda-1)/2}}}\left(\int_T^{\min{(T+1, T_*\,R^{{(\lambda-1)/2}} )}}
||\Psi(s)||_{H^\sigma_X( {3/4}, {5/4})}ds \right)^{1/2}\,\,\le \nonumber \\
\le\,\, C \sup_{0\le t \le T^*}|||\varphi|||_{3/2, (3+\lambda)/2}
+ C ||\nu||_{Y^\sigma_{3/2, \, 2+\delta}(T^*)}
\eear
(actually in an interval  slightly smaller than $( {3/4}, {5/4})$, for example: $(7/8, 9/8)$). \\
We deduce,

\bear
\label{S3regularityrmayor}
\sup_{0\le t_0\le T^*}\sup_{R>1} \left(R^{(3+\lambda)/2}N_{2;\, \sigma}(\varphi; R, t_0) \right)\le C \sup_{0\le t \le T^*}|||\varphi(t)|||_{3/2, (3+\lambda)/2}+
C\,  ||\nu||_{Y^\sigma_{3/2, \, 2+\delta}(T^*)}
\eear

We consider now the case where $0<R\le 1$. We rescale the equation for $x\in (3R/4, 5R/5)$ and $R<1$. The new variables are now
$x=XR$, $y=YR$, and $\varphi(x, t)=R^{-3/2}\, \Psi(X, t)$.  Arguing as above, the function $\Psi$ satisfies now:
\bean
\label{S3EPsiRmenor}
&&\frac{\partial \Psi}{\partial \tau}  =  R^{\frac{\lambda-1}{2}}T_{1-\theta,\, R}\left(M_{\lambda/2} \Psi\right)+Q
\eean
\bear
\label{S3Erestoenorme}
&& Q= R^{3/2}\, \nu(R\, X, \tau)+R^{(\lambda-1)/2}\left((1-\theta) \int_0^{X/2}\left((X-Y)^{-3 /2}-X^{-3/2}\right)Y^{\lambda /2}\Psi(Y)dY -\right. \nonumber\\
&&\left.-(1-\theta) X^{-3 /2}\int_{X/2}^{\infty }Y^{\lambda /2}\Psi(Y)dY -2(1-\theta)\sqrt 2 X^{(\lambda -1)/2}\Psi(X)
 -\right. \nonumber\\
 &&\hskip 7cm \left.-(1-\theta) X^{-3 /2}\int_{X/2}^{\infty }Y^{\lambda /2}\Psi(Y)dY\right) \nonumber\\
 &&+
 \theta\,R^{\lambda+1}\int_0^{X/2}\left((X-Y)^{\lambda/2}\Psi(X-Y)-X^{\lambda/2}\Psi(X)\right){y^{\lambda/2}}f_0(R\, y)\, dY \nonumber\\
 &&+\theta R^{\lambda+1}\int_0^{X/2}\left((X-Y)^{\lambda/2}f_0(R(X-Y))-X^{\lambda/2}f_0(R X) \right)Y^{\lambda/2}\Psi(Y)dY \nonumber\\
 &&-\theta \, R^{\lambda/2}\,X^{\lambda/2}\Psi(X)\int_{0}^{\infty}y^{\lambda/2}f_0(y)dy+
 \theta \, R^{\lambda/2}\,X^{\lambda/2}\Psi(X)\int_{0}^{X/2}Y^{\lambda/2}\, f_0(R\, Y)\, dY \nonumber\\
 &&-\theta\, R^{\lambda/2}\, f_0(R\, X)\, X^{\lambda/2}\, \int_0^\infty y^{\lambda/2}\, \varphi(y, t)\, dy
 -(1-\theta)\,X^{-3/2}\int_{0}^\infty \varphi(y, t)y^{\lambda/2}dy \nonumber\\
 &&+R^{\frac{\lambda-1}{2}}X^{-3/2}\int_0^{X/2}\Psi(Y)Y^{\lambda/2}dY.
\eear
Where we have used that:
\bean
x^{-3/2}\int_{x/2}^\infty g(y)y^{\lambda/2}dy & = & x^{-3/2}\int_{0}^\infty g(y)y^{\lambda/2}dy-
x^{-3/2}\int_0^{x/2}g(y)y^{\lambda/2}dy\\
& = & a(t)x^{-3/2}-x^{-3/2}\int_0^{x/2}g(y)y^{\lambda/2}dy.
\eean
By Theorem \ref{S8T3-101} with $\kappa= R^{(\lambda-1)/2}$
\bean
||\Psi||_{L^2_t(0, T_*;H^\sigma(3/4, 5/4))} & \le & C\, ||Q||_{ L^2_t(0, T_*;H_x^\sigma(1/2, 2))}
\eean
We now have:
\bean
||Q||_{H^\sigma(3/4, 5/4)}\le C ||\Psi||_{H^{(\sigma-1)_+}(1/2, 2)}+\sup_{0<t<T_*}|||\varphi(t)|||_{3/2, (3+\lambda/2)}+C\  ||\nu||_{Y^\sigma_{3/2, \, 2+\delta}(T)}
\eean
As before, a bootstrap argument as in the case $R>1$ gives
\bean
||\Psi||_{L^2_t(0, T_*;H^\sigma(3/4, 5/4))} \le C\sup_{0\le t \le T^*}|||\varphi(t)|||_{3/2, (3+\lambda)/2}+C\, ||\nu||_{Y^\sigma_{3/2, \, 2+\delta}(T)}
\eean
and then, rewriting this estimate in the original variables
\bear
\label{S3regularityrmenor}
R^{3/2}\, M_{2;\, \sigma}(\varphi, R)\le 
C\left(\sup_{0<t<T_*}|||\varphi(t)|||_{3/2, (3+\lambda)/2}+ ||\nu||_{Y^\sigma_{3/2, \, 2+\delta}(T)}\right).
\eear
Combined with (\ref{S3regularityrmayor}) we deduce
\bear
\label{S3terrible}
\sup_{0\le t_0\le T^*}\sup_{R>1} \left(R^{(3+\lambda)/2}N_{2;\, \sigma}(\varphi; R, t_0) \right)+
\sup_{0<R\le 1}R^{3/2}\, M_{2;\, \sigma}(\varphi, R) \nonumber \\
\le 
C\left(\sup_{0<t<T_*}|||\varphi(t)|||_{3/2, (3+\lambda)/2}+ ||\nu||_{Y^\sigma_{3/2, \, 2+\delta}(T)} \right),
\eear
and then
\bean
|||\varphi|||\le C\sup_{0<t<T_*}|||\varphi(t)|||_{3/2, (3+\lambda/2)}+C ||\nu||_{Y^\sigma_{3/2, \, 2+\delta}(T^*)} .
\eean
We use now
\bean
\varphi(t)=\theta\int_0^tG(t-s)\, \left(\mathcal L -L \right)(\varphi)(s)\, ds+\int_0^t G(t-s)\, \nu(s)\, ds
\eean
which yields
\bean
|||\varphi(t)|||_{3/2, (3+\lambda)/2}\le C\, \int_0^t |||G(t-s)\, \left(\mathcal L -L \right)(\varphi)(s)|||_{3/2, (3+\lambda)/2}\, ds+\\
+\int_0^t |||G(t-s)\, \nu(s)|||_{3/2, (3+\lambda)/2}\, ds.
\eean
By Lemma \ref{S8T4} and Lemma \ref{lemcor}:
\bean
|||\varphi(t)|||_{3/2, (3+\lambda)/2}\le C\, {T^*}^{\beta}|||\varphi|||+(T^*+{T^*}^{2\delta(\lambda-1)})^\beta ||\nu||_{X_{3/2, \, 2+\delta}(T^*)}
\eean
and the result follows taking $T^*$ small enough.
\qed

\section{Proof of Theorem \ref{MainTheo}.} 
 \setcounter{equation}{0}
\setcounter{theo}{0}

 We introduce the auxiliary operators $L_\varepsilon$, for $\varepsilon>0$, defined as follows:

\bear
\label{S1Eopelepsilon}
L_\varepsilon(g)&=&\int_0^{x/2}\left((x-y)^{-3 /2}-x^{-3
/2} \right)y^{\lambda /2}g(y)dy \nonumber \\ \nonumber \\ 
& + &\int_0^{x/2}\left((x-y)^{\lambda /2}g(x-y)-x^{\lambda /2}g(x)
\right)\frac{dy}{y^{3/2}+\varepsilon^{3/2}\, x^{3/2}}\nonumber \\ \nonumber \\
& - &x^{-3/2}\int_{x/2}^{\infty }y^{\lambda /2}g(y)dy-2\sqrt 2 x^{(\lambda -1)/2}g(x).
\eear
For all $\varepsilon>0$ the operator $L_\varepsilon$ is more regular than $L$. Notice in particular that $g$ and $L_\varepsilon g$ have the same regularity.

\begin{lem}
\label{S4lemmauno}
Let $0\le T \le 1$. Then, there exists a constant $C>0$ such that, for any $\varphi \in  \mathcal E_{T;\sigma}$, for all $t_0\in (0, T)$ and $\varepsilon \ge 0$:
\bear
R^2\, N_{2;\, \sigma-1/2}\left((\mathcal L -L_\varepsilon)( \varphi);\, R,\, t_0 \right)\le C\, ||| \varphi|||,\,\,\,\forall R>1,
\label{S4lemmaunoE1}\\
R^{2-\lambda/2}\, M_{2;\, \sigma-1/2}\left((\mathcal L -L_\varepsilon)( \varphi);\, R \right)\le C\, |||\varphi|||,\,\,\,\forall\,\,\, 0< R<1.
\label{S4lemmaunoE2}
\eear
\end{lem}
\textbf{Proof of Lemma \ref{S4lemmauno}. } Notice that the operator 
$(\mathcal L -L_\varepsilon)$ may be written as $A_1+A_{2, \varepsilon}$ where $A_1$ and $A_{2, \varepsilon}$ are defined in (\ref{S8E15}) and (\ref{S8E15-4bis})-(\ref{S3EHdefbis}). Lemma \ref{S4lemmauno} then follows using Lemma \ref{S3estA1reesc} and Lemma \ref{S8Tprov}. \qed

\begin{lem}
\label{S4Lemma2}
(i) There exists a constant $C>0$ such that,  for all $\varepsilon \in (0, 1]$, $\theta \in [0, 1)$, $\varphi\in \mathcal E_{T; \sigma}$ and $u\in \mathcal E_{T; \sigma}$ satisfying:
\bean
\partial_t \varphi =(1-\theta)\, L(\varphi)+\theta\, \mathcal L (\varphi)+\left(\mathcal L-L_\varepsilon \right)(u)
\eean
there holds:
\bean
|||\varphi|||\le  C\sup_{0\le t \le T^*}|||\varphi|||_{3/2, (3+\lambda)/2}+ \frac{C}{1-\theta}|||u|||.
\eean
\end{lem}
\textbf{Proof of Lemma \ref{S4Lemma2}} The proof of this Lemma is similar to that of Lemma \ref{regularity}. The difference comes from the fact that we must use the regularising effect of the operator $T_{1-\theta, R}$ of Theorem \ref{S8T3-101}. We then start by scaling  the variables. 

In the case $R>1$  and  for $x\in (3R/4, 5R/5)$  we define the new variables:
$x=XR$, $y=YR$, $t=(\tau/ R^{(\lambda-1)/2})$ and $\varphi(x, t)=R^{-(3+\lambda)/2}\, \Psi(X, \tau)$. Since $t\in (0, T_*)$, $\tau\in (0, T_*\,R^{(\lambda-1)/2})$.  The function $\psi(X, \tau)$ satisfies  equations  (\ref{S3EpsiUno})-(\ref{S3EpsiTres}) with $\widetilde \nu _1$ and $\widetilde \nu _2$ are defined as in (\ref{S3nutildeuno}), (\ref{S3nutildedos}) but where $\nu$ is now given by
\bear
\label{S4EDefdenu}
\nu=\left(\mathcal L-L_\varepsilon \right)(u).
\eear
Using Lemma \ref{S4lemmauno} and Theorem \ref{S8T3-101} with $\varepsilon=1-\theta$,  we obtain , arguing as in the proof of (\ref{S3estboots}), 
\bear
\label{S4Lemma2E1}
\sup_{0\le T\le T_*\,R^{{(\lambda-1)/2}}}\left(\int_T^{\min{(T+1, T_*\,R^{{(\lambda-1)/2}} )}}
||\Psi(s)||^2_{H^\sigma_X( {3/4}, {5/4})}ds \right)^{1/2}\le \nonumber \\
\le C \sup_{0\le t \le T^*}|||\varphi|||_{3/2, (3+\lambda)/2}+ \frac{C}{1-\theta}|||u|||
\eear
Notice that the only difference between the proof of (\ref{S4Lemma2E1}) and that of (\ref{S3estboots}) comes from the control of the term $\nu $ defined in (\ref{S4EDefdenu}). However that term is estimated as the term $P$ in (\ref{S8T3-101E1}) with $\kappa=1$, and $\varepsilon=1-\theta$ combined with (\ref{S4lemmaunoE1}).\\
We consider now the range $R\in (0, 1)$ and rescale the equation for $x\in (3R/4, 5R/5)$. The new variables are now
$x=XR$, $y=YR$, $\varphi(x, t)=R^{-3/2}\, \Psi(X, t)$ and $u(x, t)=R^{-3/2}U(X, t)$.  Arguing as above, the function $\Psi$ satisfies now the same equation
(\ref{S3EPsiRmenor}) where the term $Q$ is defined in (\ref{S3Erestoenorme}) where here again $\nu$ is given by (\ref{S4EDefdenu}). The term $R^{3/2}\nu(R, X, t)$ in (\ref{S3Erestoenorme}) is rewritten using (\ref{S1Eopelepsilon}) as follows:
\bean
R^{3/2}\nu(R, X, t) & = & R^{(\lambda-1)/2}\left(\mathcal L-L_\varepsilon\right) (U)\\
& = & \mathcal Q _0(X, t)+\mathcal Q_1(X, t)
\eean
where,
\bean
\mathcal Q _0(X, t)& = & R^{(\lambda-1)/2}\int_0^{X/2}\left((X-Y)^{\lambda /2}U(X-Y)-X^{\lambda /2}U(X)
\right)\frac{dY}{Y^{3/2}+\varepsilon^{3/2}\, X^{3/2}}\\
\mathcal Q_1(X, t)
&=&R^{(\lambda-1)/2}\int_0^{X/2}\left((X-Y)^{-3 /2}-X^{-3
/2} \right)Y^{\lambda /2}U(Y, t)dY \nonumber \\ \nonumber\\
& - &X^{-3/2}\int_{R\, X/2}^{\infty }y^{\lambda /2}u(y, t)dy-2\sqrt 2R^{(\lambda-1)/2} X^{(\lambda -1)/2}U(X).
\eean
and $\mathcal Q _1(X, t)$ satisfies,
\bean
\left|M_{2, \sigma}(\mathcal Q_1; 1) \right|\le C\, |||u|||.
\eean
Using now Theorem \ref{S8T3-101} with $\varepsilon=1-\theta$ and $\kappa =R^{(\lambda-1)/2}$ and estimating all the remaining terms as in the proof of (\ref{S3regularityrmenor})  we obtain
\bear
\label{S4Lemma2E7}
R^{3/2}M_{2, \sigma}(\varphi; R)
\le C \sup_{0\le t \le T^*}|||\varphi|||_{3/2, (3+\lambda)/2}+ \frac{C}{1-\theta}|||u|||.
\eear
Combining (\ref{S4Lemma2E1}) and (\ref{S4Lemma2E7}) the Lemma follows.\qed

\begin{lem}
\label{S4Lematres}
Let $0\le T \le 1$. Then for any $\varphi \in  \mathcal E_{T;\sigma}$, for all $t_0\in (0, T)$ :
\bean
\lim_{\varepsilon \to 0}N_{2;\, \sigma-1/2}\left(( L -L_\varepsilon)( \varphi);\, R,\, t_0 \right) & = & 0,\,\,\,\forall R>1,\\
\lim_{\varepsilon \to 0} M_{2;\, \sigma-1/2}\left(( L -L_\varepsilon)( \varphi);\, R \right) & = & 0,\,\,\,\forall\,\,\,0< R<1.
\eean
\end{lem}
\textbf{Proof of Lemma \ref{S4Lematres}.} After rescaling the variables  $x=R\, X$, $t=t_0+R^{-(\lambda-1)/2}\tau$ and $\varphi(x, t)=\psi(X, \tau)$, the two identities reduce to:
\bear
\lim_{\varepsilon\to 0}\int_0^{\tau^*}||( L -L_\varepsilon)(\psi)(\tau) ||^2_{H_X^\sigma(1/2, 2}\, d\tau =0
\eear
with $0<\tau^* <1$. Using (\ref{S1Eopele}) and (\ref{S1Eopelepsilon}) we have
\bean
( L -L_\varepsilon)(\psi)=(\mathcal W_\infty-\mathcal W_{\infty,\, \varepsilon})(\psi)
\eean
where $\mathcal W_\infty$ and $\mathcal W_{\infty,\, \varepsilon}$ are defined in (\ref{S2Eopelredondainfty}) and (\ref{S2Eopelredondaepsilon}). Therefore the Lemma follows combining (\ref{S2Elimites}) and the Lebesgue convergence Theorem.
\qed
\\ \\
\textbf{End of the proof of Theorem \ref{MainTheo}.}

Our goal is to solve (\ref{S8E10-2}) for $\theta=1$. To this end we use a continuation argument starting at $\theta=0$.

For $\theta=0$  equation (\ref{S8E10-2}) has a solution $\varphi \in \mathcal E_{T; \sigma}$. This is a consequence of the results of \cite{EV} and of  Lemma \ref{regularity} in Section 3 with $\theta=0$.

Then, we define:
\bear
\theta^*=\sup \left\{\theta \ge 0;\, \hbox{for all}\,\,\nu \in Y^\sigma_{3/2, 2+\delta}(T), \hbox{there exists}\,\,\varphi \in \mathcal E_{T; \sigma}\,\,\hbox{solution of}\,\,(\ref{S8E10-2})\,\, \right\}
\eear 
The Lemmas \ref{lemcor} and \ref{regularity} show that there exists a constant $C>0$ such that, for any $\theta<\theta^*$ and for all $\nu \in Y^{\sigma}_{3/2, 2+\delta}(T) $ there exists a function  $\varphi\in \mathcal E_{T; \sigma }$ such that 
\bean
|||\varphi|||\le C  ||\nu||_{Y^\sigma_{3/2, \, 2+\delta}(T)}.
\eean
Suppose that $\theta^*<1$. We will show that for all $\theta>\theta^*$ with $\theta-\theta^*$ sufficiently small and all $\nu\in  Y^\sigma_{3/2, 2+\delta}(T)$ there exists a function $\varphi\in \mathcal E_{T; \sigma}$ and solving (\ref{S8E10-2}). This would give a contradiction. 

To this end we use a fixed point argument.

Given $\widetilde\varphi \in \mathcal E_{T; \sigma}$  and $\nu \in Y^\sigma_{3/2, 2+\delta}(T)$
we define $\varphi_{\varepsilon, n}\in \mathcal E_{T; \sigma}$ as the solution of
\bear
\label{S4E2}
\partial_t \varphi_ {\varepsilon,n} =(1-\theta_n)\, L(\varphi_  {\varepsilon,n})+\theta_n\, \mathcal L (\varphi_ {\varepsilon,n})+(\theta-\theta_n)\left(\mathcal L-L_\varepsilon \right)(\widetilde \varphi)+\nu
\eear
where $\theta_n$ is a sequence such that $\theta_n<\theta^*$, $\theta_n \to \theta^*$ as $n\to +\infty$.
The functions $\varphi_ {\varepsilon,n}$ are well defined since $\theta_n<\theta_*$ and 
$\left(\mathcal L-L_\varepsilon \right)(\widetilde \varphi)\in Y^\sigma_{3/2, 2+\delta}(T)$.
Combining Lemma \ref{regularity}  and Lemma \ref{S4Lemma2} we obtain:
\bear
\label{S4estimateone4}
|||\varphi_{\varepsilon, n}|||\le  C(\theta-\theta_n)\,\sup_{0\le t \le T^*}|||\varphi|||_{3/2, (3+\lambda)/2}+C\, \frac{\theta-\theta_n}{1-\theta_n}|||\widetilde \varphi|||+||\nu||_{Y^\sigma_{3/2, 2+\delta}}.
\eear
Since $\varphi_{\varepsilon, n}$ satisfies equation (\ref{S4E2}) we have:
\bear
\label{S4estimateone2}
\varphi_{\varepsilon, n}=\int_0^tG(t-s)\left[\theta_n(\mathcal L-L)(\varphi_{\varepsilon, n})
+(\theta-\theta_n) (\mathcal L-L_\varepsilon)(\widetilde \varphi)+\nu\right]ds.
\eear
Using now Lemma \ref{S8T4} and Lemma \ref{lemcor} we obtain:
\bear
\label{S4estimateone3}
\sup_{0\le t \le T}||\varphi_{\varepsilon, n}||_{3/2, (3+\lambda)/2}\le C T^{\beta}
\left(|||\varphi_{\varepsilon, n}|||+(\theta-\theta_n)|||\widetilde \varphi|||+||\nu||_{X_{3/2, 2+\delta}} \right)
\eear
Therefore, using (\ref{S4estimateone4}) and (\ref{S4estimateone3}) for $T$ small we obtain
\bear
\label{S4estimateone}
|||\varphi_{\varepsilon, n}|||\le  C\, \frac{\theta-\theta_n}{1-\theta_n}|||\widetilde \varphi|||+||\nu||_{Y^\sigma_{3/2, 2+\delta}}.
\eear

Moreover, given $\widetilde\varphi \in \mathcal E_{T; \sigma}$, $\widetilde\varphi' \in \mathcal E_{T; \sigma}$ and denoting the corresponding solutions as $\varphi_{\varepsilon, n}$ and $\varphi'_{\varepsilon, n}$ a similar argument yields:
\bear
\label{S4estimatedos}
|||\varphi_{\varepsilon, n}-\varphi'_{\varepsilon, n}|||\le C \frac{\theta-\theta_n}{1-\theta_n}
|||\widetilde\varphi-\widetilde\varphi'|||
\eear
By Lemma \ref{S4Lematres} we deduce that
\bean
\lim_{\varepsilon,\,\, \varepsilon' \to 0}N_{2;\, \sigma-1/2}\left(( L_{\varepsilon'} -L_\varepsilon)(\widetilde \varphi);\, R,\, t_0 \right) & = & 0,\,\,\,\forall R>1,\,\,\forall t_0\in (0, T),\\
\lim_{\varepsilon,\,\,\varepsilon' \to 0} M_{2;\, \sigma-1/2}\left(( L_{\varepsilon'} -L_\varepsilon)( \widetilde\varphi);\, R \right) & = & 0,\,\,\,\forall\,\,\,0< R<1,\,\,\forall t_0\in (0, T).
\eean
We use now the regularising effects obtained in Theorem \ref{S8T3-101} combined with the rescaling argument that have already been used in the proof of Lemma \ref{regularity}, to obtain:
\bean
\lim_{\varepsilon,\,\, \varepsilon' \to 0}N_{2;\, \sigma}\left( \varphi_{\epsilon, n}-\varphi_{\varepsilon', n};\, R,\, t_0 \right) & = & 0,\,\,\,\forall R>1,,\,\,\forall t_0\in (0, T),\\
\lim_{\varepsilon,\,\,\varepsilon' \to 0} M_{2;\, \sigma}\left( \varphi_{\epsilon, n}-\varphi_{\varepsilon', n};\, R \right) & = & 0,\,\,\,\forall\,\,\,0< R<1,\,\,\forall t_0\in (0, T)\\
\lim_{\varepsilon,\,\, \varepsilon' \to 0}||\varphi_{\epsilon, n}-\varphi_{\varepsilon', n}||_{L^\infty ([0, T]\times[R/2, 2R])} & = & 0\,\,\forall R>0\,\,\forall t\in (0, T).
\eean
There exists then a function $\varphi_n$ defined in $\RR^+\times [0, T]$ such that
\bear
\lim_{\varepsilon,\,\, \varepsilon' \to 0}N_{2;\, \sigma}\left( \varphi_{\epsilon, n}-\varphi_n;\, R,\, t_0 \right) & = & 0,\,\,\,\forall R>1,,\,\,\forall t_0\in (0, T), \label{S4Convcajasuno}\\
\lim_{\varepsilon,\,\,\varepsilon' \to 0} M_{2;\, \sigma}\left( \varphi_{\epsilon, n}-\varphi_n;\, R \right) & = & 0,\,\,\,\forall\,\,\,0< R<1,\,\,\forall t_0\in (0, T),\label{S4Convcajasdos}\\
\lim_{\varepsilon,\,\, \varepsilon' \to 0}||\varphi_{\epsilon, n}-\varphi_{n}||_{L^\infty ([0, T]\times[R/2, 2R])} & = & 0\,\,\forall R>0,\,\,\forall t\in (0, T).\label{S4Convcajastres}
\eear
By (\ref{S4estimateone}) we have:
 \bean
\label{S4estimateonecajas}
R^{(3+\lambda)/2}\,N_{2;\, \sigma}\left( \varphi_{\epsilon, n};\, R,\, t_0 \right)\le  C\, \frac{\theta-\theta_n}{1-\theta_n}|||\widetilde \varphi|||+C||\nu||_{Y^\sigma_{3/2, 2+\delta}},\,\,\,\forall R>1,,\,\,\forall t_0\in (0, T),\\
R^{3/2}\,M_{2;\, \sigma}\left( \varphi_{\epsilon, n};\, R,\, t_0 \right)\le  C\, \frac{\theta-\theta_n}{1-\theta_n}|||\widetilde \varphi|||+C||\nu||_{Y^\sigma_{3/2, 2+\delta}},\,\,\,\forall \,\,0< R<1,\,\,\forall t_0\in (0, T)\\
\max\left\{x^{3/2}, x^{(3+\lambda)/2}\right\}|\varphi_{\epsilon, n}(x, t)| 
\le  C\, \frac{\theta-\theta_n}{1-\theta_n}|||\widetilde \varphi|||+C||\nu||_{Y^\sigma_{3/2, 2+\delta}},\,\,\forall x\in (R/2, 2R),\nonumber \\
\forall R>0\,\,\forall t\in (0, T).
\eean
Taking limits as $\varepsilon \to 0$:
 \bean
\label{S4estimateonecajas20087}
R^{(3+\lambda)/2}\, N_{2;\, \sigma}\left( \varphi_{n};\, R,\, t_0 \right)\le  C\, \frac{\theta-\theta_n}{1-\theta_n}|||\widetilde \varphi|||+C||\nu||_{Y^\sigma_{3/2, 2+\delta}},\,\,\,\forall R>1,,\,\,\forall t_0\in (0, T),\\
R^{3/2}\, M_{2;\, \sigma}\left( \varphi_{ n};\, R,\, t_0 \right)\le  C\, \frac{\theta-\theta_n}{1-\theta_n}|||\widetilde \varphi|||+C||\nu||_{Y^\sigma_{3/2, 2+\delta}},\,\,\,\forall \,\,0< R<1,\,\,\forall t_0\in (0, T),\\
\max\left\{x^{3/2}, x^{(3+\lambda)/2}\right\}|\varphi_{n}(x, t)| 
\le  C\, \frac{\theta-\theta_n}{1-\theta_n}|||\widetilde \varphi|||+C||\nu||_{Y^\sigma_{3/2, 2+\delta}},\,\,\forall x\in (R/2, 2R),\nonumber \\
\forall R>0\,\,\forall t\in (0, T).
\eean
whence $\varphi_n \in \mathcal E_{T; \sigma}$ and,
\bean
|||\varphi_n|||\le  C\, \frac{\theta-\theta_n}{1-\theta_n}|||\widetilde \varphi|||+||\nu||_{Y^\sigma_{3/2, 2+\delta}}.
\eean
A similar argument yields

\bean
\label{S4contractividad}
|||\varphi_n-\varphi'_n|||\le  C\, \frac{\theta-\theta_n}{1-\theta_n}|||\widetilde \varphi-\widetilde \varphi'|||.
\eean

Notice that $\varphi_n\in L^2(0, T; H_{loc}^\sigma(\RR^+))$. Moreover, passing to the weak limit in the equation (\ref{S4E2}) as $\varepsilon \to 0$ we obtain that $\varphi_n$ solves
\bear
\label{S4E5weak}
\partial_t \varphi_ {n} =(1-\theta_n)\, L(\varphi_  {n})+\theta_n\, \mathcal L (\varphi_ {n})+(\theta-\theta_n)\left(\mathcal L-L \right)(\widetilde \varphi)+\nu
\eear
in the sense of distributions. Then, $\varphi_n\in H^1(0, T; H_{loc}^\sigma(\RR^+))$. 

Formula (\ref{S4contractividad}) implies that the application $\widetilde \varphi \mapsto \varphi_n$ has a fixed point for any $\nu \in Y^\sigma_{3/2, 2+\delta}$,  $n$ sufficiently large and $\theta-\theta^*>0$ sufficiently small. Let us denote by $\varphi$ such a fixed point that satisfies:
\bear
\label{S4E5weakweak}
\partial_t \varphi =(1-\theta_n)\, L(\varphi)+\theta_n\, \mathcal L (\varphi)+(\theta-\theta_n)\left(\mathcal L-L \right)(\varphi)+\nu
\eear
whence,
\bear
\label{S4E5weakweak}
\partial_t \varphi =(1-\theta)\, L(\varphi)+\theta\, \mathcal L (\varphi)+\nu
\eear
and since $\theta > \theta^*$ this yields a contradiction. It then follows that $\theta^*=1$. We prove now the solvability of the equation for $\theta =1$.\\ 
 To this end we consider  a sequence $\theta_k\to 1$ and the corresponding sequence of solutions $\varphi_k\in \mathcal E_{T; \sigma}$to :
 \bear
\label{S4E5weakweakk}
\partial_t \varphi_k =(1-\theta_k)\, L(\varphi_k)+\theta_k\, \mathcal L (\varphi_k)+\nu.
\eear
By Lemma \ref{regularity} we have
\bear
\label{S4estuniffi}
|||\varphi_k|||\le C||\nu||_{Y^\sigma_{3/2, 2+\delta}}.
\eear
Therefore,
\bear
R^{(3+\lambda)/2}\, N_{2;\, \sigma}\left( \varphi_{k};\, R,\, t_0 \right)\le  C||\nu||_{Y^\sigma_{3/2, 2+\delta}},\,\,\,\forall R>1,,\,\,\forall t_0\in (0, T),\label{S4estimateonecajas1}\\
R^{3/2}\, M_{2;\, \sigma}\left( \varphi_{k};\, R,\, t_0 \right)\le  C||\nu||_{Y^\sigma_{3/2, 2+\delta}},\,\,\,\forall \,\,0< R<1,\,\,\forall t_0\in (0, T)\label{S4estimateonecajas2}\\
\max\left\{x^{3/2}, x^{(3+\lambda)/2}\right\}|\varphi_{k}(x, t)| 
\le  C||\nu||_{Y^\sigma_{3/2, 2+\delta}},\,\,\forall x\in (R/2, 2R),\nonumber \\
\forall R>0\,\,\forall t\in (0, T).\label{S4estimateonecajas3}
\eear
The sequence $\{\varphi_k\}_{k\in \NN}$ is then weakly compact in $L^2(t_0, t_0+R^{-{(\lambda-1)/2}, T}; H^\sigma (R/2, 2R))$ for all $R>0$ and $t_0\in (0, T]$. Therefore, using a diagonal procedure,  there exists a sub sequence, still denoted $\{\varphi_k\}_{k\in \NN}$, and a function $\varphi$ defined in all $\RR^+\times (0, T]$ such that $\varphi_k$ converges to $\varphi$ weakly in $L^2((0, T); H^\sigma (R_1, R_2))$ for all $R_2>R_1>0$. Since the left hand sides in the inequalities (\ref{S4estimateonecajas1})-(\ref{S4estimateonecajas3}) are all of them convex functions of $\varphi_k$, these inequalities are preserved under weak limits.
Therefore
\bear
\label{S4estimateonecajaskaes}
&&R^{(3+\lambda)/2}\, N_{2;\, \sigma}\left( \varphi;\, R,\, t_0 \right)\le  C||\nu||_{Y^\sigma_{3/2, 2+\delta}},\,\,\,\forall R>1,\,\,\hbox{for a. e. }\, t_0\in (0, T),\\
&&R^{3/2}\, M_{2;\, \sigma}\left( \varphi;\, R,\, t_0 \right)\le  C||\nu||_{Y^\sigma_{3/2, 2+\delta}},\,\,\,\forall \,\,0< R<1,\,\,
\hbox{for a. e. }\, t_0 \in (0, T)\\
&&\max\left\{x^{3/2}, x^{(3+\lambda)/2}\right\}|\varphi(x, t)| 
\le  C||\nu||_{Y^\sigma_{3/2, 2+\delta}},\,\,\forall x\in (R/2, 2R), \nonumber \\
&&\hskip 7.5cm \forall R>0,\,\,\hbox{for a. e. }\, t\in (0, T)
\eear
whence $\varphi\in \mathcal E_{T; \sigma}$. 

 On the other hand, it is possible to pass to the limit in the equation (\ref{S4E5weakweakk}) in the weak sense of $L^2(0, T; H^\sigma (R_1, R_2))$ for any $R_2>R_1>0$ to obtain that $\varphi\in 
L^2(0, T; H_{loc}^\sigma (\RR^+))\cap H^1(0, T; H_{loc}^{\sigma-1/2} (\RR^+))$ is a solution of 
\bear
\label{S4E5weakweakks}
\partial_t \varphi =\mathcal L (\varphi)+\nu.
\eear
in the sense of distributions.

Finally, in order to prove uniqueness let us assume that $\varphi_1$ and $\varphi_2$ are two solutions of
(\ref{S4E5weakweakks}). Then, the function $\psi= \varphi_1-\varphi_2$ satisfies,
\bean
\label{S4Euniqu}
\partial_t \psi =\mathcal L (\psi)\\
\psi(x, 0)=0.
\eean
and Lemma \ref{regularity} for $\theta=1$ and $\nu=0$ shows that $\psi=0$ and uniqueness holds.\qed

\section{Proof of Theorem \ref{Main2Theo}}
\setcounter{equation}{0}
\setcounter{theo}{0}
Consider the function $F_{R, t_0}(X, \tau)$ defined in (\ref{S2lafuncionErre}). The function $\Psi(X, \tau)=R^{(3+\lambda)/2}F_{R, t_0}(X, \tau)$ satisfies equation (\ref{S3EpsiDos}) with $\theta=1$. Then, using (\ref{S8T3-101E143bisbis}) we obtain

\bean
R^{(3+\lambda)/2}
\left(\int_{t_0}^{\min(t_0+R^{-(\lambda-1)/2}, T)}
 \int_\RR |\widehat F_{R, t_0}(k, \tau)|^2\, |k|^{2\,\sigma}
\min \{|k|,\, R \}\, dk, dt
\right)^{1/2}\\
\le C\left(|||\varphi|||+||\nu||_{Y_{3/2, 2+\delta}} \right)
\eean
whence Theorem \ref{Main2Theo} follows.\qed

\noindent
\textit{Acknowledgements.} JJLV is supported by Grant MTM2007-61755. He thanks Universidad Complutense for its hospitality. ME is supported by Grants MTM2008-03541 and IT-305-07.



\footnotesize

\begin{thebibliography}{AAA}

\bibitem{BZ} A. M. Balk \& V. E. Zakharov, {\it Stability of Weak-Turbulence
Kolmogorov Spectra\/} in Nonlinear Waves and Weak Turbulence, V. E. Zakharov ed.,  A. M. S.
Translations Series 2, Vol. 182, 1998, 1-81.

\bibitem{D} L. Desvillettes, {\it About the regularizing properties of the non-cut-off Kac equation\/},  Comm. Math. Phys. 168, 2 (1995), 417Ð440. 

\bibitem{EZH} M. H. Ernst, R. M. Ziff \&  E. M. Hendriks, {Coagulation processes with a phase transition}, J. of Colloid and Interface Sci. {\bf 97}, (1984), 266--277.

\bibitem {EV} M. Escobedo \& J. J. L. Velazquez, {\it On the Fundamental Solution of a Homogeneous Linearized Coagulation Equation.} Preprint.

\bibitem{GT} D. Gilbarg \& N. S. Trudinger, Elliptic partial differential equations of second order. 2nd Edition. Springer 1983.

\bibitem{LU} O. A. Ladyzhenskaja and N. N. Ural'tseva, Linear and quasilinear ellitic equations. Academic Press 1968.

\bibitem{Le} F. Leyvraz, {\it Scaling Theory and Exactly Solved 
Models in the Kintetics of Irreversible Aggregation},  Phys. Reports
\textbf{383} (2003) Issues 2-3, 95--212.


\bibitem{McL} J. B. McLeod, {\it On the scalar transport equation} Proc. London Math. Soc. \textbf{14},  N.3, (1964), pp 445-458.



\bibitem{MP1} G. Menon \& R. L. Pego, {\it Approach to self-similarity in Smoluchowski's coagulation equations} Comm. Pure. Appl. Math.  \textbf{57}, (2004), pp. 1197-1232.

\bibitem{MP2} G. Menon \& R. L. Pego, {\it Dynamical scaling in Smoluchowski's coagulation equation: uniform convergence} SIAM J. Math. Anal. \textbf{36}, (2005), pp. 1629-1651.


\bibitem{vDE} P. G. J. van Dongen \& M.H. Ernst,  {\it Cluster size
distribution in irreversible aggregation at large times}
J. Phys. A \textbf{18} (1985) 2779--2793.

\bibitem{V} C. Villani,  {\it A review of mathematical topics in 
collisional kinetic theory\/}, in  Handbook of 
Mathematical Fluid Dynamics (Vol. 1), S. Friedlander and 
D. Serre ed., Elsevier Science (2002). 



\bibitem{vW} W. von Wahl, {\it The Continuity or stability method for nonlinear eliptic and parabolic equations and systems.\/} Milan J. of Math.  \textbf{62}, N.1, (1992), pp. 157-183.



\end{thebibliography}
\end{document}